\documentclass{article}

\usepackage[english]{babel}

\usepackage[a4paper,top=2cm,bottom=2cm,left=3cm,right=3cm,marginparwidth=2cm]{geometry}

\usepackage{setspace}
\usepackage{amsmath}
\usepackage{graphicx}
\usepackage{setspace}
\usepackage{multirow}
\usepackage{longtable}
\usepackage[english]{babel}
\usepackage[utf8]{inputenc}
\usepackage{algorithm}
\usepackage[noend]{algpseudocode}
\usepackage{csquotes}
\usepackage{subcaption}
\usepackage{graphicx}
\usepackage{tikz}
\usepackage{lipsum}
\usepackage{tcolorbox}
\usepackage{tabularx}
\usepackage{booktabs}
\usepackage{listings}
\usepackage{titlesec}
\usepackage{longtable}
\usepackage{rotating}
\usepackage{pdflscape}
\usepackage{rotating}
\newlength{\RoundedBoxWidth}
\newsavebox{\GrayRoundedBox}
   {\setlength{\RoundedBoxWidth}{\dimexpr#1}
    \begin{lrbox}{\GrayRoundedBox}
       \begin{minipage}{\RoundedBoxWidth}}%
   {   \end{minipage}
    \end{lrbox}
    \begin{center}
    \begin{tikzpicture}%
       \draw node[draw=black,fill=black!10,rounded corners,%
             inner sep=2ex,text width=\RoundedBoxWidth]%
             {\usebox{\GrayRoundedBox}};
    \end{tikzpicture}
    \end{center}}

\usepackage{placeins} 
\newcolumntype{Y}{>{\raggedleft\arraybackslash}X}% raggedleft column X

\usepackage[colorlinks=true, allcolors=blue]{hyperref}

\usepackage{amsfonts}
\usepackage[citestyle=bwl-FU]{biblatex}
\graphicspath{ {./graphs/} }

\newcounter{noteMCctr} 

\newcounter{noteYZctr} 

\title{\textbf{Dynamic Time Warping for Lead-Lag Relationships in Lagged Multi-Factor Models}}

\author{Yichi Zhang$^{1,4,\footnote{Corresponding author; Email: yichi.zhang@stats.ox.ac.uk; The remaining authors are listed in alphabetical order.}}$, Mihai Cucuringu$^{1,2,4,7}$, Alexander Y. Shestopaloff$^{5,6}$, Stefan Zohren$^{3,4,7}$\\}

\date{
\textit{$^{1}$Department of Statistics, University of Oxford\\
        $^{2}$Mathematical Institute, University of Oxford\\
        $^{3}$Department of Engineering, University of Oxford\\
        $^{4}$Oxford-Man Institute of Quantitative Finance, University of Oxford\\
        $^{5}$School of Mathematical Sciences, Queen Mary University of London\\
        $^{6}$Department of Mathematics and Statistics, Memorial University of Newfoundland\\
        $^{7}$The Alan Turing Institute, London\\}
        }

\begin{document}
\maketitle
\centerline{\today}
\vspace{1cm}
\begin{abstract}
In multivariate time series systems, lead-lag relationships reveal dependencies between time series when they are shifted in time relative to each other. Uncovering such relationships is valuable in downstream tasks, such as control, forecasting, and clustering. By understanding the temporal dependencies between different time series, one can better comprehend the complex interactions and patterns within the system.
We develop a cluster-driven methodology based on dynamic time warping for robust detection of lead-lag relationships in lagged multi-factor models. We establish connections to the multireference alignment problem for both the homogeneous and heterogeneous settings. 
Since multivariate time series are ubiquitous in a wide range of domains, we demonstrate that our algorithm is able to robustly detect lead-lag relationships in financial markets, which can be subsequently leveraged in trading strategies with significant economic benefits. 

    
\end{abstract} 

\newpage

\tableofcontents
\newpage

\section{INTRODUCTION}

%\subsection{Background}

Natural physical systems often produce high-dimensional, nonlinear time series data, which are prevalent across various domains. Numerous contributions have been made to analyze such time series from different perspectives [\cite{cao2020neural, cartea2018enhancing, cont2001empirical, cui2021state, drinkall2022forecasting,lu2022trade, lu2023co,sokolov2022assessing,vuletic2023fin,qi2023correlation}]. %\MC{Please update all list references to 
% \cite{cao2020neural,cartea2018enhancing,cont2001empirical} so multiple citations inside the same  "cite" command}
%
For example, \cite{cont2001empirical} explored financial time series and emphasized various statistical properties such as distributional characteristics, tail properties, and extreme fluctuations.

High-dimensional time series can offer valuable insights through the discovery of latent structures, such as lead-lag relationships. These relationships are commonly observed and play a significant role in the field of finance [\cite{albers2021fragmentation,bennett2022lead,buccheri2021high,ito2020direct,li2021dynamic,miori2022returns,tolikas2018lead,yao2020time,zhang2023robust}], the environment \cite{de2021detecting,wu2010detecting}, and biology \cite{runge2019detecting}. For instance, \cite{bennett2022lead} created a directed network to capture pairwise lead-lag relationships among equity prices in the US market. The analysis revealed clusters with significant directed flow imbalance.
 
 Dynamic time warping (DTW) is an algorithm to quantify similarities between two time series, even when they exhibit variations in speed, enabling the calculation of an optimal alignment between them [\cite{bellman1959adaptive,berndt1994using,keogh2005exact,senin2008dynamic}]. The versatility of DTW is evident in its application to diverse domains, such as financial markets [\cite{gupta2020examining,howard2022lead,ito2020direct,stubinger2022using}], bioinformatics [\cite{aach2001aligning,gavrila1995towards}], and robotics \cite{schmill1999learned}. For example, \cite{ito2020direct} introduced Multinomial Dynamic Time Warping (MDTW) to explore lead-lag relationships in FX market data. \cite{howard2022lead} applied DTW and found that global market price discovery oscillates between S\&P 500 futures and FTSE 100 futures. Despite the widely recognized importance and potentially high impact of the problem, limited progress has been made in robustly using DTW for the inference of lead-lag relationships in lagged multi-factor models.

\begin{tcolorbox}

\textbf{Summary of main contributions}.
\begin{enumerate}
\item We introduce a computationally scalable framework for lead-lag detection in high-dimensional time series based on DTW, with clustering used as a denoising step. 
\item We show it is capable of reliably detecting lead-lag relationships in a variety of factor model-based simulated high-dimensional time series. 
\item In financial markets, we leverage the detected lead-lag relationships to construct a profitable trading strategy and demonstrate that our algorithm outperforms the benchmark in most cases. 
\item Our algorithm is also faster than the benchmark in terms of computing time by a factor of 10.
\end{enumerate}

\end{tcolorbox}

\textbf{Paper outline}. This paper is organized as follows. Section \ref{sec:dtw} and Section \ref{sec: model} introduce the definition of DTW and the {\em lagged multi-factor model}. Section \ref{sec: methodology} describes our proposed DTW algorithm. In Section \ref{sec: synthetic}, we validate our algorithm on synthetic data sets from the lagged multi-factor model, and explore lead-lag relationships in the US equity, ETF, and futures markets in Section \ref{sec: financial}. We then move on to a robustness analysis in Section \ref{sec: robustness}. Finally, we summarize our findings, and discuss possible future research directions in Section \ref{sec: conclusion}.

% \MC{need to emphasize here the lead-lag literature to which we are closest... and in general financial application, as this is a finance venue}
% \MC{also need to make it clear what gap(s) in the literature we are aiming to fill}
% \MC{need some citations in this paragraph to DTW. Actually I see there is not a single citation in this whole section.}
\vspace{-1mm}

\section{DYNAMIC TIME WARPING}
\label{sec:dtw}

\vspace{-1mm}

In this section, we introduce the DTW algorithm. Suppose we have two time series, denoted as $A$ and $B$ with lengths $n$ and $m$, respectively, shown in Figure \ref{tab:step} Step 1.  %\MC{could try to push that figure on the first page if we can and it fits.. always nice to have a figure on page 1, more visible. No worries if it won't fit - I moved it earlier but doesn't seem to fit given the current other stuff in Sec 1}
\vspace{-0.01cm}
\begin{equation}
\begin{aligned}
A = a_1, a_2, \ldots, a_i, \ldots, a_n \\
B = b_1, b_2, \ldots, b_j, \ldots, b_m 
\end{aligned}
\end{equation}

In order to align two time series utilizing DTW, the process involves constructing an $n \times m$ matrix, where the $\left(i^{\text {th }}, j^{\text {th }}\right)$ element of the matrix contains the Euclidean distance $d\left(a_i, b_j\right)$ between the two points $a_i$ and $b_j$ from $A$ and $B$, respectively, i.e. $d\left(a_i, b_j\right)=\left(a_i-b_j\right)^2$. Each matrix element $(i, j)$ corresponds to the alignment between the points $a_i$ and $b_j$. Figure \ref{tab:step} Step 2 illustrates an example of the distance matrix.

A warping path, denoted as $W$, represents a consecutive set of matrix elements that capture the mapping between time series $A$ and $B$. The index of the DTW can be expressed as follows
\vspace{-0.01cm}
\begin{equation}
W= \{w_1, w_2, \ldots, w_k, \ldots, w_{\mathrm{K}} \} \quad \max (m, n) \leq \mathrm{K}<m+n-1
\end{equation}

\noindent
where the $k^{\text {th }}$ element of $W$ is denoted as $w_k=(i, j)_k$.

The warping path must adhere to several constraints:

\begin{itemize}
\item \textbf{Boundary}: The warping path starts at $w_1 = (1, 1)$ and ends at $w_K = (m, n)$. This ensures that the warping path initiates from the bottom-left corner cell and terminates at the top-right corner cell of the matrix.

\item \textbf{Continuity}: For a given $w_k = (x, y)$, the preceding element $w_{k-1} = (x', y')$, where $x - x' \leq 1$ and $y - y' \leq 1$. This restriction allows only adjacent cell transitions within the warping path.

\item \textbf{Monotonicity}: For a given $w_k = (x, y)$, the preceding element $w_{k-1} = (x', y')$, where $x - x' \geq 0$ and $y - y' \geq 0$. This constraint ensures that the points in $W$ exhibit monotonically increasing indices over time.

\end{itemize}

\begin{table}[htbp]
  \centering
  \vspace{-2mm}
    \begin{tabular}{p{6cm}p{6cm}}

      \includegraphics[width=\linewidth,trim=0cm 0cm 0cm 0cm,clip]{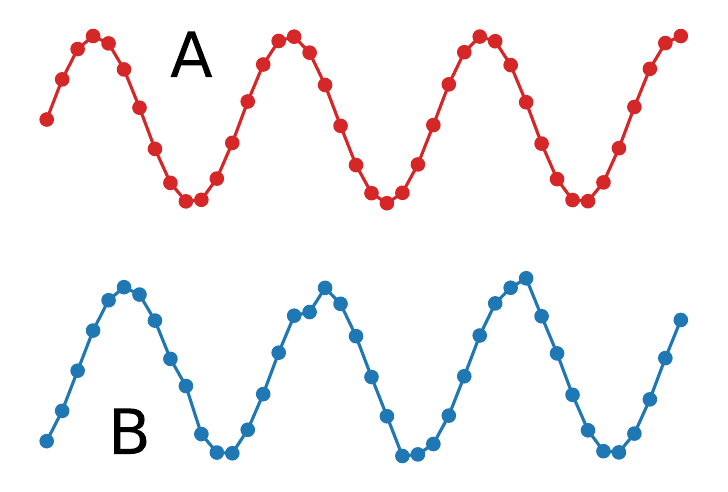}
    \captionof{figure}*{Step 1}
    \vspace{0.5cm}
    \includegraphics[width=\linewidth,trim=0cm 0cm 0cm 0cm,clip]{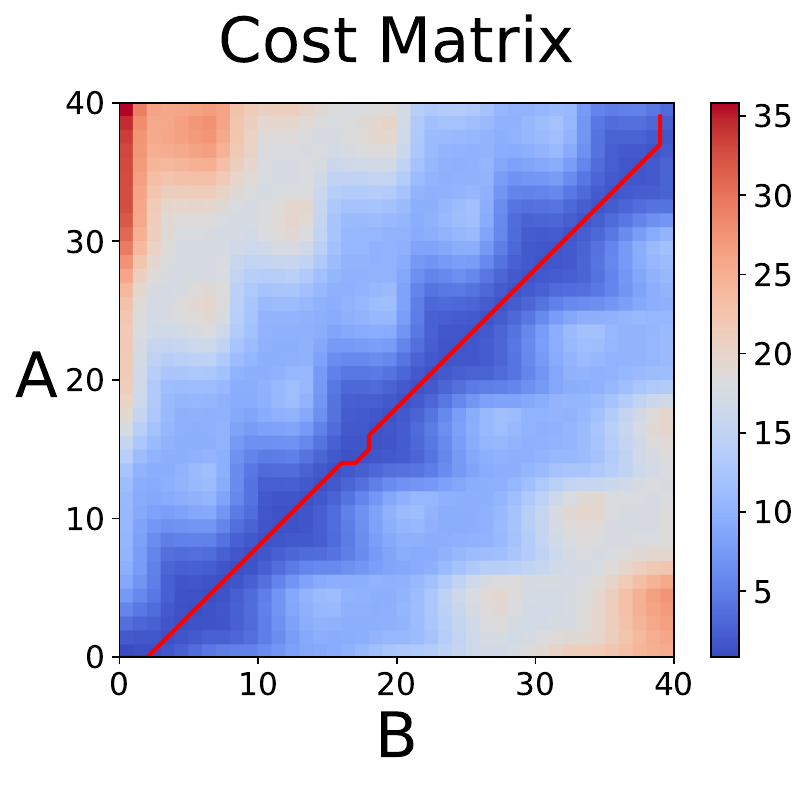}
    \vspace{-0.8cm}
    \captionof{figure}*{Step 3}
%\vspace{-2cm}
    &
        \includegraphics[width=\linewidth,trim=0cm 0cm 0cm 0cm,clip]{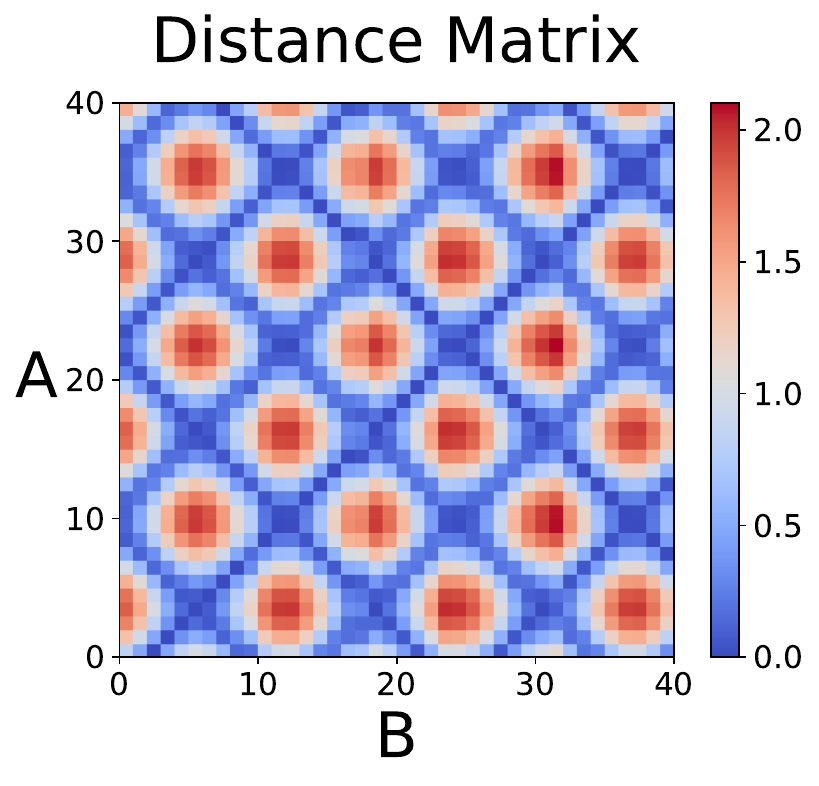}
        \captionof{figure}*{Step 2}
        \vspace{0.5cm}
      \includegraphics[width=\linewidth,trim=0cm 0cm 0cm 0cm,clip]{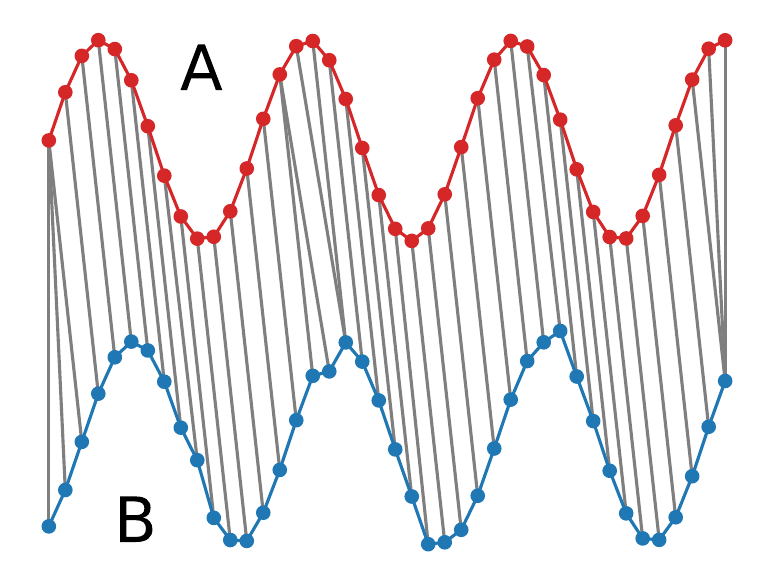}
      \vspace{0.5cm}
     \captionof{figure}*{Step 4}
%\vspace{-2cm}
    \\
    
    \end{tabular}
\vspace{-7mm}
\captionof{figure}{Step 1: Two time series $A$ and $B$ are similar, but they are out of phase with each other. Step 2: The distance matrix, which the $(i^{\text {th }}, j^{\text {th }})$ element of the matrix has the value of the Euclidean distance $d\left(a_i, b_j\right)$ from $A$ and $B$, respectively. Step 3: The cost matrix, which contains the optimal warping path, is visually indicated by the red line. Step 4: The resulting alignment between the two time series, $A$ and $B$, is shown.}
\label{tab:step}
\end{table}

Among the various warping paths that adhere to the aforementioned constraints, our objective is to identify the path that minimizes the warping cost

%\vspace{-5mm}
\begin{equation}
\operatorname{DTW}(A, B)=\min \left\{\sqrt{\sum_{k=1}^{\mathrm{K}} w_k} \right..
\end{equation}
\vspace{-3mm}

To find this optimal path, dynamic programming techniques are employed to evaluate the following subsequent recurrence relation, which defines the cumulative distance $c(i, j)$ as the distance $d(i, j)$ located in the current cell as well as the minimum of the cumulative distances of the adjacent elements
%\vspace{-0.01cm}
\begin{equation}
c(i, j)=d\left(a_i, b_j\right)+\min \{c(i-1, j-1), c(i-1, j), c(i, j-1)\}
\end{equation}

In Figure \ref{tab:step} Step 3, the optimal path is depicted by a red line. The resulting alignment can be seen in Figure \ref{tab:step} Step 4. In conclusion, the complete DTW algorithm is illustrated below

\begin{algorithm}[htp]
\caption{Dynamic Time Warping (DTW)}
\label{dtw_algo}
%\begin{algorithmic}[1]

\textbf{Input:} Time series $A$ with length $n$, time series $B$ with length $m$.
\begin{algorithmic}[1]
\Procedure{DTW}{$A:$ array $[1..n], B:$ array $[1..m]$} 
\State $DTW \gets$ array $[0..n, 0..m]$
\For{$i \gets 0$ \textbf{to} $n$}
    \For{$j \gets 0$ \textbf{to} $m$}
        \State $DTW[i, j] \gets \infty$
    \EndFor
\EndFor
\State $DTW[0, 0] \gets 0$

\For{$i \gets 1$ \textbf{to} $n$}
    \For{$j \gets 1$ \textbf{to} $m$}
        \State $cost \gets d(A[i], B[j])$
        \State $DTW[i, j] \gets cost + \min(DTW[i-1, j],$
        \State \hspace{1.8cm}$DTW[i, j-1], DTW[i-1, j-1])$
    \EndFor
\EndFor

\State \textbf{return} $DTW[n, m]$
\EndProcedure
\end{algorithmic}
\end{algorithm}

Note that the Euclidean distance between two time series can be considered as a specific case of DTW, where the warping path $W$ is constrained such that $w_k = (i, j)_k$ with $i = j = k$. In other words, the window size $S$ is set to 0. This constraint is applicable only when the two time series have the same length. A visual comparison of the Euclidean distance and DTW is presented in Figure \ref{tab:euclid_vs_dtw}.

\begin{table}[htbp]
  \centering
    \begin{tabular}{p{7cm}|p{7cm}}

      \includegraphics[width=\linewidth,trim=0cm 8cm 0cm 8cm,clip]{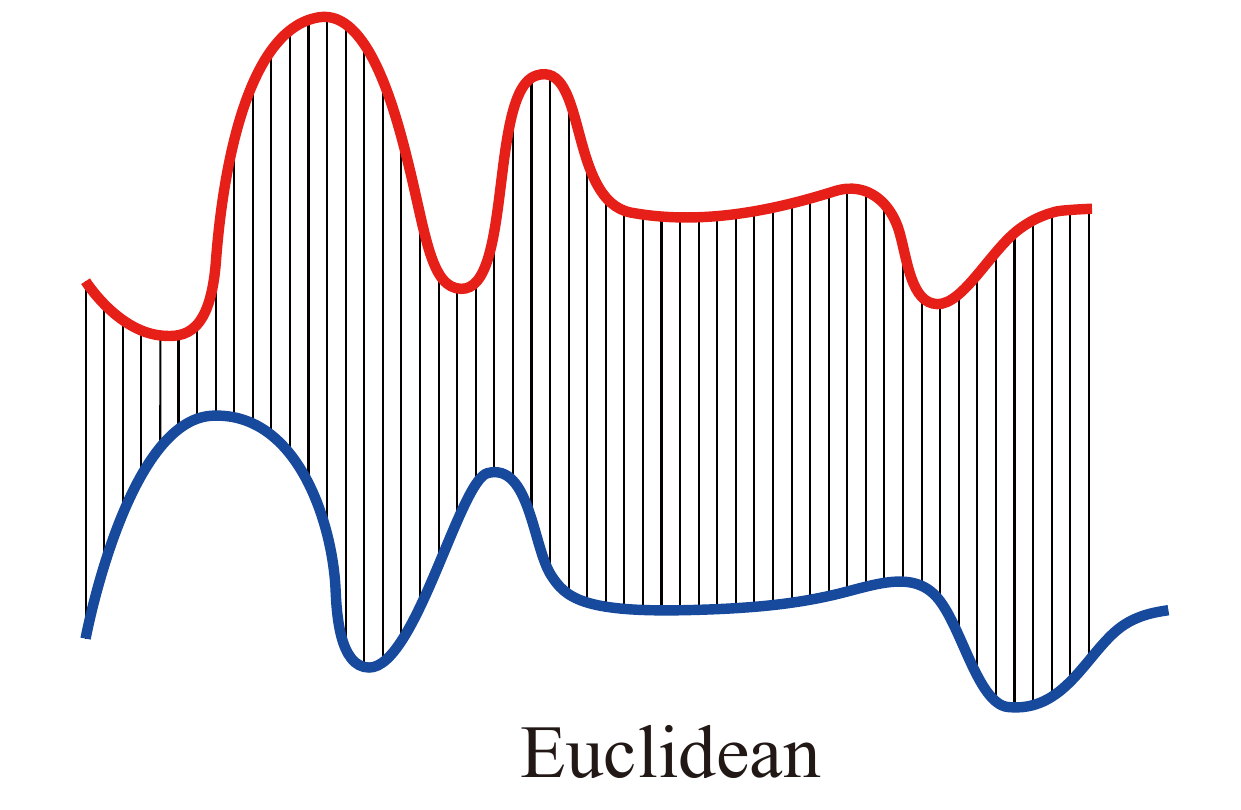}
\vspace{-2cm}
    &
      \includegraphics[width=\linewidth,trim=0cm 8cm 0cm 8cm,clip]{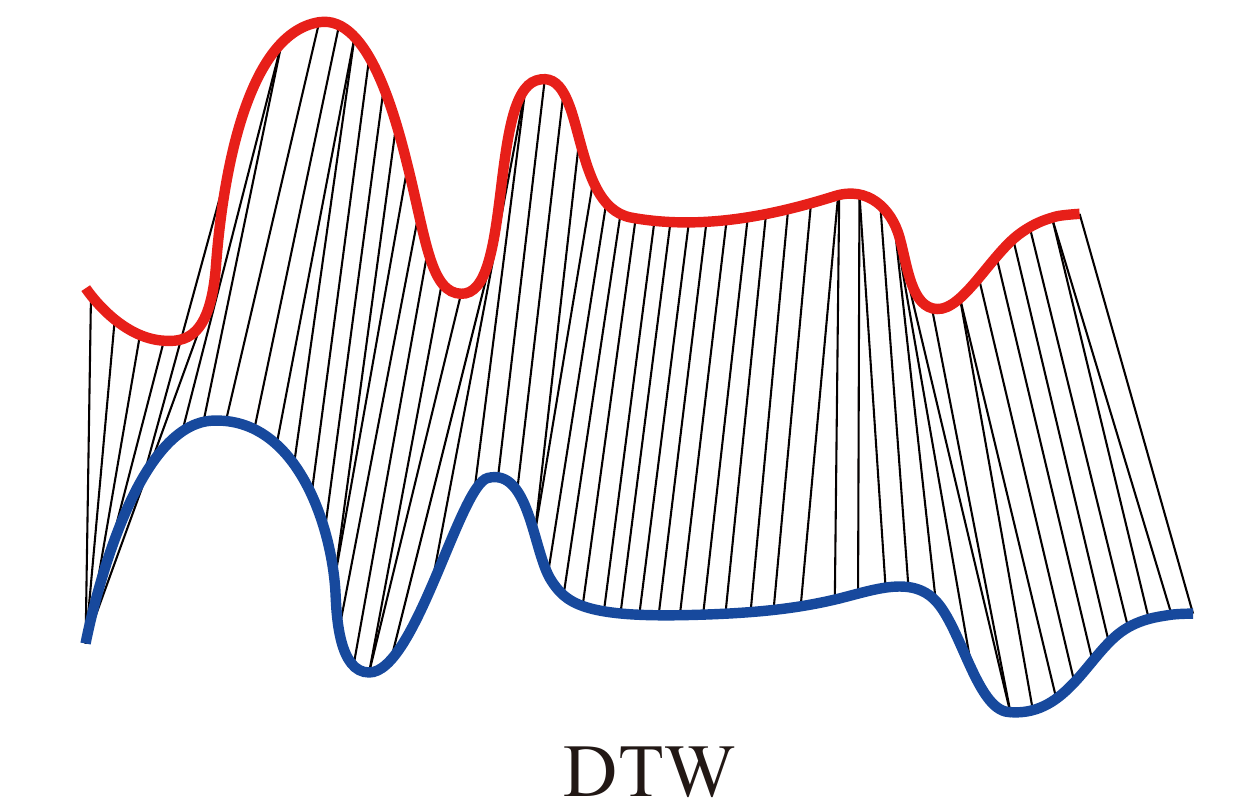}
\vspace{-2cm}
    \\
    \end{tabular}
\captionof{figure}{Left: Euclidean distance measures direct the $i^{\text {th}}$ point in one time series aligns with the $i^{\text {th}}$ point in another time series, assuming the same length, which will yield a pessimistic dissimilarity measure. Right: DTW accounts for variations in speed and calculates optimal alignment between two time series, allowing for different lengths, which is more flexible in calculating the intuitive distance measure.}%\MC{these figures are nice! I think we discussed this before - did you create them from scratch? otherwise we need to cite the source.}
\label{tab:euclid_vs_dtw}
\end{table}

\section{Model setup} \label{sec: model}

In this section, we will introduce both the standard and lagged versions of the {\em multi-factor model}, which we will adopt as the underlying model for our time series data. Specifically, we will employ the lagged multi-factor model to validate our algorithm using synthetic data before applying it to real-world scenarios. The fundamental idea behind these models is to represent a time series as a (noisy) combination of factors, with each factor exhibiting different levels of exposure. The model we propose for the detection of lead-lag relationships exhibits significant parallels with the problem of multireference alignment (MRA), which is concerned with the estimation of a single signal from a set of $n$ cyclically and noisily shifted copies of itself, as shown in \cite{bandeira2014multireference}.

\subsection{Description}
Let us begin by revisiting the standard multi-factor model for a multivariate time series
%\vspace{-0.01cm}
\begin{equation} \label{eq:multi-factor_model}
X_{i}^{t} = \sum_{j=1}^{k} B_{ij}f_{j}^{t} + \epsilon_{i}^{t} \hspace{0.4cm} i = 1,\ldots,n;\quad t = 1,\ldots,T,
\end{equation}

\noindent
where $X_{i}^{t}$ is the time series $i$ (e.g., the excess return of a financial asset) at time $t$, $k$ is the number of factors, $B_{ij}$ is the exposure of time series $i$ to factor $j$, $f_{j}^{t}$ is the factor $j$ at time $t$, and $\epsilon_{i}^{t}$ is the noise at time $t$, with variance $\sigma^{2}$.  Furthermore, we have $n$ as the total number of time series, and $T$ as the total number of time steps.

In this paper, our emphasis is on the lagged version of the multi-factor model, which can be expressed as follows
%\vspace{-0.02cm}
\begin{equation} \label{eq:lagged_multi-factor_model}
X_{i}^{t} = \sum_{j=1}^{k} B_{ij}f_{j}^{t-L_{ij}} + \epsilon_{i}^{t} \hspace{0.4cm} i = 1,\ldots,n;\quad t = 1,\ldots,T, 
\end{equation}

\noindent
where the primary distinction in the lagged multi-factor model, in comparison to the standard multi-factor model, is the inclusion of $L_{ij}$, representing the lag at which time series $i$ is exposed to factor $j$. Consequently, $f_{j}^{t-L_{ij}}$ corresponds to the value of factor $j$ at time $t-L_{ij}$.

In the lagged multi-factor model (\ref{eq:lagged_multi-factor_model}), we consider %\MC{maybe here, and elsewhere where needed, we could not say "introduce" as we introduced this in our first paper.. but more like "consider".. and then say smth "in line with the models introduced in \cite{our_first_paper}" } 
two key settings, which are first introduced in \cite{zhang2023robust}

\begin{itemize}
  \item \textbf{Single Membership:} Each time series has a lagged exposure to a single factor. We consider the following two main categories. 
  \begin{itemize}
    \item \textbf{Homogeneous Setting:} The model only has one factor, i.e. $k=1$.
    
    \item \textbf{Heterogeneous Setting:} The model has more than one factor, i.e. $k \geq 2$. However, each time series is exposed only to a single factor.
  \end{itemize}
  \item \textbf{Mixed Membership:} Each time series is permitted to have a lagged exposure to more than one factor, resulting in a mixed configuration. Therefore, the model comprises at least two factors, indicated by $k \geq 2$.
  
\end{itemize}

In this paper, our primary objective is to perform inference on the lag values $L_{ij}$ in the lagged multi-factor model, specifically focusing on the single membership setting. We do not place emphasis on the inference of the unknown coefficient matrix $B$ and factors $f$. As shown later, the estimation of $L_{ij}$ alone holds practical significance in specific applications, such as finance. The investigation of the mixed membership setting is left 
for future work.
% research endeavors.

\section{Methodology}
\label{sec: methodology}

In this section, we propose a robust algorithm for detecting lead-lag relationships using the combination of DTW and K-Medoids (DTW\_KMed), and use the DTW library from \cite{meert_wannes_2020_7158824}.

We consider a set of time series denoted as $X_{n \times T}$ as our input. Initially, we employ DTW to compute pairwise distances between every pair of time series from $X_{n \times T}$. Subsequently, we apply K-Medoids clustering to group similar time series into clusters based on the DTW distance matrix. The pairs of time series $i$ and $j$ are denoted as $\{X_{i}, X_{j}\}$. For each cluster $\phi_{d}$ ($d = 1,\ldots, K$), we record the path $W\{X_{i}, X_{j}\}$ by performing DTW on $\{X_{i}, X_{j}\}$. Then, we calculate the difference for each index pair in the path $W$, denoted as $\Delta(W\{X_{i}, X_{j}\})$. The calculation can be expressed as follows
%\MC{there is some strange white space gap between text and equations, throughout the paper.. we should check and fix this at the end}
%\vspace{-0.01cm}
\begin{equation}
\Delta(W\{X_{i}, X_{j}\}) = \{\Delta(w_1), \Delta(w_2), \ldots, \Delta(w_k), \ldots, \Delta(w_K)\} 
\end{equation}

\noindent
where $\max (m, n) \leq \mathrm{K}<m+n-1$, and $\Delta(w_K) = \Delta((i,j)_k)=i-j$.

Then, the value of relative lags of $\{X_{i}, X_{j}\}$ in $\phi_{d}$ (estimated by mode or median) can be expressed as
\vspace{-0.01cm}

\begin{equation}
\gamma\{X_{i}, X_{j}\} = \left\{
     \begin{array}{@{}l@{\thinspace}l}
        \text{Mode}(\Delta(W\{X_{i}, X_{j}\})) \hspace{0.6cm} \text{Mode
estimation} \\
        \text{Median}(\Delta(W\{X_{i}, X_{j}\})) \hspace{0.35cm} \text{Median
estimation}\\
     \end{array}
   \right.
\end{equation}

\begin{figure}[!htbp]
\centering
\includegraphics[width=0.6\textwidth,trim=0cm 0cm 0cm 0cm,clip]{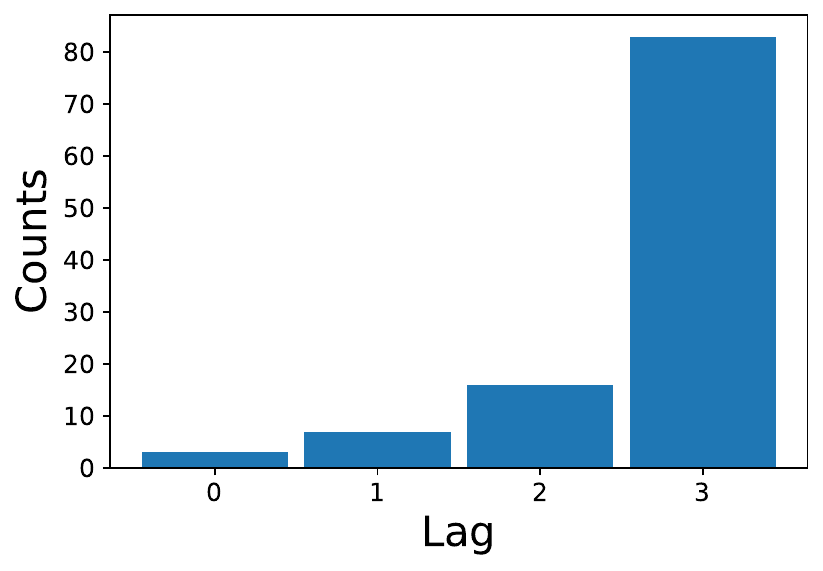}

\captionof{figure}{Histogram of the relative lags from two time series.}
\label{fig:dtw_hist_lags}
\end{figure}

For instance, once we have applied K-medoids clustering using the DTW distance matrix computed from $X_{n \times T}$, let's consider two time series, $X_1$ and $X_2$, that contain $100$ data points and belong to the same cluster, with a known ground truth lag value of $3$. Subsequently, we calculate the relative lags of $\Delta(W\{X_{1}, X_{2}\})$. Figure \ref{fig:dtw_hist_lags} illustrates the relative lags obtained from $\{X_{1}, X_{2}\}$.

% \begin{table}[htbp]
%   \centering
%   %\resizebox{\linewidth}{!}{
%   \caption{The relative lags from two time series.}
%     \begin{tabular}{p{2cm}||p{1cm}p{1cm}p{1cm}p{1cm}}
%         \toprule
%             \textbf{Lag} & 0& 1 & 2 & 3 \\
    
%             \textbf{Counts} &2&13&24&75\\
%         \bottomrule
%     \end{tabular}
% \label{tab: dtw_lags}
% \end{table}

By employing mode or median estimation of $\Delta(W\{X_{1}, X_{2}\})$, denoted as $\gamma\{X_{1}, X_{2}\}$, with a value of 3, as evident from Figure \ref{fig:dtw_hist_lags}, we can conclude that the result remains consistent and robust with respect to the ground truth value. Despite the presence of outliers, such as $\{0,1,2\}$, in $\Delta(W\{X_{i}, X_{j}\})$, the estimated lag value aligns well with the ground truth value of $3$.

Next step, the construction of the lead-lag matrix $\Gamma_{n \times n}$ is accomplished by

%\vspace{-4mm}
\begin{equation}
\Gamma_{ij} = \left\{
     \begin{array}{@{}l@{\thinspace}l}
       \gamma\{X_{i}, X_{j}\}  & \hspace{1cm}\text{if}\hspace{0.3cm} \text{$X_{i}$ and $X_{j}$ are in the same cluster}\\
       0  & \hspace{1cm}\text{otherwise}\\
     \end{array}
   \right.
\end{equation}

We summarize the above procedures in Algorithm \ref{dtw_lead-lag_detection}.

\begin{algorithm}[htp] 
\caption{\textbf{\small DTW\_KMed for Lead-lag Relationship Detection}}
\label{dtw_lead-lag_detection}
%\begin{algorithmic}[1]
\textbf{Input:} Time series matrix $X_{n \times T}$. 
\textbf{Output:} Lead-lag matrix $\Gamma_{n \times n}$.
\begin{algorithmic}[1]
\State Compute the DTW distance matrix from $X_{n \times T}$.
\State Apply K-medoids clustering using the DTW distance matrix.
\State For each cluster $\phi_{d}$ ($d = 1,\ldots, K$): 
\begin{itemize}
    \item Record path $W$ from every pair of time series $\{X_{i}, X_{j}\}$.
    \item Calculate the difference $\Delta(W\{X_{i}, X_{j}\})$ for every index pair in the path $W$
    \item Record lag $\gamma\{X_{i}, X_{j}\}$ by taking \textit{mode} or \textit{median} estimation of the difference $\Delta(W\{X_{i}, X_{j}\})$.
\end{itemize}
\State Calculate the lead-lag matrix $\Gamma_{n \times n}$ by setting $\gamma\{X_{i}, X_{j}\}$, where $\{X_{i}, X_{j}\}$ is in the same cluster. Otherwise, the entry is set to $0$.
\end{algorithmic}
\end{algorithm}

\section{Synthetic data experiments}
\label{sec: synthetic}
The synthetic data experiments serve the purpose of simulating data generated by a multi-factor model with a known ground truth lead-lag matrix $L$. The main objective is to evaluate and validate the performance of our proposed algorithms under various scenarios.

\subsection{Synthetic data generating process}

As noted earlier, our focus is the single membership setting. We generate synthetic data from the lagged multi-factor model (\ref{eq:lagged_multi-factor_model}) with $k = \{1,2\}$ factors. Furthermore, we specify the maximum number of lags as $M=5$ and the length of the time series as $T=100$. The factors $f$ and noise $\epsilon$ are drawn from $\mathcal{N}(0,1)$. We define $B$ and $L$ as follows:

\begin{table}[htbp]
  \centering
  \captionof{table}{Top row: Loading matrix $B$. Bottom row: Lag matrix $L$.}
    \begin{tabular}{p{0.3cm}p{4.5cm}|p{4.5cm}p{4.5cm}}
      \multicolumn{1}{c}{} &
      \multicolumn{1}{c}{\textbf{Homogeneous Setting}}    &  \multicolumn{2}{c}{\textbf{Heterogeneous Setting}}  \\

\multirow{8}{*}{\textit{B}}
\multirow{24}{*}{}
\multirow{24}{*}{}
\multirow{24}{*}{}
\multirow{18}{*}{\textit{L}}
    
    &

    \begin{equation*}
    \left[\begin{array}{ccc} 
    	1 \\
    	1 \\
    	1 \\
    	1 \\
    	1 \\
    	1 \\
    \end{array}\right] 
    \end{equation*}
    
    \begin{equation*}
    \left[\begin{array}{ccc} 
    	0 \\
    	1 \\
    	2 \\
    	3 \\
    	4 \\
    	5 \\
    \end{array}\right] 
    \end{equation*}

    &
    \begin{equation*}
    \left[\begin{array}{ccc} 
    	1 & 0 \\
    	1 & 0 \\
    	1 & 0 \\
    	0 & 1 \\
    	0 & 1 \\
    	0 & 1 \\
    \end{array}\right] 
    \end{equation*}
    
    \begin{equation*}
    \left[\begin{array}{ccc} 
    	0 & 0 \\
    	2 & 0 \\
    	4 & 0 \\
    	0 & 0 \\
    	0 & 2 \\
    	0 & 4 \\
    \end{array}\right] 
    \end{equation*}

    &
    \begin{equation*}
    \left[\begin{array}{ccc} 
    	1 & 0 & 0\\
    	1 & 0 & 0 \\
    	0 & 1 & 0\\
    	0 & 1 & 0 \\
    	0 & 0 & 1\\
    	0 & 0 & 1 \\
    \end{array}\right] 
    \end{equation*}
    
    \begin{equation*}
    \left[\begin{array}{ccc} 
    	0 & 0 & 0\\
    	3 & 0 & 0 \\
    	0 & 0 & 0\\
    	0 & 3 & 0 \\
    	0 & 0 & 0\\
    	0 & 0 & 3 \\
    \end{array}\right] 
    \end{equation*}
    \\
    \multicolumn{1}{c}{} &\multicolumn{1}{c}{$k=1$} & \multicolumn{1}{c}{$k=2$} & \multicolumn{1}{c}{$k=3$}  \\
    \end{tabular}
\end{table}

For validation purposes, we set the number of time series as $n=120$ to demonstrate the effectiveness of our algorithms. When estimating the lead-lag matrix, we use a sliding window of length $l=21$ and a shift of $s=1$. After estimating the lead-lag matrix, we calculate the error matrix $E$ to evaluate the performance, which we denote as 
\begin{equation} \label{eq:error_matrix}
E_{n \times n} = \Gamma_{n \times n} - \Psi_{n \times n},
\end{equation}
where $\Gamma_{n \times n}$ is the estimated lead-lag matrix, and $\Psi_{n \times n}$ is the ground truth lead-lag matrix, which can be obtained from $L_{n \times k}$.

%\subsection{Results}

% We set the number of clusters $K$ to be the ground truth value $k$ for all synthetic data experiments. Then, we apply K-medoids with Euclidean distance and DTW to cluster the synthetic data described above with $k=3$. When clustering by using Euclidean distance, the clusters often become confused due to the variability of the time axis. Fortunately, DTW offsets the variability in the time axis and performs much better by grouping clusters correctly. The result is shown in Figure \ref{tab:euclid_vs_dtw_cluster}.

% \begin{table}[htbp]
%   \centering
%     \begin{tabular}{p{4cm}|p{4cm}}

%       \includegraphics[width=\linewidth,trim=7.5cm 1cm 1cm 1cm,clip]{plots/K-medoids_euclidean.pdf}
% \vspace{-2cm}
%     &
%       \includegraphics[width=\linewidth,trim=7.5cm 1cm 1cm 1cm,clip]{plots/K-medoids_DTW.pdf}
% \vspace{-2cm}
%     \\
%     \end{tabular}
% \captionof{figure}{Left: K-medoids clustering with Euclidean distance. Right: K-medoids clustering with DTW.}
% \label{tab:euclid_vs_dtw_cluster}
% \end{table}

\subsection{Simulation results}

In the homogeneous setting ($k = 1$), as shown in Figure \ref{tab:ari_sigma_dtw}, all five algorithms (KM as K-Means, Euc\_KMed as Euclidean + K-Medoids, Man\_KMed as Manhattan + K-Medoids, Cos\_KMed as Cosine + K-Medoids, and DTW\_KMed as DTW + K-Medoids) demonstrate optimal performance with an Adjusted Rand Index (ARI) of 1. This indicates that they successfully detect the lead-lag relationships with high accuracy when there is only one underlying factor, and this result is expected based on our experimental setup. However, in the heterogeneous setting ($k = 2$), we observe a general decrease in ARI as the noise level $\sigma$ increases. Despite this trend, our proposed DTW\_KMed algorithm achieves a consistently high ARI value within the range of $\sigma$ from 0 to 1.5. This performance surpasses the other algorithms, which maintain relatively lower ARI values across the noise levels. This observation highlights that DTW excels in capturing intricate lead-lag patterns with a higher level of robustness.

\begin{table}[htbp]
  \centering
    \begin{tabular}{p{4.5cm}|p{4.5cm}p{4.5cm}}
      \multicolumn{1}{c}{\textbf{Homogeneous Setting}}    &  \multicolumn{2}{c}{\textbf{Heterogeneous Setting}}  \\

      \includegraphics[width=\linewidth,trim=0cm 0cm 0cm 0cm,clip]{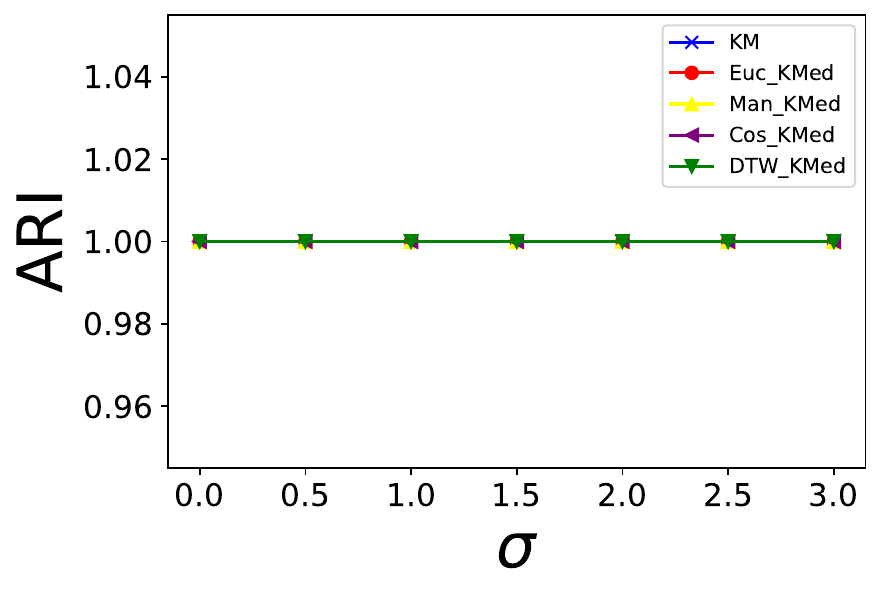}

    &
      \includegraphics[width=\linewidth,trim=0cm 0cm 0cm 0cm,clip]{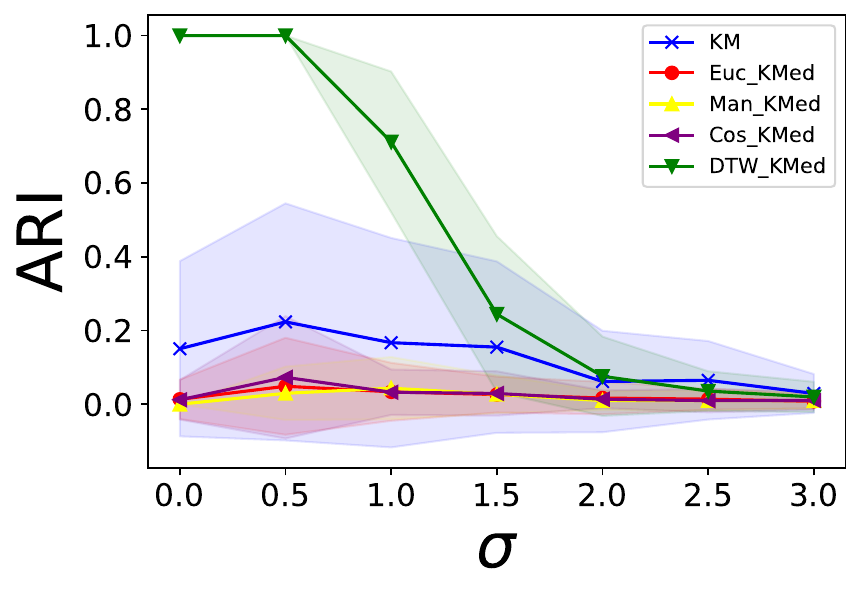}
      
    &
      \includegraphics[width=\linewidth,trim=0cm 0cm 0cm 0cm,clip]{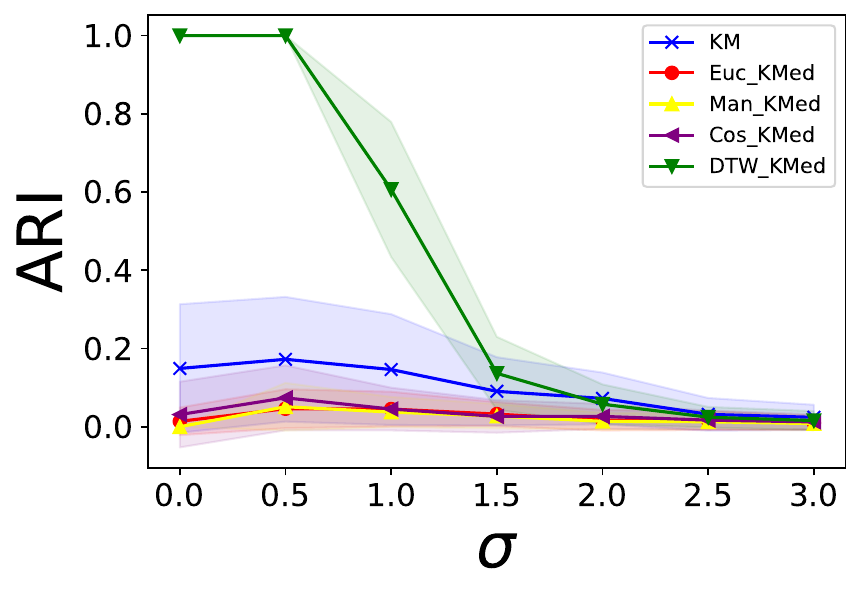}

        \\
    \multicolumn{1}{c}{$k=1$} & \multicolumn{1}{c}{$k=2$} & \multicolumn{1}{c}{$k=3$}  \\
    \end{tabular}

\captionof{figure}{Average and confidence interval for the ARI with different $\sigma$ levels based on 100 simulations for every iteration.}
\label{tab:ari_sigma_dtw}
\end{table}

In DTW, the window size $S$ defines the maximum allowed shifts from the two diagonals smaller than this number. %When no warping is allowed ($S = 0$), it may lead to mapping the top of a spike in the time series to a lower spot on another time series. Conversely, with unlimited warping allowed (no restriction on shifts), it becomes possible to map the top of a spike in the time series to a very distant spike on another time series, potentially resulting in incorrect clustering.
Hence, selecting the appropriate window size $S$ is of critical importance in DTW. The choice of $S$ affects how much temporal distortion is allowed between time series, thereby influencing the alignment and capturing the underlying lead-lag relationships effectively. Properly tuning $S$ enables DTW to strike a balance between capturing complex patterns while avoiding excessive warping that might lead to misclassification.

Across the synthetic experiments, we set the true lag equal to 5. In Figure \ref{tab:ari_window_dtw}, we observe similar results to the homogeneous setting in Figure \ref{tab:ari_sigma_dtw}. However, in the heterogeneous setting, the ARI tends to be relatively lower when the window size ranges from 0 to 5. Conversely, when the window size equals or exceeds 5, the ARI increases significantly.

This behaviour aligns with the rationale that DTW requires a window size that equals or exceeds the true lag of 5. This enables DTW to calculate the optimal alignment with enough flexibility to capture the actual lead-lag relationship effectively. When the window size is too restrictive, it may not allow for sufficient temporal distortion, leading to suboptimal alignment and lower ARI values in the presence of multiple factors. As such, selecting an appropriate window size is crucial to ensure accurate lead-lag relationship detection with DTW.

\begin{table}[htbp]
  \centering
    \begin{tabular}{p{4.5cm}|p{4.5cm}p{4.5cm}}
      \multicolumn{1}{c}{\textbf{Homogeneous Setting}}    &  \multicolumn{2}{c}{\textbf{Heterogeneous Setting}}  \\

      \includegraphics[width=\linewidth,trim=0cm 0cm 0cm 0cm,clip]{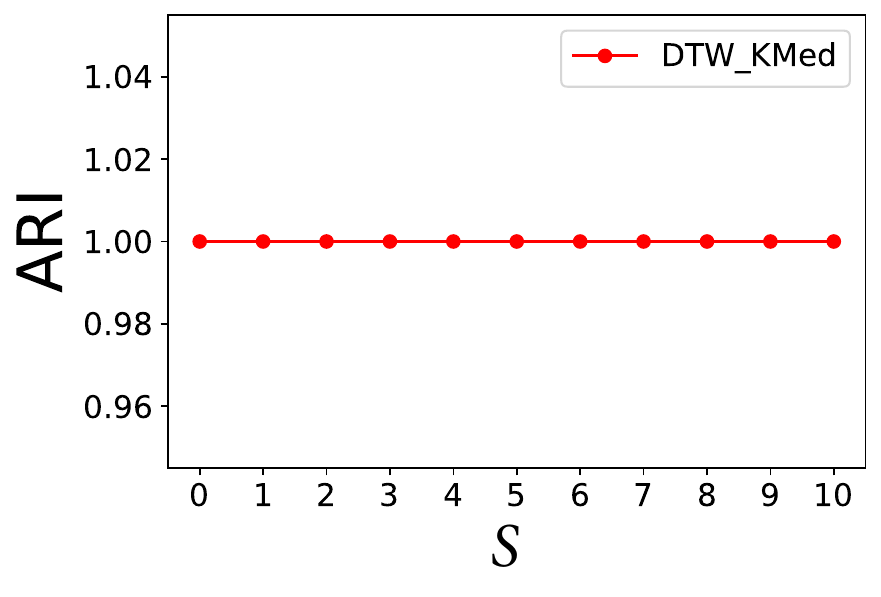}
\vspace{-5cm}
    &
      \includegraphics[width=\linewidth,trim=0cm 0cm 0cm 0cm,clip]{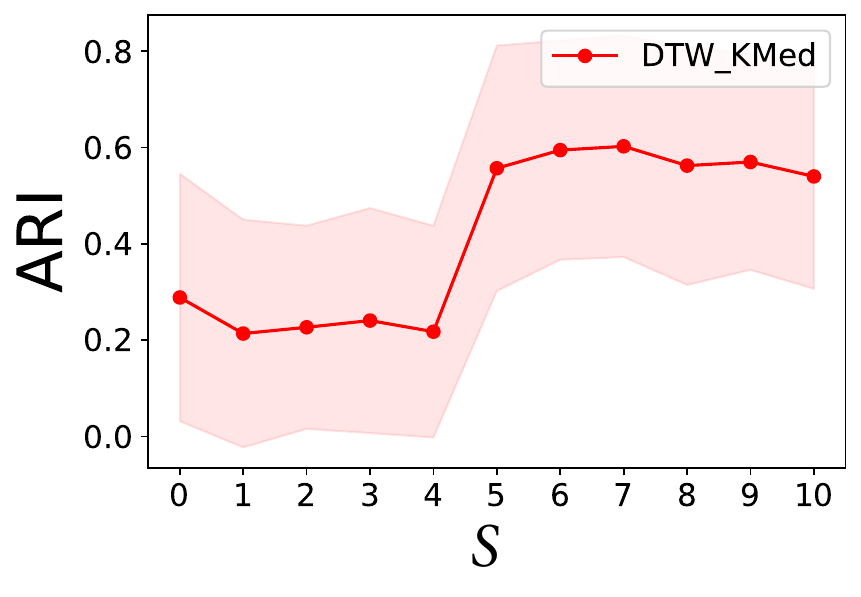}

        &
      \includegraphics[width=\linewidth,trim=0cm 0cm 0cm 0cm,clip]{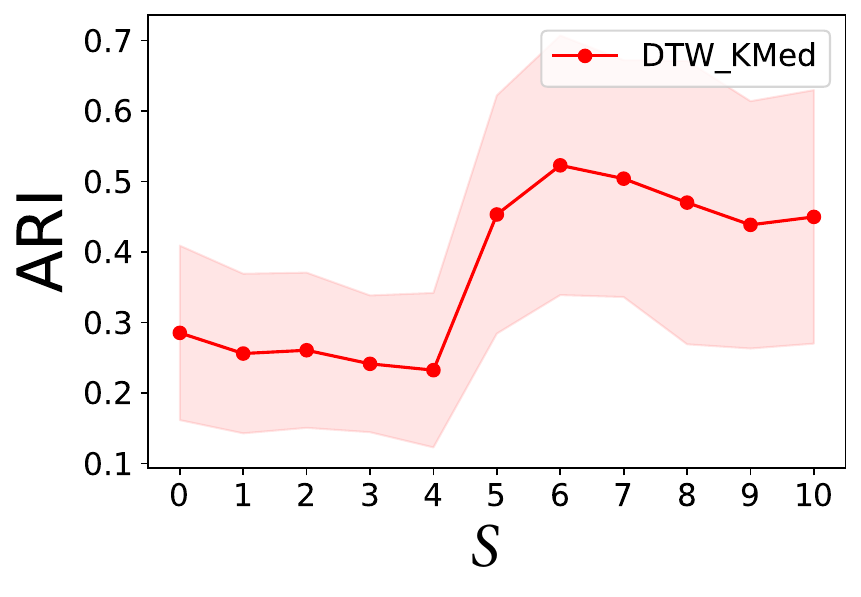}
    \\
        \multicolumn{1}{c}{$k=1$} & \multicolumn{1}{c}{$k=2$} & \multicolumn{1}{c}{$k=3$}  \\

    \end{tabular}

\captionof{figure}{Average and confidence interval for the ARI with different window sizes $S$ based on 100 simulations for every iteration (the true lag is 5).}
\label{tab:ari_window_dtw}
\end{table}

As depicted in Figure \ref{tab:mse_sigma_dtw}, for both the homogeneous and heterogeneous settings, the Mean Squared Error (MSE) remains close to 0 when $\sigma$ ranges from $0.0$ to $0.5$. However, as $\sigma$ increases beyond $0.5$, the MSE rises significantly for both algorithms. Overall, both algorithms demonstrate acceptable performance for low noise levels, but the DTW\_KMed mode estimation (DTW\_KMed\_Mod) exhibits better performance across varying levels of noise than DTW\_KMed median estimation (DTW\_KMed\_Med).

\begin{table}[htbp]
  \centering
    \begin{tabular}{p{4.5cm}|p{4.5cm}p{4.5cm}}
      \multicolumn{1}{c}{\textbf{Homogeneous Setting}}    &  \multicolumn{2}{c}{\textbf{Heterogeneous Setting}}  \\

      \includegraphics[width=\linewidth,trim=0cm 0cm 0cm 0cm,clip]{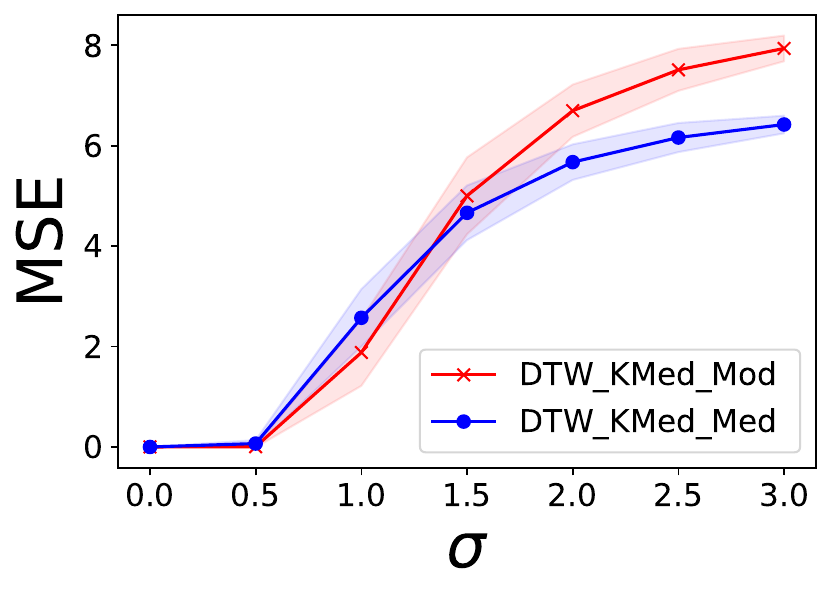}

    &
      \includegraphics[width=\linewidth,trim=0cm 0cm 0cm 0cm,clip]{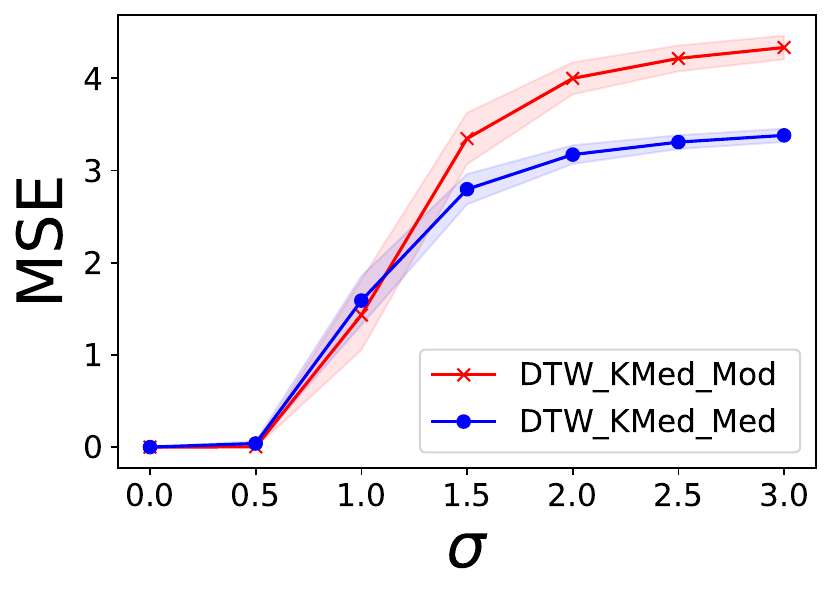}
        &
      \includegraphics[width=\linewidth,trim=0cm 0cm 0cm 0cm,clip]{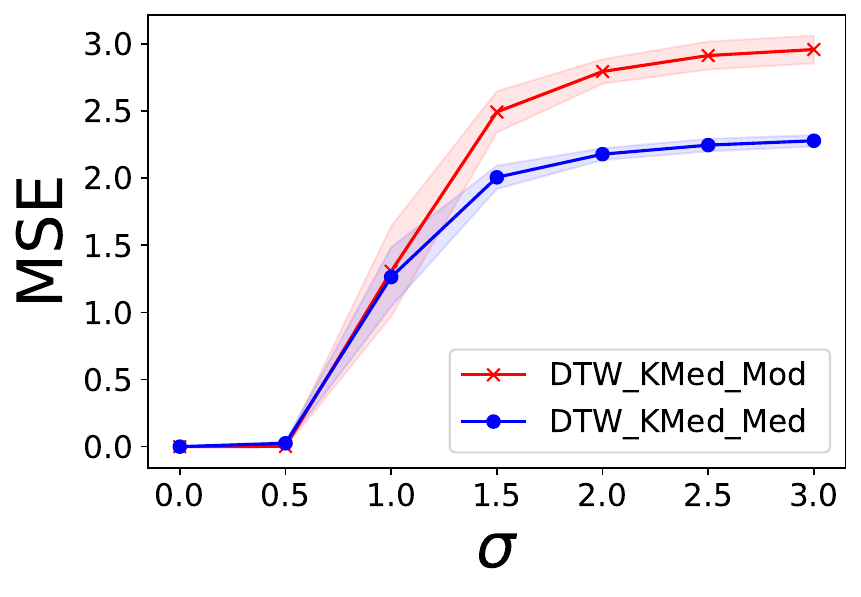}

    \\

            \multicolumn{1}{c}{$k=1$} & \multicolumn{1}{c}{$k=2$} & \multicolumn{1}{c}{$k=3$}  \\
    \end{tabular}

\captionof{figure}{Average and confidence interval for the MSE with different $\sigma$ levels based on 100 simulations for every iteration.}
\label{tab:mse_sigma_dtw}
\end{table}

With the true lag set as 5 and $\sigma$ as 1, %\MC{what do we mean by the alignment?} 
the DTW\_KMed\_Mod algorithm outperforms the DTW\_KMed\_Med algorithm in the homogeneous setting ($k = 1$). In the heterogeneous setting ($k = 2$), when the window size $S$ ranges from 0 to 5, the MSE is higher for both algorithms. However, as the window size $S$ exceeds 5, both algorithms achieve a lower MSE. This behaviour aligns with the understanding that DTW requires a $S$ that equals or exceeds the true lag of 5 to effectively calculate the optimal alignment. Thus, when the $S$ is less than 5, the alignment may not fully capture the true lead-lag relationship, resulting in higher MSE values. However, when the $S$ is larger than or equal to 5, both algorithms achieve better alignment, leading to reduced MSE values.

\begin{table}[htbp]
  \centering
    \begin{tabular}{p{4.5cm}|p{4.5cm}p{4.5cm}}
      \multicolumn{1}{c}{\textbf{Homogeneous Setting}}    &  \multicolumn{2}{c}{\textbf{Heterogeneous Setting}}  \\

      \includegraphics[width=\linewidth,trim=0cm 0cm 0cm 0cm,clip]{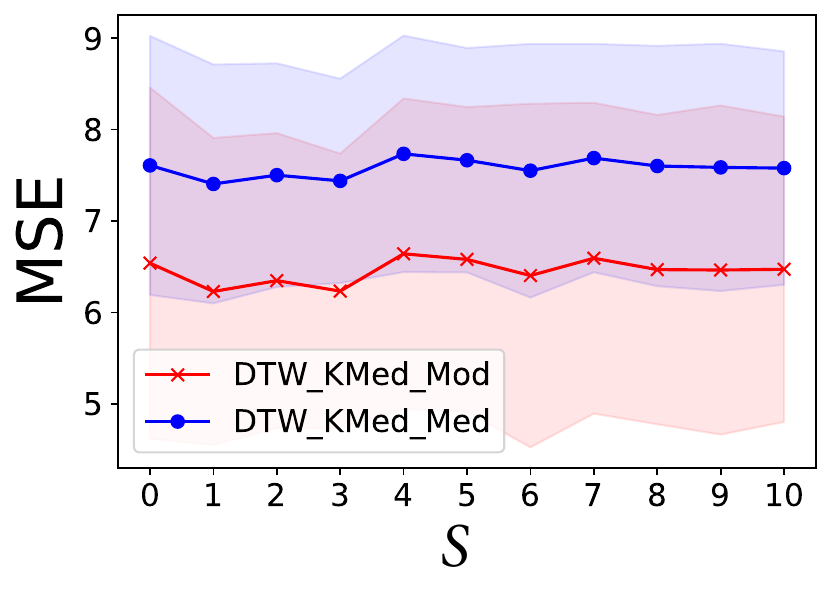}

    &
      \includegraphics[width=\linewidth,trim=0cm 0cm 0cm 0cm,clip]{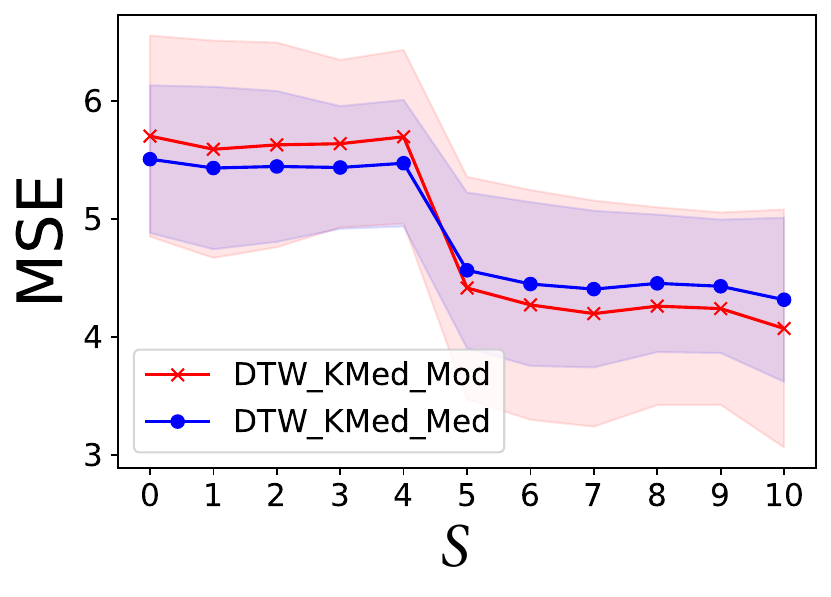}

    &
      \includegraphics[width=\linewidth,trim=0cm 0cm 0cm 0cm,clip]{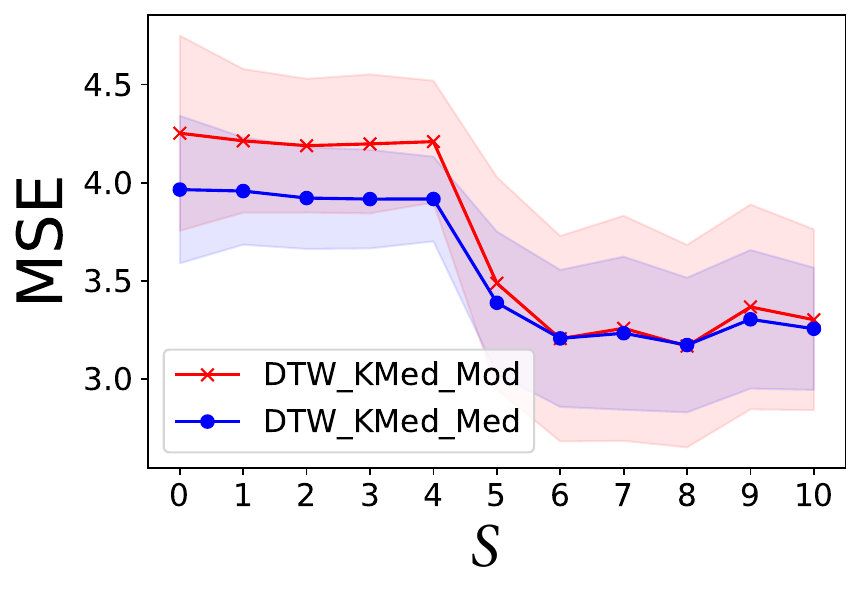}
    \\
          \multicolumn{1}{c}{$k=1$} & \multicolumn{1}{c}{$k=2$} & \multicolumn{1}{c}{$k=3$}  \\
    \end{tabular}

\captionof{figure}{Average and confidence interval for the MSE with different window sizes $S$ based on 100 simulations for every iteration (the true lag is 5, and $\sigma$ is 1).}
\label{tab:mse_window_dtw}
\end{table}

\section{Financial data experiments}
\label{sec: financial}

\subsection{Data description}

In this section, we conduct a large-scale experiment using financial data to apply our algorithms. As mentioned earlier, this is a context where lead-lag relationships naturally occur. For our financial data experiments, we consider three different data sets, each varying in terms of the number and type of assets, as well as the number of days included in the data set. All the data sets are considered at a daily frequency. The summarized details of the data sets are presented in Table \ref{tab: dataset}, with additional information about the Pinnacle Data Corp CLC data set available in the Appendix \ref{sec:pinnacle} Tables [\ref{tab: Grains}, \ref{tab: Meats}, \ref{tab: Foodfibr}, \ref{tab: Oils}, \ref{tab: Metals}, \ref{tab: Indexes}, \ref{tab: Bonds}, \ref{tab: Currency}].
% \begin{itemize}
% \item The first dataset consists of US equities sourced from \href{https://wrds-www.wharton.upenn.edu/pages/about/data-vendors/center-for-research-in-security-prices-crsp/}{Wharton’s CRSP} data set. This dataset contains a total of 679 equities and spans 5211 trading days, starting from January 3, 2000, to December 31, 2019.
% \item The second dataset is derived from the same \href{https://wrds-www.wharton.upenn.edu/pages/about/data-vendors/center-for-research-in-security-prices-crsp/}{Wharton’s CRSP} data set, but it focuses on Exchange Traded Funds (ETFs). This dataset includes 14 ETFs and covers 3324 trading days, spanning from April 12, 2006, to July 1, 2019.

% \item The third dataset is obtained from the \href{https://pinnacledata2.com/clc.html}{Pinnacle Data Corp CLC}, and consists of 52 futures contracts covering various asset classes, such as commodities, fixed income, and currency futures. It spans 5166 trading days, ranging from January 5, 2000, to October 16, 2020. Notably, all futures contracts in this dataset are adjusted for rolling effects.

% \end{itemize}
% All the datasets are considered at a daily frequency. The summarized details of the datasets are presented in Table \ref{tab: dataset}, with additional information about the Pinnacle Data Corp CLC dataset available in the Appendix \ref{sec:pinnacle} Tables [\ref{tab: Grains}, \ref{tab: Meats}, \ref{tab: Foodfibr}, \ref{tab: Oils}, \ref{tab: Metals}, \ref{tab: Indexes}, \ref{tab: Bonds}, \ref{tab: Currency}].

\begin{table}[htbp]
  \centering
\caption{Summary of the three financial data sets considered in the numerical experiments.}
    \begin{tabular}  {p{3cm}p{1cm}p{1cm}p{1.9cm}p{1.8cm}p{1.8cm}p{1.7cm}}
        \toprule
            \textbf{Data source} & \textbf{Type} & \textbf{Freq} & \textbf{\#\ of assets} & \textbf{Start date} & \textbf{End date} & \textbf{\#\ of days}\\
        \midrule
            Wharton’s CRSP     & Equity  & Daily & 679 & 2000/01/03 & 2019/12/31 & 5211\\
            Wharton’s CRSP     & ETF     & Daily & 14  & 2006/04/12 & 2019/07/01 & 3324\\
            Pinnacle Data Corp & Futures  & Daily & 52  & 2000/01/05 & 2020/10/16 & 5166\\ 
        \bottomrule
    \end{tabular}

\label{tab: dataset}
\end{table}

\subsection{Data pre-processing}

With regard to the US equity and ETF data sets, we use the close-to-close adjusted daily returns from Wharton’s CRSP. Due to the large number of NaNs in the equity data set, we drop the days for which more than 10\% of the equities have zero returns as well as the equities for which more than 50\% of days have zero returns. Instead of working with raw returns, we consider the \textit{market excess returns}, a standard measure of how well each equity performed relative to the broader market. For both of these data sets, the return of the S\&P Composite Index is selected to compute the market excess returns by subtracting it from the return of each asset (i.e., for simplicity, we assume each asset has $\beta=1$ exposure to the market). Also, we winsorize the extreme value of excess returns for which any value is larger than 0.15 or smaller than -0.15.

For the futures data set, we use the close-to-close price series from the Pinnacle Data Corp CLC data set, and discard the days for which more than 10\% of the futures have zero prices in the respective dates, and drop the futures for which more than 160 days have zero prices. Afterwards, we first use forward-fill, then backward-fill to fill out the zero prices. Lastly, we compute the log-return from the close-to-close price. The remainder of the data pre-processing is the same as above.

\subsection{Benchmark}

In order to evaluate our proposed methodology, we also introduce a benchmark to detect lead-lag relationships without the use of clustering. It is very common to compute a sample cross-correlation function (CCF) between two time series. A CCF between time series $X_i$ and $X_j$ evaluated at lag $m$ is given by

\begin{equation} \label{eq:CCF}
\textrm{CCF}^{ij}(m) = \text{CORR}(\{X_{i}^{t-m}\},\{X_{j}^{t}\}),
\end{equation}

\noindent
where CORR() denotes a choice of the CCF. The corresponding lead-lag matrix $\Gamma_{n \times n}$ is estimated by computing the signed normalized area under the curve of CCF, given by

\begin{equation} \label{eq:CCF-auc}
\Gamma_{ij} =\frac{\text{MAX} (I(i, j), I(j, i)) \cdot \text{SIGN}(I(i, j)-I(j, i))}{I(i, j)+I(j, i)},
\end{equation}

\noindent
where $I(i, j)=\sum_{m=1}^M\left|\textrm{CCF}^{ij}(m)\right|$ for a user-specified maximum lag $M$.

We summarize the CCF procedures in Algorithm \ref{benchmark}.

\begin{algorithm}[htp] 
\caption{\textbf{\small: CCF Algorithm}}
\label{benchmark}
\textbf{Input:} Time series matrix $X_{n \times T}$. \\
\textbf{Output:} Lead-lag matrix $\Gamma_{n \times n}$.
\begin{algorithmic}[1]
\State Calculate CCF for every pair of time series $\{X_{i}, X_{j}\}$.
\State Calculate the lead-lag matrix $\Gamma_{n \times n}$ by computing the signed normalized area under the curve of CCF.
\end{algorithmic}
\end{algorithm}

Also, we use four algorithms (KM\_Mod, KM\_Med, SP\_Mod, and SP\_Med) as our benchmark from \cite{zhang2023robust}.

\subsection{Trading strategies}

In this section, we present the trading strategies employed in this paper. Our approach involves a series of steps applied to a dataset consisting of $n$ time series, each having a length of $T$. Firstly, we extract the data by implementing a sliding window approach with a fixed length of $l = 21$. Subsequently, we employ the DTW\_KMed algorithm to detect the lead-lag relationship, which is further validated through a synthetic data experiment.
Once the lead-lag matrix is obtained, We then utilize the lead-lag matrix to rank the time series from the most leading to the most lagging using the \textit{RowSum} ranking [\cite{gleich2011rank,huber1963pairwise}], in order to then group the time series into leaders and laggers, where the leaders are employed to forecast the behaviour of the laggers.

% \begin{table}[htbp]
%   \centering
%     \begin{tabular}{p{4cm}|p{4cm}}

%       \includegraphics[width=\linewidth,trim=2cm 5.5cm 2cm 5.3cm,clip]{plots/lag.pdf}

%     &
%       \includegraphics[width=\linewidth,trim=2cm 5.5cm 2cm 5.3cm,clip]{plots/lead.pdf}

%     \\
%     \multicolumn{1}{c}{$G_\beta$} & \multicolumn{1}{c}{$D_\alpha$} \\
%     \end{tabular}
% \captionof{figure}{$G_\beta$ strategy: Use $D_\alpha$ predict $G_\beta$ (left). $D_\alpha$ strategy: Use $D_\alpha$ predict $D_\alpha$ (right).}
% \label{tab:ranking}
% \end{table}

% %\vspace{-0.5cm}

\begin{figure}[!htbp]
\centering
\includegraphics[width= 0.5\textwidth,trim=0.5cm 5cm 0.5cm 5cm,clip]{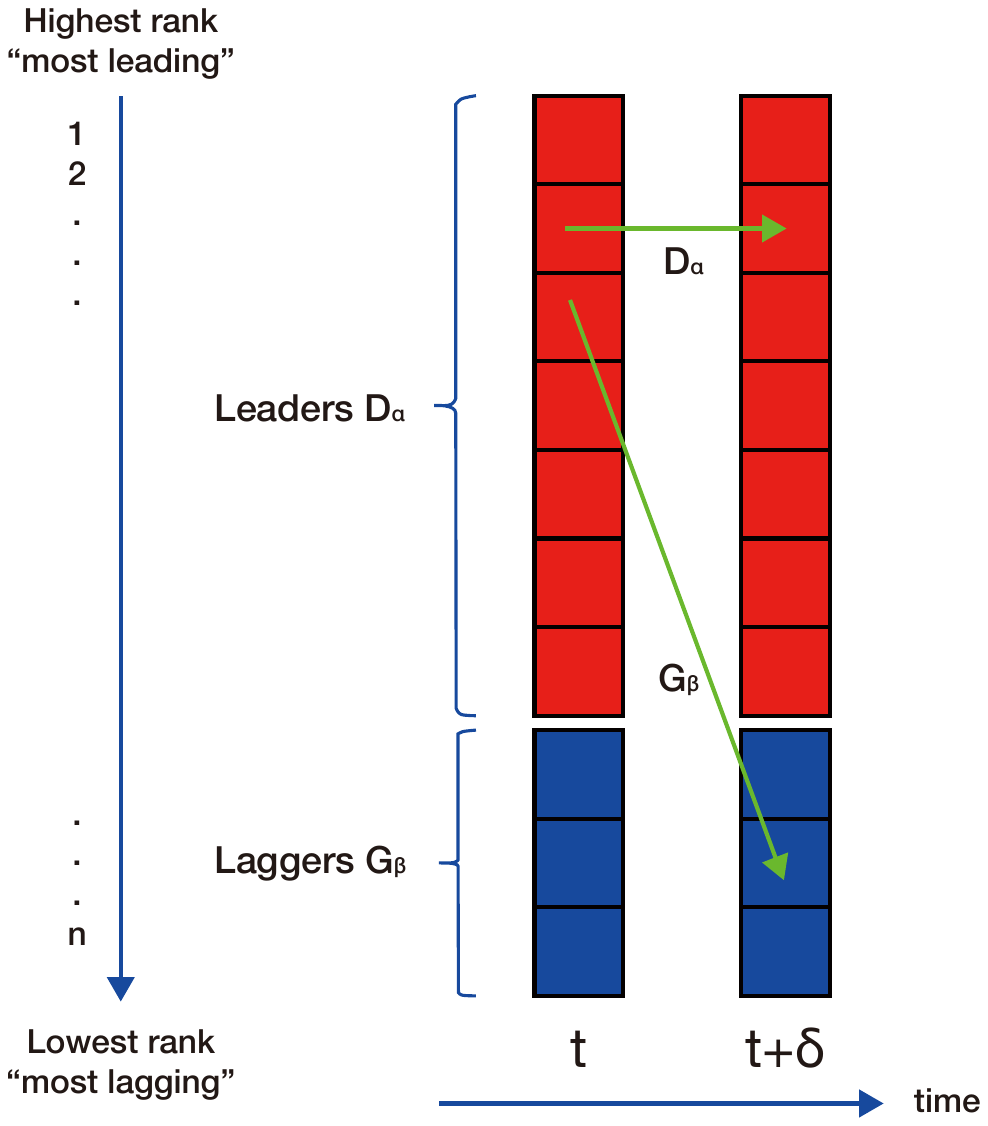}

\caption{$G_\beta$ strategy: Use $D_\alpha$ predict $G_\beta$. $D_\alpha$ strategy: Use $D_\alpha$ predict $D_\alpha$.}
\label{fig: ranking}
\end{figure}

%  %\vspace{-0.3cm}

% \YZ{add lit review}

\begin{figure}[!htbp]
\centering
\includegraphics[width=0.8\textwidth,trim=0.5cm 10cm 0.5cm 7cm,clip]{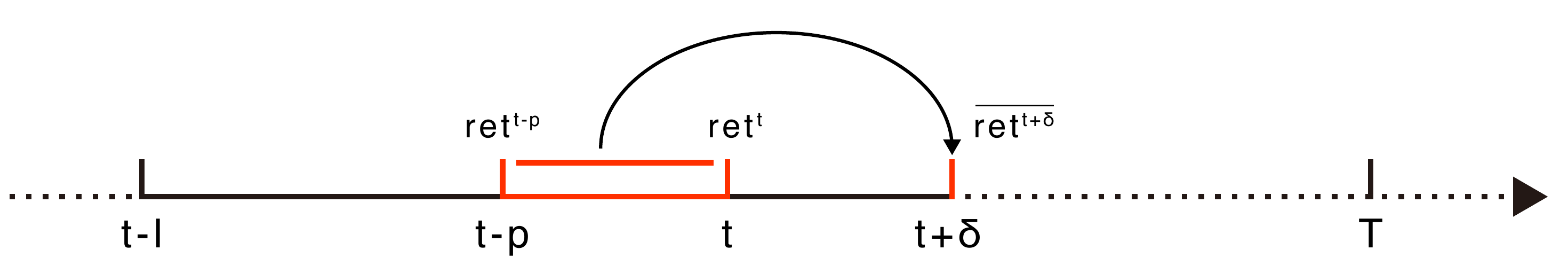}

\caption{Illustration of the trading pipeline at time $t$, given the EWMA of past $p$ days of average winsorized time series excess returns to predict future $\delta$ days average excess returns.} %\MC{maybe add more detail in the caption to make it self-contained}}
\label{fig:sliding_window}
\end{figure}

Momentum, a well-studied phenomenon in finance literature [\cite{jegadeesh2022momentum, jegadeesh2001profitability, lim2019enhancing,poh2022transfer,tan2023spatio,wood2021trading,wood2021slow,roberts2023network,zohren2023learning}], refers to the tendency of assets that have exhibited strong performance in the recent past to continue their performance in the near future, and vice versa. 
In our trading strategy, we identify the top $\alpha = 0.75$ fraction of the time series as Leaders $D_\alpha$, while the remaining bottom fraction $\beta = 1 - \alpha$ is classified as Laggers $G_\beta$.
To predict the future performance, we employ the exponentially weighted moving average (EWMA) signal, considering the past $p = \{1,3,5,7\}$ days of average winsorized time series excess returns from $D_\alpha$. This prediction aims to estimate the average excess returns of $G_\beta$ and $D_\alpha$ in the subsequent $\delta = \{1,3,5,7\}$ days. We assume that $G_\beta$ can catch up with $D_\alpha$, while $D_\alpha$ provides the necessary momentum to sustain the trend over the $\delta$ days. Figure \ref{fig: ranking} visually illustrates this concept. To ensure continuous trading, we shift the sliding window by $h = 1$ and repeat the lead-lag matrix calculation and ranking steps until the end of the time series. Figure \ref{fig:sliding_window} provides a depiction of our trading pipeline at time $t$, and we summarize the trading strategy in Algorithm \ref{strategy}.

\begin{algorithm}[htp] 
\caption{\textbf{\small: Trading strategy}}
\label{strategy}
\textbf{Input:} Time series matrix $X_{n \times T}$. \hspace{20cm}
\begin{algorithmic}[1]
\State Construct the matrix $X_{n \times T}$ by employing a sliding window of length $l$ starting from the beginning of the time series, resulting in $X_{n \times l}$.
\State Apply the \textbf{\small DTW\_KMed for Lead-lag Relationship Detection Algorithm} to $X_{n \times l}$, resulting in the computation of the lead-lag matrix $\Gamma_{n \times n}$.
\State Based on $\Gamma_{n \times n}$, rank the time series from the most leading to the most lagging using the \textit{RowSum} ranking methodology.
\State Select the top $\alpha$ fraction of the time series as Leaders $D_\alpha$, and the bottom $\beta = 1 - \alpha$ as Laggers $G_\beta$.
\State Employ the Exponentially Weighted Moving Average (EWMA) on the past $p$ days of the average winsorized time series excess returns of $D_\alpha$ to predict the average future excess returns of $G_\beta$ and $D_\alpha$ for a duration of $\delta$ days.
\State Shift the sliding window by $h$, and repeat Steps 1-5 until the end of the time series.
\end{algorithmic}
\end{algorithm}

\subsection{Performance evaluation}

%\YZ{Add performance section}
When assessing the effectiveness of various trading strategies, we rely on the following metrics to evaluate their performance. 
We compute the Profit and Loss (PnL) of $G_\beta$ on a given day $t+\delta$ as 

    \begin{equation}
    \text{PnL}_{G_\beta}^{t+\delta} = \text{sign}(\text{EWMA}(ret_{D_\alpha}^{t-p}:ret_{D_\alpha}^{t}))\cdot\overline{ret_{G_\beta}^{t+\delta}}, t=l,\ldots,T-\delta,
    \end{equation}
since the strategy makes profits whenever the sign of the forecast agrees with the sign of the future return. Correspondingly, the PnL of $D_\alpha$ on a given day $t+\delta$ is given by
%\vspace{-3mm}
\begin{equation}
\text{PnL}_{D_\alpha}^{t+\delta} = \text{sign}(\text{EWMA}(ret_{D_\alpha}^{t-p}:ret_{D_\alpha}^{t}))\cdot\overline{ret_{D_\alpha}^{t+\delta}}, t=l,\ldots,T-\delta,
    \end{equation}

\noindent
where $ret_{D_\alpha}^{t-p}$ and $ret_{D_\alpha}^{t}$ are the excess return of $D_\alpha$ at $t-p$ and $t$, respectively, while $\text{EWMA}(ret_{D_\alpha}^{t-p}:ret_{D_\alpha}^{t})$ denotes the exponentially weighted moving average from the excess return of $D_\alpha$ from $t-p$ to $t$. Furthermore, $\overline{ret_{G_\beta}^{t+\delta}}$ depicts the mean of the excess return of $G_\beta$ at $t+\delta$, and $\overline{ret_{D_\alpha}^{t+\delta}}$ is the mean of the excess return of $D_\alpha$ at $t+\delta$.
%Rescaled to Target Volatility

We rescale the PnL by their volatility to target
equal risk assignment, and set our annualized volatility target $\sigma_{\text{tgt}}$ to be 0.15.

\begin{equation}
\text{PnL}_{\text{rescaled}} = \frac{\sigma_{\text{target}}}{\text{STD(PnL)}\cdot\sqrt{252}}\cdot\text{PnL}.
\end{equation}

Based on $\text{PnL}_{\text{rescaled}}$, we proceed to calculate the following annualized metrics, in line with the works of [\cite{lim2019enhancing, wood2022slow, tan2023spatio, poh2022transfer, wood2021trading, wood2021slow, zhang2023robust,liu2023multi,liu2023deep}], and more detailed information can be found in the Appendix \ref{sec:performance}.
%\MC{no appendix!!}

\begin{itemize}
  \item \textbf{Profitability}: cumulative PnL, annualized expected excess return (E[Returns]), hit rate.
  \item \textbf{Risk}: volatility, downside deviation, maximum drawdown.
  \item \textbf{Performance}: Sortino ratio, Calmar ratio, average profit / average loss, PnL per trade, Sharpe ratio, P-value.
\end{itemize}

% \MC{the types of tables/plots we should report: sharpes/pnl tables across methods, figures with cumsum pnl plots (a few of them would be good), we could aggregate the lagging index by sector membership for the equity data; if there is room, we could show performance at the yearly level; stability of the lead/lag score - just to give the reader an idea about persistence across time of leaders and laggers}

\subsection{Results}

In the case of the equity data set, comprehensive results with various tuning settings are available in Supplemental material [\cite{zhang2023DTWsupplemental}] B.1. Our findings indicate that utilizing the EWMA on the past seven days of the average winsorized time series excess returns of the $D_\alpha$, with $\alpha = 0.75$, for predicting the average future seven days of the excess return of the $G_\beta$ and $D_\alpha$ consistently yields favourable performance across all algorithms. Figure \ref{tab:equity_pnl} presents a comparison of cumulative PnL for the $G_\beta$ strategy (left) and $D_\alpha$ strategy (right). In the $G_\beta$ strategy, before 2008, the Sharpe ratio (SR) of DTW\_KMed\_Mod and DTW\_KMed\_Med outperformed the other algorithms. However, after 2008, except for CCF, all algorithms displayed a substantial growth trend and eventually achieved similar performance levels. On the other hand, in the $D_\alpha$ strategy, before 2008, all algorithms performed at roughly the same level. However, after 2008, DTW\_KMed\_Med emerged as the most profitable strategy, clearly outperforming others with an SR of 0.93. Additionally, Tables [\ref{tab:equity_metric_lag}, \ref{tab:equity_metric_lead}] present the performance of DTW\_KMed\_Mod, DTW\_KMed\_Med, and other algorithms for the $G_\beta$ strategy and $D_\alpha$ strategy based on various metrics (rescaled to target volatility).

Tables [\ref{tab:etf_metric_lag}, \ref{tab:etf_metric_lead}] along with Figure \ref{tab:etf_pnl} present the results for the ETF data set using the same settings as for the equity data. In the $G_\beta$ strategy, we do not find evidence of consistently detecting lead-lag relationships that lead to a profitable outcome. However, in the $D_\alpha$ strategy, the SR of SP\_Med and DTW\_KMed\_Mod are leading with SR values of 0.8 and 0.78, respectively. Full results across all tuning settings are reported in Supplemental material [\cite{zhang2023DTWsupplemental}] B.2.

Results for the futures data, using the same settings as for the equity data, are presented in Tables [\ref{tab:futures_metric_lag}, \ref{tab:futures_metric_lead}], along with Figure \ref{tab:futures_pnl}. For this data set, we do not observe the ability to consistently detect profitable lead-lag relationships for any of the strategies. Full results across all tuning settings are available in Supplemental material [\cite{zhang2023DTWsupplemental}] B.3.

In addition to the real data results, we have also included the results of the synthetic data experiments in the Appendix \ref{sec:synthetic}.

\begin{table}[!htbp]
  \centering
    \begin{tabular}{p{7cm}p{7cm}}
      \multicolumn{1}{c}{\textbf{$G_\beta$ strategy}}    &  
      \multicolumn{1}{c}{\textbf{$D_\alpha$ strategy}}  \\

      \includegraphics[width=\linewidth,trim=0cm 0cm 0cm 0cm,clip]{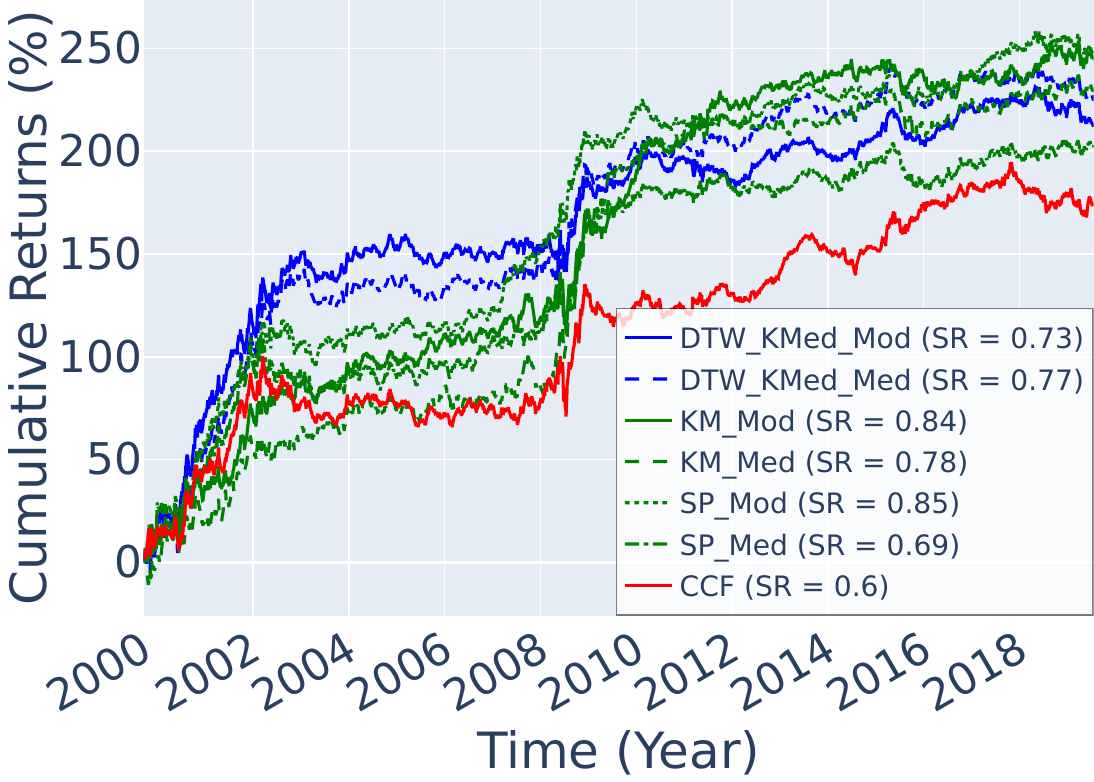}
      &
      \includegraphics[width=\linewidth,trim=0cm 0cm 0cm 0cm,clip]{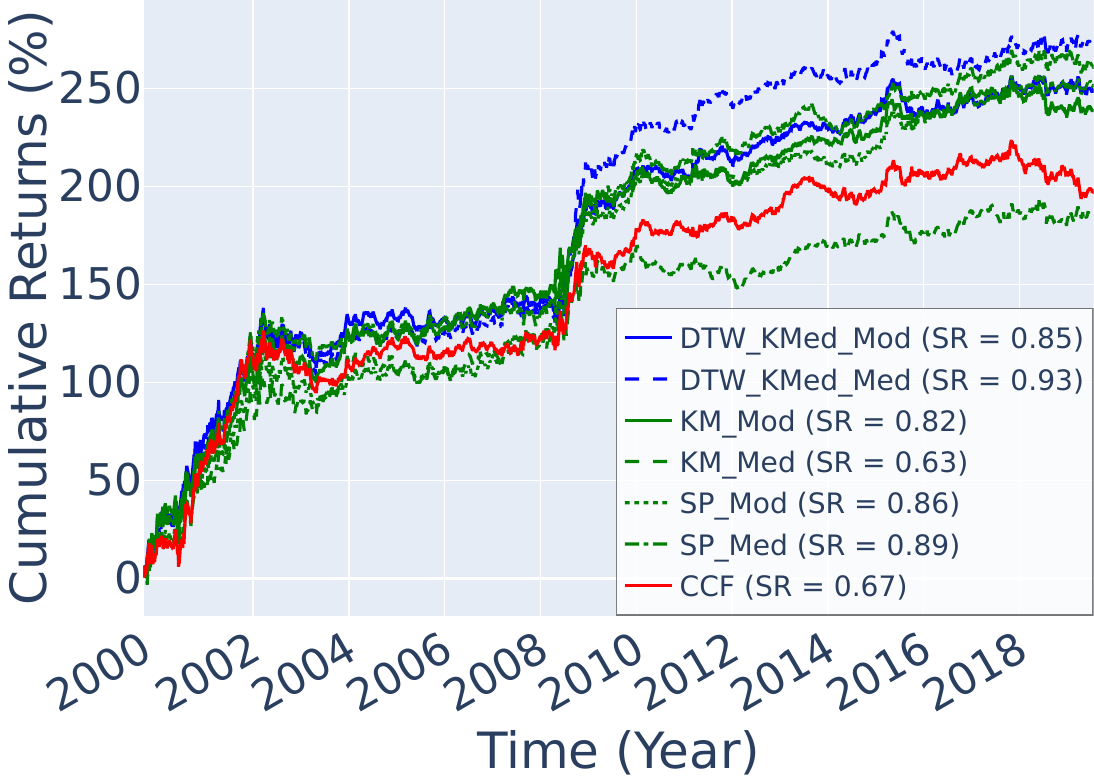}
    \end{tabular}

\captionof{figure}{Equity data set: cumulative PnL for $G_\beta$ strategy (left) and $D_\alpha$ strategy (right) - rescaled to target volatility. The experiment has been set with the values $p = 7$, $\delta = 7$, $\alpha = 0.25$, and $K=5$.}

\label{tab:equity_pnl}
\end{table}

%\vspace{-0.5cm}

\begin{table}[htbp]
  \centering
  \resizebox{\linewidth}{!}{
    \begin{tabular}{p{3.4cm}p{1cm}p{1.1cm}p{1.1cm}p{1.1cm}p{1.1cm}p{2.5cm}p{2.5cm}}
        \toprule
        \multicolumn{1}{c}{\textbf{$G_\beta$ strategy}}  &\multicolumn{5}{c}{\textbf{Benchmark}}  &  \multicolumn{2}{c}{\textbf{Proposed}}  \\
        \cmidrule(lr){2-6} \cmidrule(lr){7-8}
         & CCF & KM\_Mod & KM\_Med & SP\_Mod & SP\_Med &DTW\_KMod\_Mod&DTW\_KMod\_Med\\
        \midrule
        E[Returns]       &0.089  & 0.126 & 0.118  & \textbf{0.127*}  & 0.104  & 0.109  & 0.115  \\
        Volatility        &0.15   & 0.15  & 0.15   & 0.15   & 0.15   & 0.15   & 0.15   \\
        Downside deviation      &0.105  & 0.103 & \textbf{0.101*}  & 0.103  & 0.105  & 0.106  & 0.107  \\
        Maximum drawdown    &-0.313 & -0.26 & -0.215 & -0.214 & -0.287 & -0.205 & \textbf{-0.188*} \\
        Sortino ratio  &0.85   & 1.222 & 1.166  & \textbf{1.237*}  & 0.988  & 1.026  & 1.084  \\ 
        Calmar ratio       &0.285  & 0.484 & 0.548  & 0.594  & 0.362  & 0.531  & \textbf{0.614*}  \\
        Hit rate        &0.499  & \textbf{0.521*} & 0.51   & 0.516  & 0.519  & 0.514  & 0.512  \\
        Avg. profit / avg. loss   &\textbf{1.117*}  & 1.068 & 1.107  & 1.091  & 1.051  & 1.08   & 1.1    \\
        PnL per trade   &3.542  & 4.996 & 4.672  & \textbf{5.041*}  & 4.119  & 4.317  & 4.581  \\
        Sharpe ratio &0.595  & 0.839 & 0.785  & \textbf{0.847*}  & 0.692  & 0.725  & 0.77   \\
        P-value     &0.009  & \textbf{0*}     & \textbf{0*}      & \textbf{0*}      & 0.002  & 0.001  & 0.001 \\
        \bottomrule
    \end{tabular}}
\caption{Equity data set: performance metrics for $G_\beta$ strategy - rescaled to target volatility. The experiment has been set with the values $p = 7$, $\delta = 7$, $\alpha = 0.25$, and $K=5$.}
\label{tab:equity_metric_lag}
\end{table}

%\vspace{-1cm}

\begin{table}[htbp]
  \centering
 \resizebox{\linewidth}{!}{
    \begin{tabular}{p{3.4cm}p{1cm}p{1.1cm}p{1.1cm}p{1.1cm}p{1.1cm}p{2.5cm}p{2.5cm}}
        \toprule
        \multicolumn{1}{c}{\textbf{$G_\beta$ strategy}}  &\multicolumn{5}{c}{\textbf{Benchmark}}  &  \multicolumn{2}{c}{\textbf{Proposed}}  \\
        \cmidrule(lr){2-6} \cmidrule(lr){7-8}
         & CCF & KM\_Mod & KM\_Med & SP\_Mod & SP\_Med &DTW\_KMod\_Mod&DTW\_KMod\_Med\\
        \midrule
        E[Returns]       &0.101  & 0.122  & 0.095 & 0.129  & 0.134  & 0.128  & \textbf{0.14*}   \\
        Volatility        &0.15   & 0.15   & 0.15  & 0.15   & 0.15   & 0.15   & 0.15   \\
        Downside deviation      &0.105  & 0.107  & 0.108 & 0.106  & 0.105  & 0.103  & \textbf{0.102*}  \\
        Maximum drawdown    &-0.288 & -0.251 & \textbf{-0.21*} & -0.283 & -0.227 & -0.276 & -0.297 \\
        Sortino ratio  &0.964  & 1.148  & 0.876 & 1.222  & 1.269  & 1.238  & \textbf{1.376*}  \\
        Calmar ratio       &0.352  & 0.488  & 0.452 & 0.456  & \textbf{0.59*}   & 0.463  & 0.472  \\
        Hit rate        &0.518  & 0.521  & 0.513 & 0.522  & \textbf{0.525*}  & 0.52   & 0.523  \\
        Avg. profit / avg. loss   &1.05   & 1.069  & 1.066 & 1.072  & 1.065  & \textbf{1.078*}  & \textbf{1.078*}  \\
        PnL per trade   &4.018  & 4.856  & 3.765 & 5.121  & 5.312  & 5.073  & \textbf{5.557*}  \\
        Sharpe ratio  & 0.675  & 0.816  & 0.632 & 0.86   & 0.892  & 0.852  & \textbf{0.934*}  \\
        P-value     &0.003  & \textbf{0*}      & 0.005 & \textbf{0*}      & \textbf{0*}     &\textbf{0*}      & \textbf{0*}    \\
        \bottomrule
    \end{tabular}}
\caption{Equity data set: performance metrics for $D_\alpha$ strategy - rescaled to target volatility. The experiment has been set with the values $p = 7$, $\delta = 7$, $\alpha = 0.25$, and $K=5$.}
\label{tab:equity_metric_lead}
\end{table}
%\vspace{-0.5cm}

%%%%%%%%%%%%%%%%%%%%

% Tables [\ref{tab:etf_metric_lag}] and [\ref{tab:etf_metric_lead}] along with Figure \ref{tab:etf_pnl} present the results for the ETF data set using the same settings as for the equity data. In the $G_\beta$ strategy, we do not find evidence of consistently detecting lead-lag relationships that lead to a profitable outcome. However, in the $D_\alpha$ strategy, the SR of SP\_Med and DTW\_KMed\_Mod are leading with SR values of 0.8 and 0.78, respectively. Full results across all tuning settings are reported in the supplemental material.

\begin{table}[!htbp]
  \centering
    \begin{tabular}{p{7cm}p{7cm}}
      \multicolumn{1}{c}{\textbf{$G_\beta$ strategy}}    &  
      \multicolumn{1}{c}{\textbf{$D_\alpha$ strategy}}  \\

      \includegraphics[width=\linewidth,trim=0cm 0cm 0cm 0cm,clip]{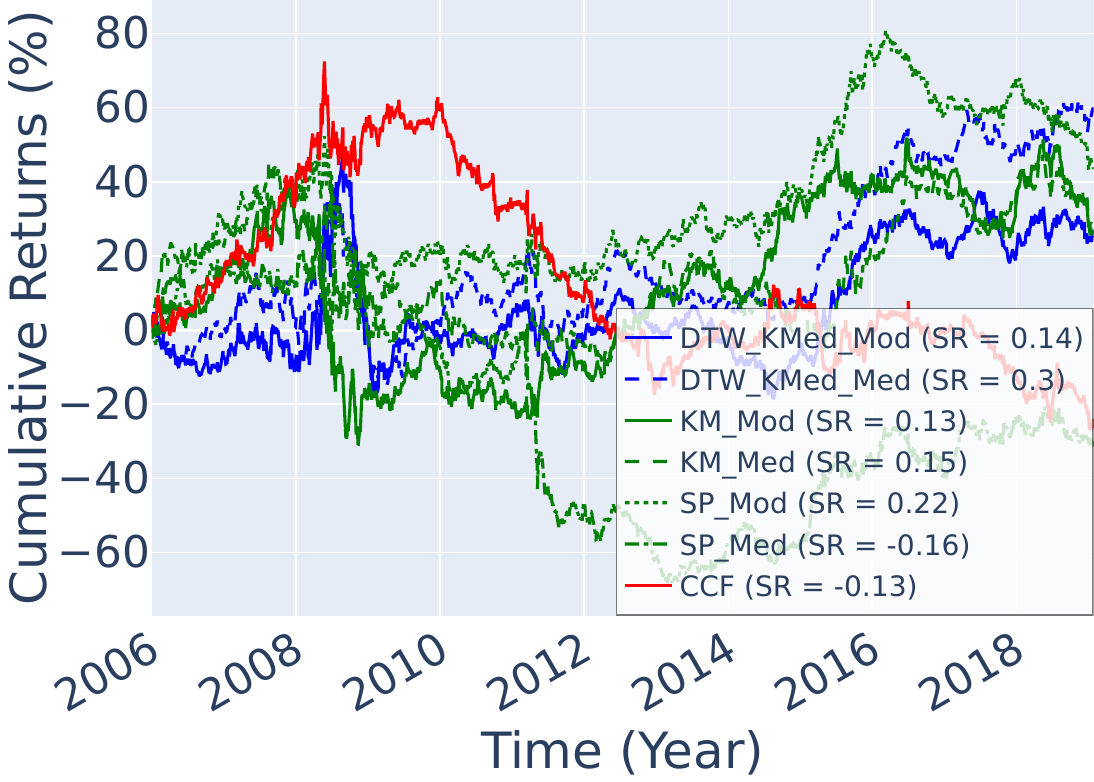}
      &
      \includegraphics[width=\linewidth,trim=0cm 0cm 0cm 0cm,clip]{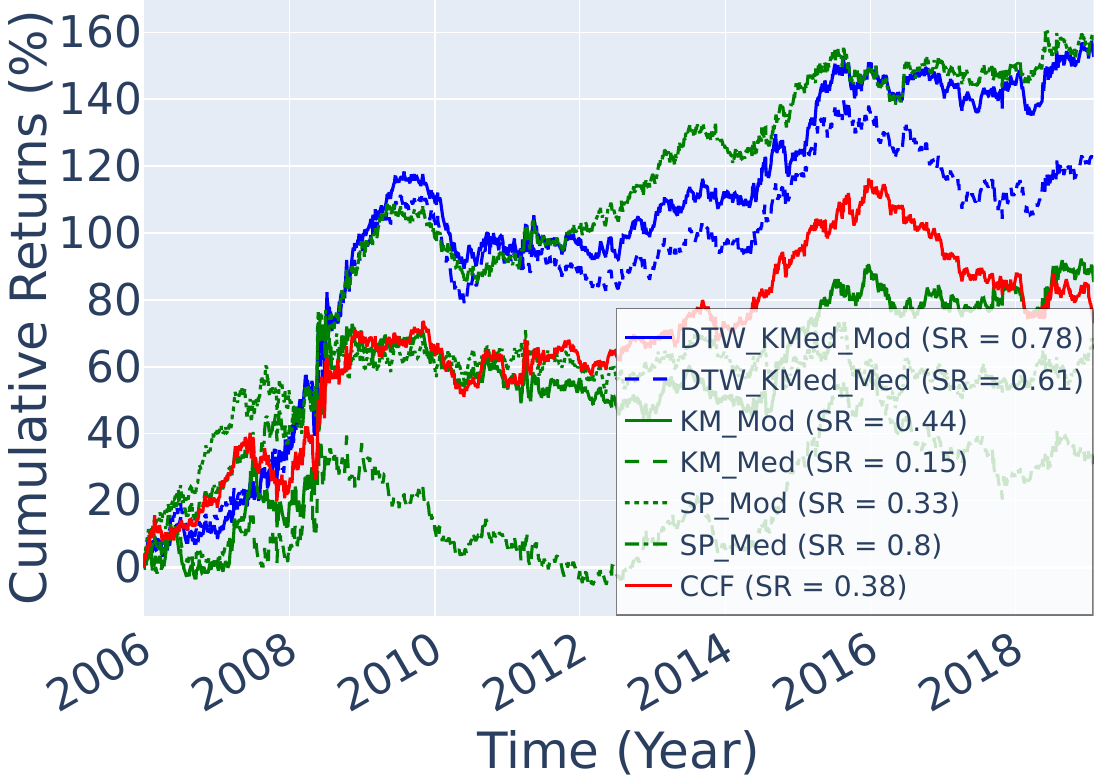}
    \end{tabular}

\captionof{figure}{ETF data set: cumulative PnL for $G_\beta$ strategy (left) and $D_\alpha$ strategy (right) - rescaled to target volatility. The experiment has been set with the values $p = 7$, $\delta = 7$, $\alpha = 0.25$, and $K=5$.}

\label{tab:etf_pnl}
\end{table}

% \vspace{-0.5cm}

\begin{table}[htbp]
  \centering
  \resizebox{\linewidth}{!}{
    \begin{tabular}{p{3.4cm}p{1cm}p{1.1cm}p{1.1cm}p{1.1cm}p{1.1cm}p{2.5cm}p{2.5cm}}
        \toprule
        \multicolumn{1}{c}{\textbf{$G_\beta$ strategy}}  &\multicolumn{5}{c}{\textbf{Benchmark}}  &  \multicolumn{2}{c}{\textbf{Proposed}}  \\
        \cmidrule(lr){2-6} \cmidrule(lr){7-8}
         & CCF & KM\_Mod & KM\_Med & SP\_Mod & SP\_Med &DTW\_KMod\_Mod&DTW\_KMod\_Med\\
        \midrule
        E[Returns]       &-0.019 & 0.02   & 0.022  & 0.033 & -0.024 & 0.021  & 0.045  \\
        Volatility        &0.15   & 0.15   & 0.15   & 0.15  & 0.15   & 0.15   & 0.15   \\
        Downside deviation      &0.116  & 0.115  & 0.116  & 0.113 & 0.123  & 0.111  & 0.115  \\
        Maximum drawdown    &-0.668 & -0.525 & -0.465 & -0.38 & -0.653 & -0.507 & -0.428 \\
        Sortino ratio  &-0.165 & 0.176  & 0.188  & 0.294 & -0.192 & 0.186  & 0.394  \\ 
        Calmar ratio       &-0.029 & 0.038  & 0.047  & 0.088 & -0.036 & 0.041  & 0.106  \\
        Hit rate        &0.492  & 0.512  & 0.499  & 0.509 & 0.509  & 0.512  & 0.513  \\
        Avg. profit / avg. loss   &1.006  & 0.978  & 1.034  & 1.009 & 0.935  & 0.98   & 1.005  \\
        PnL per trade   &-0.76  & 0.8    & 0.87   & 1.325 & -0.939 & 0.818  & 1.799  \\
        Sharpe ratio &-0.128 & 0.134  & 0.146  & 0.223 & -0.158 & 0.137  & 0.302  \\
        P-value     &0.644  & 0.628  & 0.598  & 0.421 & 0.568  & 0.619  & 0.275 \\
        \bottomrule
    \end{tabular}}
\caption{ETF data set: performance metrics for $G_\beta$ strategy - rescaled to target volatility. The experiment has been set with the values $p = 7$, $\delta = 7$, $\alpha = 0.25$, and $K=5$.}
\label{tab:etf_metric_lag}
\end{table}

% \vspace{-1cm}
\begin{table}[htbp]
  \centering
 \resizebox{\linewidth}{!}{
    \begin{tabular}{p{3.4cm}p{1cm}p{1.1cm}p{1.1cm}p{1.1cm}p{1.1cm}p{2.5cm}p{2.5cm}}
        \toprule
        \multicolumn{1}{c}{\textbf{$G_\beta$ strategy}}  &\multicolumn{5}{c}{\textbf{Benchmark}}  &  \multicolumn{2}{c}{\textbf{Proposed}}  \\
        \cmidrule(lr){2-6} \cmidrule(lr){7-8}
         & CCF & KM\_Mod & KM\_Med & SP\_Mod & SP\_Med &DTW\_KMod\_Mod&DTW\_KMod\_Med\\
        \midrule
        E[Returns]       &0.056  & 0.065  & 0.022  & 0.05   & 0.121 & 0.117  & 0.091 \\
        Volatility        &0.15   & 0.15   & 0.15   & 0.15   & 0.15  & 0.15   & 0.15  \\
        Downside deviation      &0.097  & 0.097  & 0.105  & 0.104  & 0.098 & 0.093  & 0.095 \\
        Maximum drawdown    &-0.362 & -0.267 & -0.385 & -0.298 & -0.22 & -0.256 & -0.31 \\
        Sortino ratio  &0.581  & 0.676  & 0.213  & 0.474  & 1.236 & 1.257  & 0.955 \\
        Calmar ratio       &0.156  & 0.245  & 0.058  & 0.166  & 0.548 & 0.456  & 0.293 \\
        Hit rate        &0.504  & 0.502  & 0.5    & 0.508  & 0.513 & 0.521  & 0.517 \\
        Avg. profit / avg. loss   &1.056  & 1.077  & 1.027  & 1.03   & 1.097 & 1.056  & 1.039 \\
        PnL per trade   &2.234  & 2.591  & 0.889  & 1.965  & 4.784 & 4.635  & 3.608 \\
        Sharpe ratio  & 0.375  & 0.435  & 0.149  & 0.33   & 0.804 & 0.779  & 0.606 \\
        P-value     &0.169  & 0.105  & 0.588  & 0.229  & 0.003 & 0.004  & 0.026 \\
        \bottomrule
    \end{tabular}}
\caption{ETF data set: performance metrics for $D_\alpha$ strategy - rescaled to target volatility. The experiment has been set with the values $p = 7$, $\delta = 7$, $\alpha = 0.25$, and $K=5$.}
\label{tab:etf_metric_lead}
\end{table}

% \vspace{-0.5cm}

% Results for the futures data, using the same settings as for the equity data, are presented in Tables [\ref{tab:futures_metric_lag}] and [\ref{tab:futures_metric_lead}], along with Figure \ref{tab:futures_pnl}. For this data set, we do not observe the ability to consistently detect profitable lead-lag relationships for any of the strategies. Full results across all tuning settings are available in the supplemental material.

\begin{table}[!htbp]
  \centering
    \begin{tabular}{p{7cm}p{7cm}}
      \multicolumn{1}{c}{\textbf{$G_\beta$ strategy}}    &  
      \multicolumn{1}{c}{\textbf{$D_\alpha$ strategy}}  \\

      \includegraphics[width=\linewidth,trim=0cm 0cm 0cm 0cm,clip]{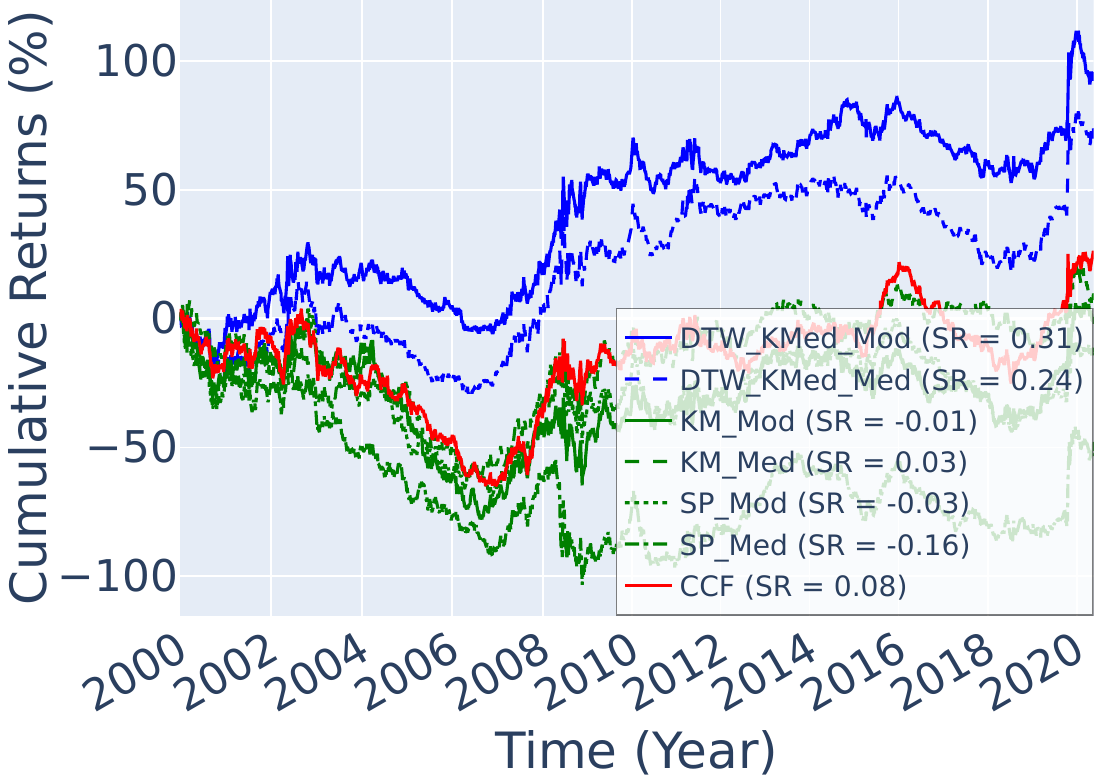}
      &
      \includegraphics[width=\linewidth,trim=0cm 0cm 0cm 0cm,clip]{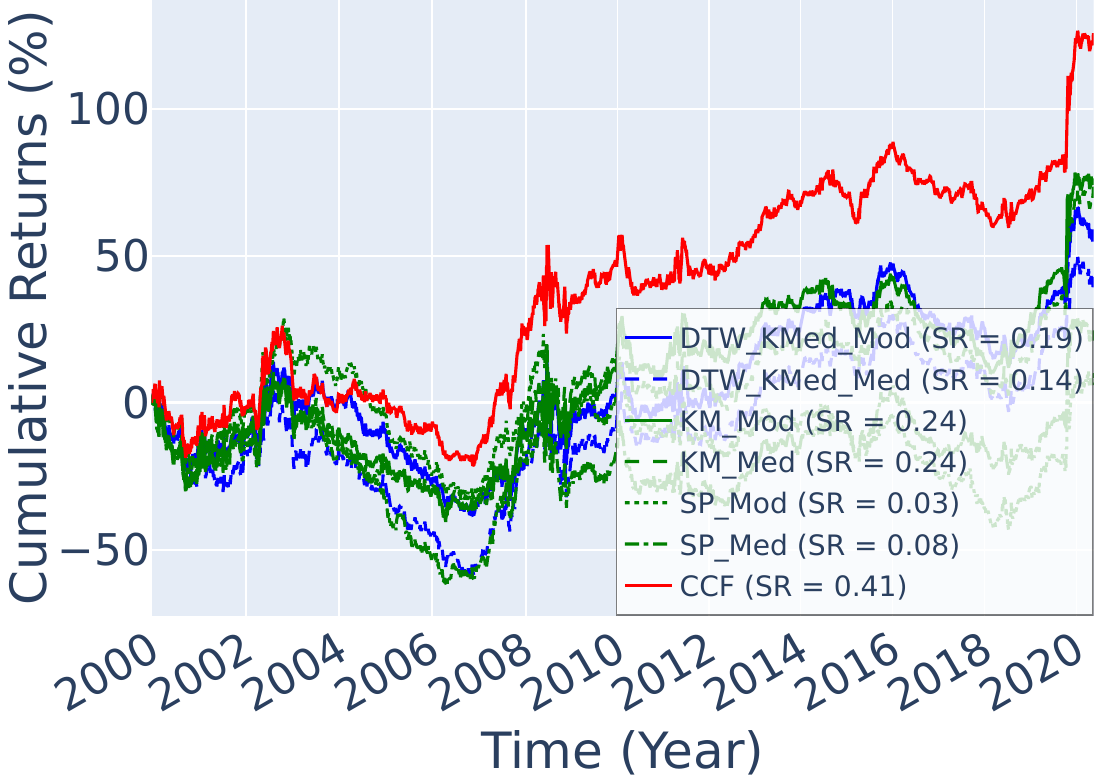}
    \end{tabular}

\vspace{-0.4cm}
\captionof{figure}{Futures data set: cumulative PnL for $G_\beta$ strategy (left) and $D_\alpha$ strategy (right) - rescaled to target volatility. The experiment has been set with the values $p = 7$, $\delta = 7$, $\alpha = 0.25$, and $K=5$.}

\label{tab:futures_pnl}
\end{table}

% \vspace{-0.5cm}

\begin{table}[htbp]
  \centering
  \resizebox{\linewidth}{!}{
    \begin{tabular}{p{3.4cm}p{1cm}p{1.1cm}p{1.1cm}p{1.1cm}p{1.1cm}p{2.5cm}p{2.5cm}}
        \toprule
        \multicolumn{1}{c}{\textbf{$G_\beta$ strategy}}  &\multicolumn{5}{c}{\textbf{Benchmark}}  &  \multicolumn{2}{c}{\textbf{Proposed}}  \\
        \cmidrule(lr){2-6} \cmidrule(lr){7-8}
         & CCF & KM\_Mod & KM\_Med & SP\_Mod & SP\_Med &DTW\_KMod\_Mod&DTW\_KMod\_Med\\
        \midrule
        E[Returns]       &0.013  & -0.001 & 0.004  & -0.005 & -0.024 & 0.047  & 0.036  \\
        Volatility        &0.15   & 0.15   & 0.15   & 0.15   & 0.15   & 0.15   & 0.15   \\
        Downside deviation      &0.104  & 0.102  & 0.103  & 0.107  & 0.109  & 0.103  & 0.104  \\
        Maximum drawdown    &-0.535 & -0.587 & -0.538 & -0.544 & -0.698 & -0.316 & -0.376 \\
        Sortino ratio  &0.121  & -0.01  & 0.042  & -0.049 & -0.217 & 0.453  & 0.347  \\
        Calmar ratio       &0.024  & -0.002 & 0.008  & -0.01  & -0.034 & 0.147  & 0.096  \\
        Hit rate        &0.502  & 0.498  & 0.494  & 0.497  & 0.497  & 0.506  & 0.505  \\
        Avg. profit / avg. loss   &1.007  & 1.008  & 1.028  & 1.006  & 0.984  & 1.033  & 1.023  \\
        PnL per trade   &0.499  & -0.039 & 0.173  & -0.206 & -0.94  & 1.848  & 1.427  \\
        Sharpe ratio &0.084  & -0.007 & 0.029  & -0.035 & -0.158 & 0.31   & 0.24   \\
        P-value     &0.705  & 0.976  & 0.896  & 0.876  & 0.476  & 0.159  & 0.278 \\
        \bottomrule
    \end{tabular}}
\caption{Futures data set: performance metrics for $G_\beta$ strategy - rescaled to target volatility. The experiment has been set with the values $p = 7$, $\delta = 7$, $\alpha = 0.25$, and $K=5$.}
\label{tab:futures_metric_lag}
\end{table}

% \vspace{-0.5cm}
\begin{table}[htbp]
  \centering
 \resizebox{\linewidth}{!}{
    \begin{tabular}{p{3.4cm}p{1cm}p{1.1cm}p{1.1cm}p{1.1cm}p{1.1cm}p{2.5cm}p{2.5cm}}
        \toprule
        \multicolumn{1}{c}{\textbf{$G_\beta$ strategy}}  &\multicolumn{5}{c}{\textbf{Benchmark}}  &  \multicolumn{2}{c}{\textbf{Proposed}}  \\
        \cmidrule(lr){2-6} \cmidrule(lr){7-8}
         & CCF & KM\_Mod & KM\_Med & SP\_Mod & SP\_Med &DTW\_KMod\_Mod&DTW\_KMod\_Med\\
        \midrule
        E[Returns]       &0.061  & 0.036  & 0.036  & 0.005  & 0.012  & 0.029  & 0.021  \\
        Volatility        &0.15   & 0.15   & 0.15   & 0.15   & 0.15  & 0.15   & 0.15  \\
        Downside deviation      &0.107  & 0.108  & 0.108  & 0.11   & 0.105  & 0.107  & 0.107  \\
        Maximum drawdown    &-0.393 & -0.412 & -0.404 & -0.581 & -0.524 & -0.424 & -0.502 \\
        
        Sortino ratio  &0.574  & 0.337  & 0.331  & 0.042  & 0.112  & 0.269  & 0.193  \\
        Calmar ratio       &0.156  & 0.088  & 0.089  & 0.008  & 0.022  & 0.068  & 0.041  \\
        Hit rate        &0.511  & 0.502  & 0.499  & 0.491  & 0.496  & 0.498  & 0.501  \\
        Avg. profit / avg. loss   &1.032  & 1.038  & 1.05   & 1.042  & 1.032  & 1.042  & 1.022  \\
        PnL per trade   &2.439  & 1.445  & 1.424  & 0.183  & 0.468  & 1.139  & 0.822  \\
        Sharpe ratio  &0.41   & 0.243  & 0.239  & 0.031  & 0.079  & 0.191  & 0.138  \\
        P-value     &0.064  & 0.272  & 0.28   & 0.889  & 0.723  & 0.387  & 0.533 \\
        \bottomrule
    \end{tabular}}
\caption{Futures data set: performance metrics for $D_\alpha$ strategy - rescaled to target volatility. The experiment has been set with the values $p = 7$, $\delta = 7$, $\alpha = 0.25$, and $K=5$.}
\label{tab:futures_metric_lead}
\end{table}

\newpage
%\vspace{-0.5cm}
\section{Robustness analysis}
\label{sec: robustness}

We test the robustness of the DTW\_KMod\_Mod and DTW\_KMod\_Med by conducting experiments with different numbers of clusters $K$. Specifically, for the equity data set, we consider $K = \{5,10,15,20\}$. In Table \ref{tab:equity_robust}, we observe that the performance of both algorithms does not change significantly while maintaining a high SR. It is worth noting that the P-values are almost all lower than 0.05, indicating that all the results are statistically significant in our experiments.

Table \ref{tab:etf_robust} presents the performance of the DTW\_KMod\_Mod and DTW\_KMod\_Med tested on the ETF data set. Due to the smaller cross-section for this data set, we only consider the $K$ values of $5$ and $10$. Both algorithms demonstrate fairly good performance for high alpha in the $D_{\alpha}$ strategy.

Table \ref{tab:futures_robust} presents the performance of the DTW\_KMod\_Mod and DTW\_KMod\_Med on the futures data set, with $K$ ranging from $5$ to $20$ in increments of $5$. It is observed that for this data set, the SR for the $D_{\alpha}$ strategy tends to be more sensitive to changes in $K$, and achieving profitability becomes challenging.

\begin{table}[htbp]
  \centering
  \resizebox{\linewidth}{!}{
    \begin{tabular}{p{3.7cm}p{1cm}p{1cm}p{1cm}p{1cm}|p{1cm}p{1cm}p{1cm}p{1cm}}
        \toprule
        \multicolumn{1}{c}{\textbf{DTW\_KMod\_Mod}}  &\multicolumn{4}{c}{\textbf{$G_\beta$ strategy}}  &  \multicolumn{4}{c}{\textbf{$D_\alpha$ strategy}}  \\
        \cmidrule(lr){2-5} \cmidrule(lr){6-9}
         \textbf{$K$} &  5 & 10 & 15 & 20 &  5 & 10 & 15 & 20 \\
        \midrule
        E[Returns]       &0.109  & 0.098  & 0.098  & 0.096  & 0.128  & 0.106 & 0.114  & 0.101  \\
        Volatility        &0.15   & 0.15   & 0.15   & 0.15   & 0.15   & 0.15  & 0.15   & 0.15   \\
        Downside deviation      &0.106  & 0.107  & 0.108  & 0.108  & 0.103  & 0.105 & 0.105  & 0.105  \\
        Maximum drawdown    &-0.205 & -0.192 & -0.198 & -0.235 & -0.276 & -0.27 & -0.221 & -0.272 \\
        
        Sortino ratio  &1.026  & 0.92   & 0.908  & 0.891  & 1.238  & 1.007 & 1.084  & 0.961  \\
        Calmar ratio       &0.531  & 0.512  & 0.497  & 0.408  & 0.463  & 0.392 & 0.514  & 0.372  \\
        Hit rate        &0.514  & 0.512  & 0.516  & 0.514  & 0.52   & 0.519 & 0.522  & 0.519  \\
        Avg. profit / avg. loss   &1.08   & 1.075  & 1.058  & 1.066  & 1.078  & 1.051 & 1.05   & 1.044  \\
        PnL per trade   &4.317  & 3.9    & 3.906  & 3.805  & 5.073  & 4.201 & 4.511  & 4.012  \\
        Sharpe ratio &0.725  & 0.655  & 0.656  & 0.639  & 0.852  & 0.706 & 0.758  & 0.674  \\
        P-value     &0.001  & 0.004  & 0.004  & 0.005  & 0      & 0.002 & 0.001  & 0.003 \\
        
        \midrule
        \midrule
        \multicolumn{1}{c}{\textbf{DTW\_KMod\_Med}}  &\multicolumn{4}{c}{\textbf{$G_\beta$ strategy}}  &  \multicolumn{4}{c}{\textbf{$D_\alpha$ strategy}}  \\
        \cmidrule(lr){2-5} \cmidrule(lr){6-9}
         \textbf{$K$} &  5 & 10 & 15 & 20 &  5 & 10 & 15 & 20 \\
         \midrule
        E[Returns]       &0.115  & 0.097  & 0.083  & 0.092 & 0.14   & 0.116  & 0.099  & 0.112  \\
        Volatility        &0.15   & 0.15  & 0.15   & 0.15   & 0.15   & 0.15   & 0.15   & 0.15   \\
        Downside deviation      &0.107  & 0.107  & 0.109  & 0.109 & 0.102  & 0.104  & 0.105  & 0.104  \\
        Maximum drawdown    &-0.188 & -0.197 & -0.226 & -0.21 & -0.297 & -0.226 & -0.241 & -0.274 \\       
        Sortino ratio  &1.084  & 0.907  & 0.76   & 0.844 & 1.376  & 1.112  & 0.946  & 1.083  \\
        Calmar ratio       &0.614  & 0.493  & 0.367  & 0.437 & 0.472  & 0.513  & 0.413  & 0.41   \\
        Hit rate        &0.512  & 0.513  & 0.509  & 0.513 & 0.523  & 0.519  & 0.514  & 0.517  \\
        Avg. profit / avg. loss   &1.1    & 1.069  & 1.07   & 1.065 & 1.078  & 1.067  & 1.067  & 1.068  \\
        PnL per trade   &4.581  & 3.851  & 3.293  & 3.646 & 5.557  & 4.596  & 3.948  & 4.459  \\
        Sharpe ratio &0.77   & 0.647  & 0.553  & 0.612 & 0.934  & 0.772  & 0.663  & 0.749  \\
        P-value     &0.001  & 0.004  & 0.015  & 0.007 & 0      & 0.001  & 0.003  & 0.001 \\
        
\bottomrule 
\end{tabular}}
\caption{Equity data set: robustness analysis for $K$ - rescaled to target volatility. The experiment has been set with the values $p = 7$, and $\delta = 7$.}
\label{tab:equity_robust}
\end{table}

\begin{table}[htbp]
  \centering
  \resizebox{0.62\linewidth}{!}{
    \begin{tabular}{p{3.7cm}p{1cm}p{1cm}|p{1cm}p{1cm}}
        \toprule
        \multicolumn{1}{c}{\textbf{DTW\_KMod\_Mod}}  &\multicolumn{2}{c}{\textbf{$G_\beta$ strategy}}  &  \multicolumn{2}{c}{\textbf{$D_\alpha$ strategy}}  \\
        \cmidrule(lr){2-3} \cmidrule(lr){4-5}
         \textbf{$K$} &  5 & 10  &  5 & 10  \\
        \midrule
        E[Returns]       & 0.021  & 0.058  & 0.117  & 0.07   \\
        Volatility        &0.15   & 0.15   & 0.15   & 0.15   \\
        Downside deviation      &0.111  & 0.116  & 0.093  & 0.098  \\
        Maximum drawdown    &-0.507 & -0.353 & -0.256 & -0.321 \\
        
        Sortino ratio  &0.186  & 0.505  & 1.257  & 0.715  \\
        Calmar ratio       &0.041  & 0.166  & 0.456  & 0.219  \\
        Hit rate        &0.512  & 0.506  & 0.521  & 0.505  \\
        Avg. profit / avg. loss   &0.98   & 1.059  & 1.056  & 1.067  \\
        PnL per trade   &0.818  & 2.321  & 4.635  & 2.783  \\
        Sharpe ratio &0.137  & 0.39   & 0.779  & 0.468  \\
        P-value     &0.619  & 0.161  & 0.004  & 0.088 \\

        \midrule
        \midrule
        \multicolumn{1}{c}{\textbf{DTW\_KMod\_Med}}  &\multicolumn{2}{c}{\textbf{$G_\beta$ strategy}}  &  \multicolumn{2}{c}{\textbf{$D_\alpha$ strategy}}  \\
         \cmidrule(lr){2-3} \cmidrule(lr){4-5}
         \textbf{$K$} &  5 & 10  &  5 & 10  \\
         \midrule
        E[Returns]       &0.045  & 0.063  & 0.091 & 0.061  \\
        Volatility        &0.15   & 0.15   & 0.15  & 0.15   \\
        Downside deviation      &0.115  & 0.113  & 0.095 & 0.098  \\
        Maximum drawdown    &-0.428 & -0.415 & -0.31 & -0.388 \\      
        Sortino ratio  &0.394  & 0.561  & 0.955 & 0.621  \\
        Calmar ratio       &0.106  & 0.153  & 0.293 & 0.157  \\
        Hit rate        &0.513  & 0.508  & 0.517 & 0.507  \\
        Avg. profit / avg. loss   & 1.005  & 1.054  & 1.039 & 1.045  \\
        PnL per trade   &1.799  & 2.514  & 3.608 & 2.413  \\
        Sharpe ratio &0.302  & 0.422  & 0.606 & 0.405  \\
        P-value     &0.275  & 0.128  & 0.026 & 0.14  \\

\bottomrule 
\end{tabular}}
\caption{ETF data set: robustness analysis for $K$ - rescaled to target volatility. The experiment has been set with the values $p = 7$, and $\delta = 7$.}
\label{tab:etf_robust}
\end{table}

\begin{table}[htbp]
  \centering
  \resizebox{\linewidth}{!}{
    \begin{tabular}{p{3.7cm}p{1cm}p{1cm}p{1cm}p{1cm}|p{1cm}p{1cm}p{1cm}p{1cm}}
        \toprule
        \multicolumn{1}{c}{\textbf{DTW\_KMod\_Mod}}  &\multicolumn{4}{c}{\textbf{$G_\beta$ strategy}}  &  \multicolumn{4}{c}{\textbf{$D_\alpha$ strategy}}  \\
        \cmidrule(lr){2-5} \cmidrule(lr){6-9}
         \textbf{$K$} &  5 & 10 & 15 & 20 &  5 & 10 & 15 & 20 \\
        \midrule
        E[Returns]       &0.047  & 0.051  & 0.064  & 0.055  & 0.029  & 0.042  & 0.048  & 0.04   \\
        Volatility        &0.15   & 0.15   & 0.15   & 0.15   & 0.15   & 0.15  & 0.15   & 0.15   \\
        Downside deviation      &0.103  & 0.102  & 0.103  & 0.104  & 0.107  & 0.105  & 0.105  & 0.105  \\
        Maximum drawdown    &-0.316 & -0.412 & -0.421 & -0.338 & -0.424 & -0.468 & -0.459 & -0.455 \\
        
        Sortino ratio  &0.453  & 0.496  & 0.618  & 0.528  & 0.269  & 0.395  & 0.457  & 0.378  \\
        Calmar ratio       &0.147  & 0.123  & 0.152  & 0.163  & 0.068  & 0.089  & 0.105  & 0.087  \\
        Hit rate        &0.506  & 0.5    & 0.501  & 0.508  & 0.498  & 0.5    & 0.5    & 0.499  \\
        Avg. profit / avg. loss   &1.033  & 1.065  & 1.08   & 1.04   & 1.042  & 1.05   & 1.058  & 1.055  \\
        PnL per trade   &1.848  & 2.015  & 2.534  & 2.19   & 1.139  & 1.649  & 1.908  & 1.577  \\
        Sharpe ratio &0.31   & 0.338  & 0.426  & 0.368  & 0.191  & 0.277  & 0.321  & 0.265  \\
        P-value     &0.159  & 0.124  & 0.053  & 0.095  & 0.387  & 0.21   & 0.147  & 0.231 \\

        \midrule
        \midrule
        \multicolumn{1}{c}{\textbf{DTW\_KMod\_Med}}  &\multicolumn{4}{c}{\textbf{$G_\beta$ strategy}}  &  \multicolumn{4}{c}{\textbf{$D_\alpha$ strategy}}  \\
        \cmidrule(lr){2-5} \cmidrule(lr){6-9}
         \textbf{$K$} &  5 & 10 & 15 & 20 &  5 & 10 & 15 & 20 \\
         \midrule
        E[Returns]       &0.036  & 0.05   & 0.061  & 0.055  & 0.021  & 0.033  & 0.053  & 0.039  \\
        Volatility        &0.15   & 0.15  & 0.15   & 0.15   & 0.15   & 0.15   & 0.15   & 0.15   \\
        Downside deviation      &0.104  & 0.104  & 0.105  & 0.106  & 0.107  & 0.105  & 0.106  & 0.105  \\
        Maximum drawdown    &-0.376 & -0.376 & -0.428 & -0.365 & -0.502 & -0.465 & -0.475 & -0.479 \\
        
        Sortino ratio  &0.347  & 0.48   & 0.583  & 0.521  & 0.193  & 0.313  & 0.505  & 0.374  \\
        Calmar ratio       &0.096  & 0.133  & 0.143  & 0.151  & 0.041  & 0.071  & 0.112  & 0.082  \\
        Hit rate        &0.505  & 0.502  & 0.505  & 0.506  & 0.501  & 0.499  & 0.504  & 0.499  \\
        Avg. profit / avg. loss   &1.023  & 1.055  & 1.057  & 1.048  & 1.022  & 1.045  & 1.052  & 1.054  \\
        PnL per trade   &1.427  & 1.978  & 2.436  & 2.184  & 0.822  & 1.305  & 2.115  & 1.556  \\
        Sharpe ratio &0.24   & 0.332  & 0.409  & 0.367  & 0.138  & 0.219  & 0.355  & 0.261  \\
        P-value     &0.278  & 0.132  & 0.063  & 0.096  & 0.533  & 0.321  & 0.108  & 0.237 \\

\bottomrule 
\end{tabular}}
\caption{Futures data set: robustness analysis for $K$ - rescaled to target volatility. The experiment has been set with the values $p = 7$, and $\delta = 7$.}
\label{tab:futures_robust}
\end{table}

\newpage
%\vspace{-10mm}
\section{CONCLUSION AND FUTURE WORK}
\label{sec: conclusion}

In this study, we introduce a Dynamic Time Warping (DTW) based approach for robustly detecting lead-lag relationships in high-dimensional multivariate time series, with a specific focus on lagged multi-factor models. Our proposed algorithms show promising Sharpe Ratios when applied to financial data sets, indicating their potential economic benefits compared to the benchmark. 

To enhance the methodology further, a possible future direction could involve exploring dynamic selection of the number of clusters $K$. Additionally, another interesting direction would be to delve into the more intricate mixed membership model described in Section \ref{sec: model}, which poses a more challenging task. Investigating intraday lead-lag relationships using, for example, minutely data, could be another fruitful area of research.

% \newpage

\printbibliography

\newpage
\appendix
\section{Appendix} \label{appendix}

\subsection{Performance metrics} \label{sec:performance}
The cumulative PnL is the sum of daily PnL across all trading days

\begin{equation}
\text{Cumulative PnL} = \sum\text{PnL}_{\text{rescaled}}.
\end{equation}

The annualized expected excess return (E[Returns]) is a measure of the excess return earned by an investment over a benchmark index during a one-year period. This value can be calculated by

\begin{equation}
\text{E[Returns]} = \text{AVG}(\text{PnL}_{\text{rescaled}})\cdot 252.
\end{equation}

The annualized volatility measures the one-year risk associated with an investment

\begin{equation}
\text{Volatility} = \text{STD}(\text{PnL}_{\text{rescaled}})\cdot \sqrt{252}.
\end{equation}

Additionally, we calculate downside deviation and maximum drawdown to measure downside risk. The Sortino ratio is often preferred by investors more concerned with downside risk than overall risk or volatility. It is derived by

\begin{equation}
\text{Sortino ratio} = \frac{\text{E[Returns]}}{\text{downside deviation}}.
\end{equation}

The Calmar ratio is frequently utilized by investors with a focus on long-term risk and downside protection. A higher Calmar ratio signifies that the strategy has produced superior returns relative to its maximum drawdown, while a lower Calmar ratio suggests underperformance considering the level of risk taken. This ratio is calculated by

\begin{equation}
\text{Calmar ratio} = \frac{\text{E[Returns]}}{\text{maximum drawdown}}.
\end{equation}

The hit rate, also known as the win rate or success rate, measures the percentage of successful trades made by the strategy. It is defined as

\begin{equation}
\text{Hit rate} = \frac{|\text{PnL}_{\text{rescaled}}^{+}|}{|\text{PnL}_{\text{rescaled}}|},
\end{equation}
where $|\text{PnL}_{\text{rescaled}}^{+}|$ is the number of profitable trades, and $|\text{PnL}_{\text{rescaled}}|$ is the total number of trades.

The average profit / average loss (avg. profit / avg. loss) ratio measures the strategy's average profit size relative to its average loss size

\begin{equation}
\text{Avg. profit / avg. loss} = \frac{\text{AVG}(\text{PnL}_{\text{rescaled}}^{+})}{\text{AVG}(\text{PnL}_{\text{rescaled}}^{-})},
\end{equation}
where $\text{AVG}(\text{PnL}_{\text{rescaled}}^{+})$ is the average profit per trade, and $\text{AVG}(\text{PnL}_{\text{rescaled}}^{-})$ is the average loss per trade.

The PnL per trade illustrates the amount earned by the strategy, in basis points, for each basket of $G_\beta$ or $D_\alpha$ traded in the markets (excluding transaction costs). It is given by

\begin{equation}
\text{PnL per trade} = \text{AVG}(\text{PnL}_{\text{rescaled}})\cdot 10^{4},
\end{equation}
where we assume that the strategy trades the same amount of notional every day (i.e.,  a constant unit bet size is used every trading day).

We also compute the annualized Sharpe ratio to quantify the profit gained per unit of risk taken
\begin{equation}
\text{Sharpe ratio} = \frac{\text{AVG}(\text{PnL}_{\text{rescaled}})}{\text{STD}(\text{PnL}_{\text{rescaled}})} \cdot\sqrt{252}.
\end{equation}

It is important to assess the statistical significance of Sharpe ratio when back-testing a sample of hypothetical strategies [\cite{bailey2014deflated}, \cite{ledoit2008robust}, \cite{michael2022option}]. We use a test with the null hypothesis $H_0: \text{Sharpe ratio}=0$, and implement the method proposed by [\cite{bailey2014deflated}] to compute the test statistic

\begin{equation}
\frac{(\text{Sharpe ratio})\cdot\sqrt{T-1}}{\sqrt{1-\gamma_{1}\cdot (\text{Sharpe ratio})+(\gamma_{2}-1)\cdot(\text{Sharpe ratio})^{2} / 4}},
\end{equation}

\noindent
where the Sharpe ratio is what we are testing, $T$ represents the length of the sample, and $\gamma_{1}$ and $\gamma_{2}$ are the skewness and kurtosis of the returns distribution for the selected strategy, respectively. Under the null hypothesis, this test statistic is assumed to follow a standard normal distribution.

To assess the predictive performance of our algorithm, we create a straightforward trading strategy. If this strategy proves profitable with a statistically significant Sharpe ratio, it signifies our ability to leverage the discovered lead-lag relationships for the prediction task.

\newpage

\subsection{Synthetic data experiments} 
\label{sec:synthetic}

Figure \ref{tab:lag_noise_pnl} presents the results of synthetic data experiments for the $G_\beta$ strategy (left) and $D_\alpha$ strategy (right). In both the Homogeneous Setting and Heterogeneous Setting, the SR of all algorithms consistently increased with the number of days. Additionally, when the noise level increased from 1 to 2, the SR of the algorithms was not significantly affected. In particular, DTW\_KMed\_Mod and DTW\_KMed\_Med  outperform other algorithms in the $G_\beta$ strategy.

In Figures \ref{tab:mod_noise_pnl} and \ref{tab:med_noise_pnl}, as the number of days increases, we observe a consistent pattern for $G_\beta$ and $D_\alpha$ in both the $G_\beta$ strategy (left) and $D_\alpha$ strategy (right), regardless of whether it is in the homogeneous setting or heterogeneous setting. Although there is a slight deviation as the level of noise increases from 1 to 2, overall, the results of both strategies align with the experimental expectations, primarily due to the structure of the lag matrix $L$.

\begin{table}[htbp]
  \centering
  \resizebox{\linewidth}{!}{
    \begin{tabular}{p{4cm}p{4cm}|p{4cm}p{4cm}}
      \multicolumn{2}{c}{\textbf{$G_\beta$ strategy}}&\multicolumn{2}{c}{\textbf{$D_\alpha$ strategy}}\\
      \cmidrule(lr){1-2} \cmidrule(lr){3-4}\\
      \multicolumn{1}{c}{\textbf{Homogeneous Setting}}    &  \multicolumn{1}{c}{\textbf{Heterogeneous Setting}}  &      \multicolumn{1}{c}{\textbf{Homogeneous Setting}}    &  \multicolumn{1}{c}{\textbf{Heterogeneous Setting}}  \\

      \includegraphics[width=\linewidth,trim=0cm 0cm 0cm 0cm,clip]{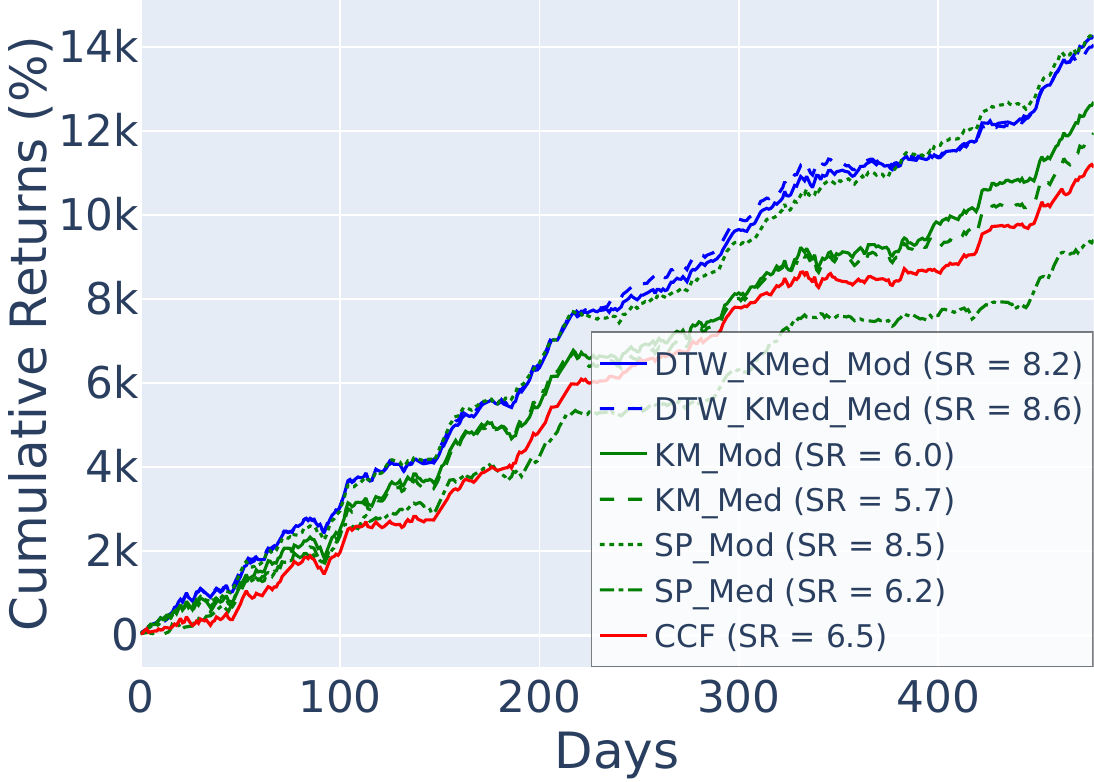}

      \includegraphics[width=\linewidth,trim=0cm 0cm 0cm 0cm,clip]{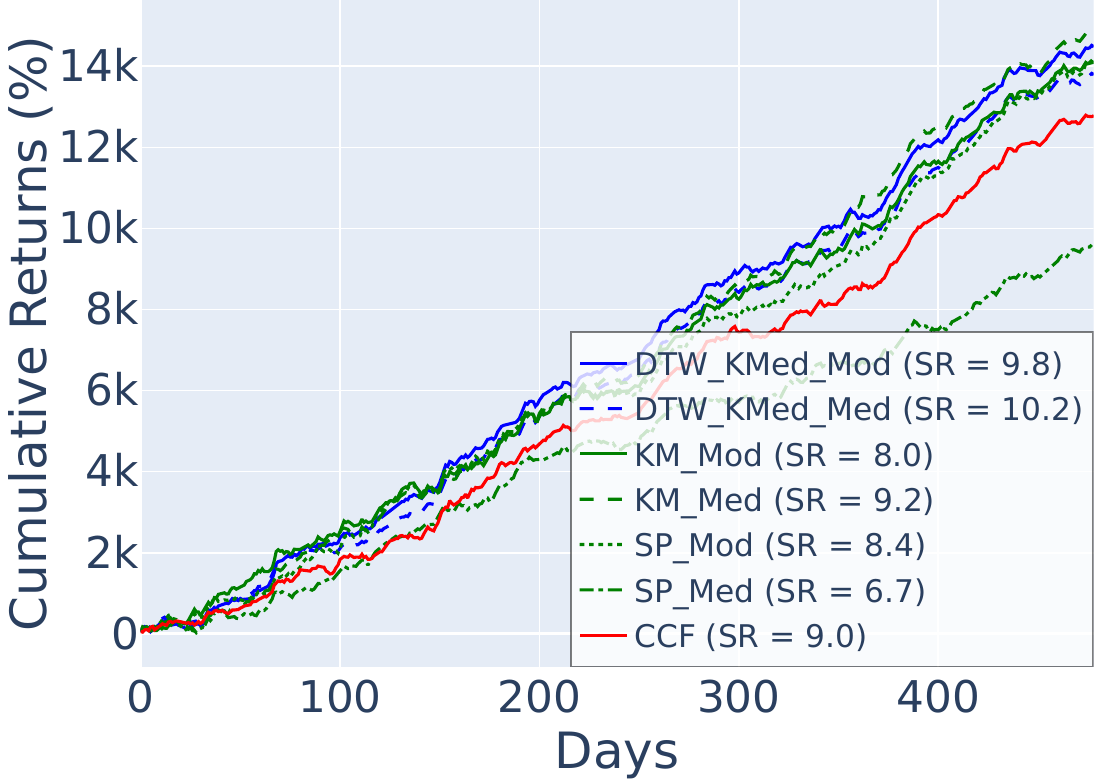}

        \includegraphics[width=\linewidth,trim=0cm 0cm 0cm 0cm,clip]{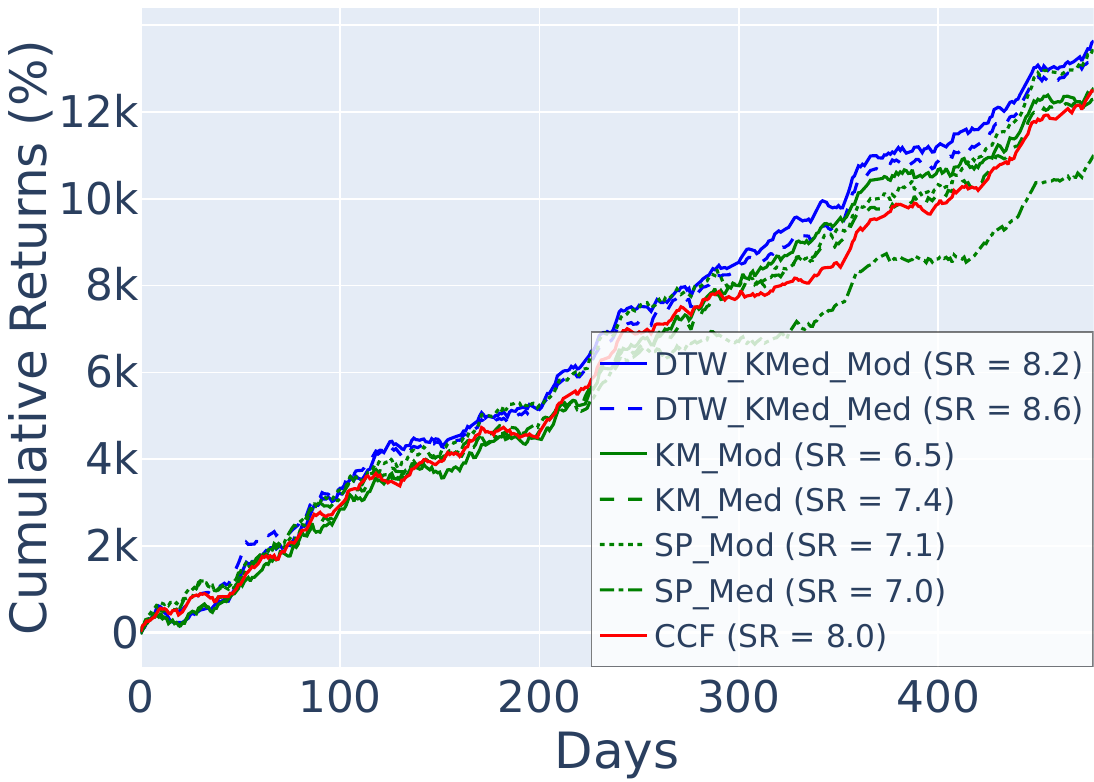}

    &
      \includegraphics[width=\linewidth,trim=0cm 0cm 0cm 0cm,clip]{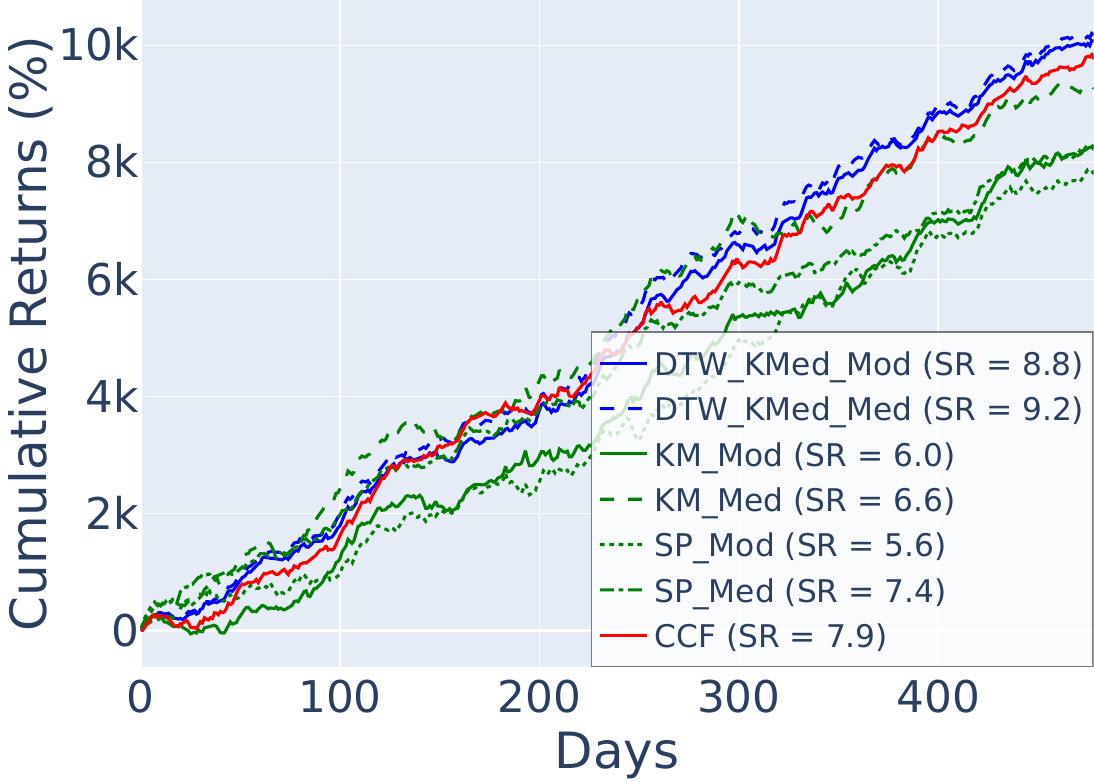}

      \includegraphics[width=\linewidth,trim=0cm 0cm 0cm 0cm,clip]{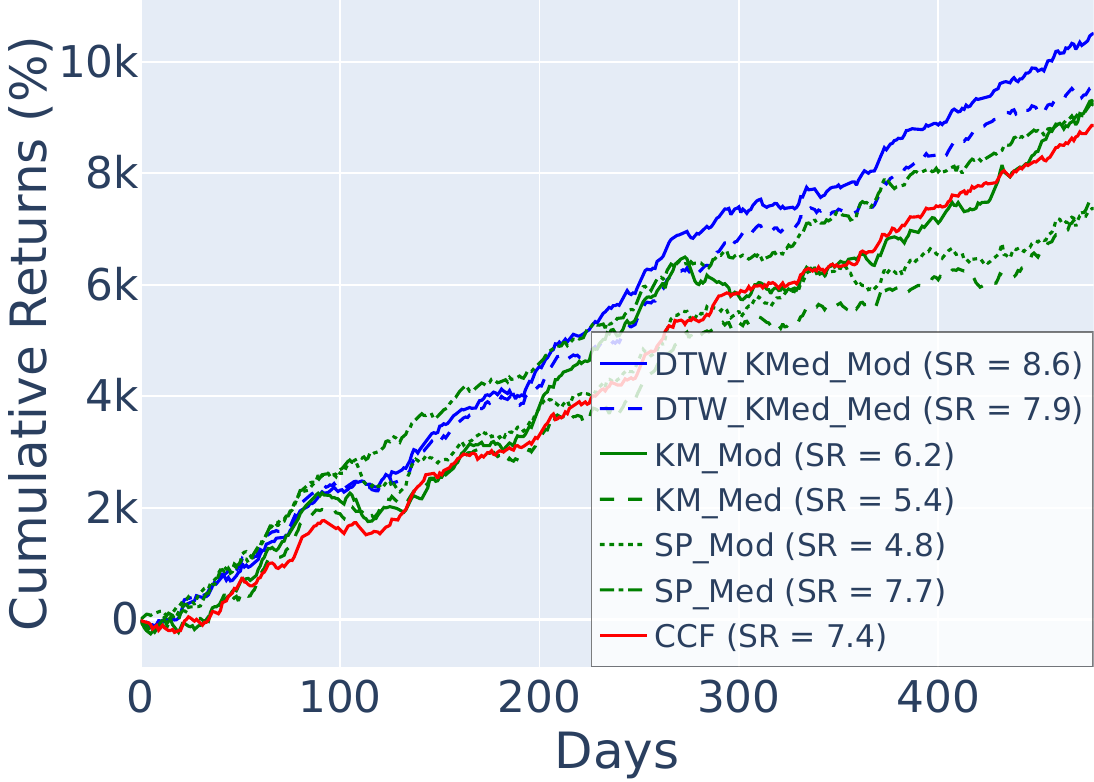}

        \includegraphics[width=\linewidth,trim=0cm 0cm 0cm 0cm,clip]{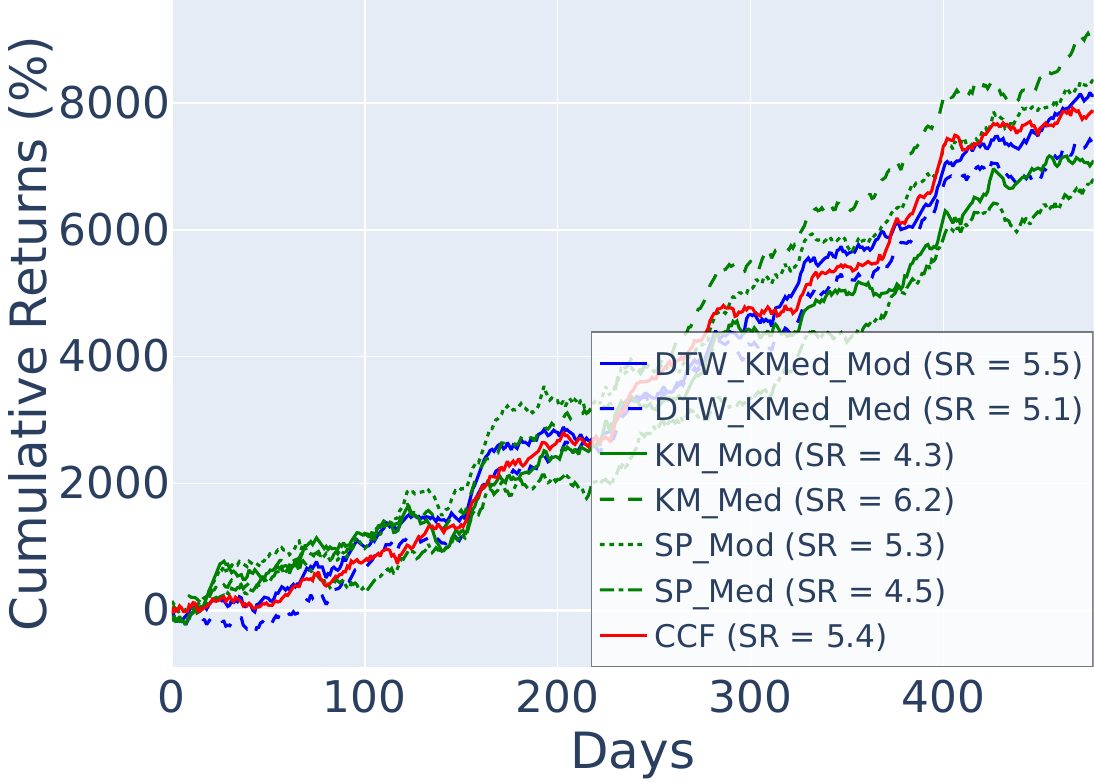}

    &
          \includegraphics[width=\linewidth,trim=0cm 0cm 0cm 0cm,clip]{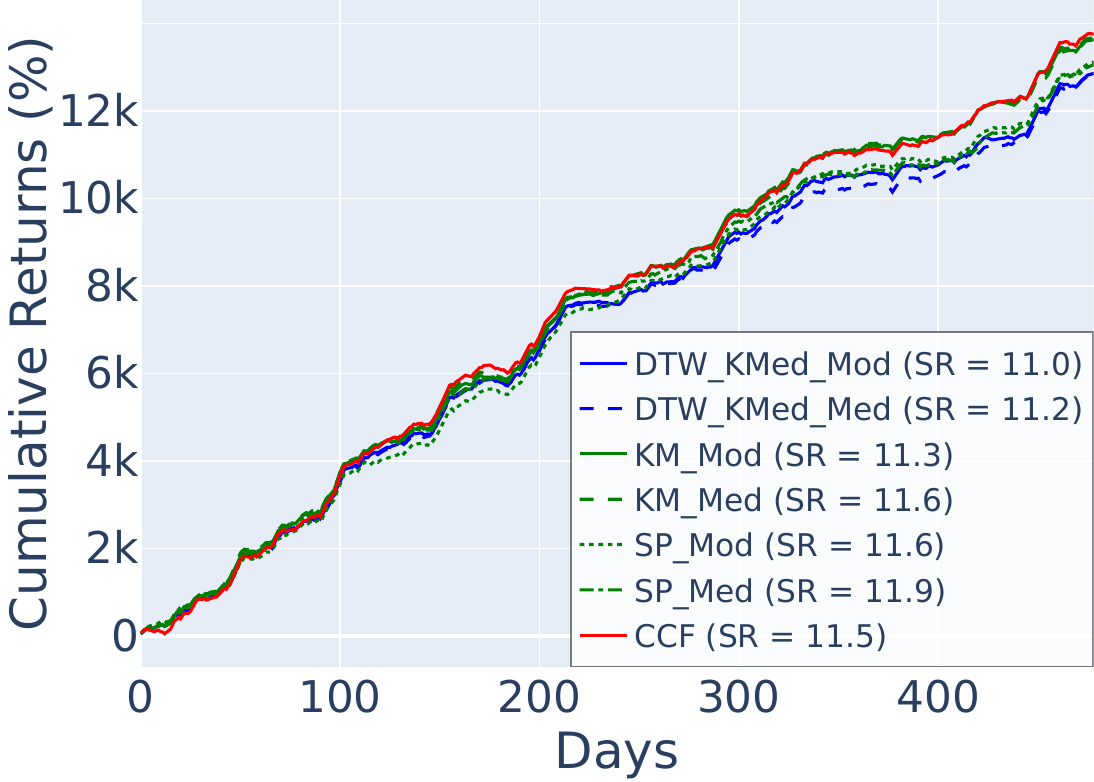}

      \includegraphics[width=\linewidth,trim=0cm 0cm 0cm 0cm,clip]{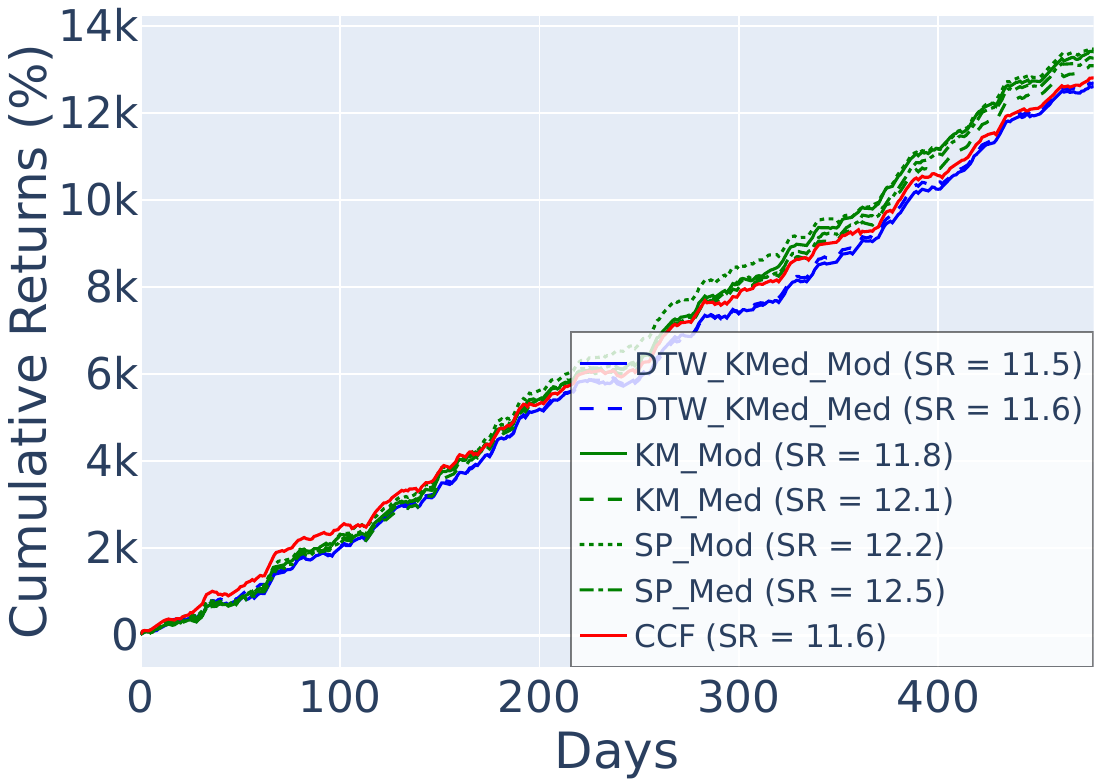}

        \includegraphics[width=\linewidth,trim=0cm 0cm 0cm 0cm,clip]{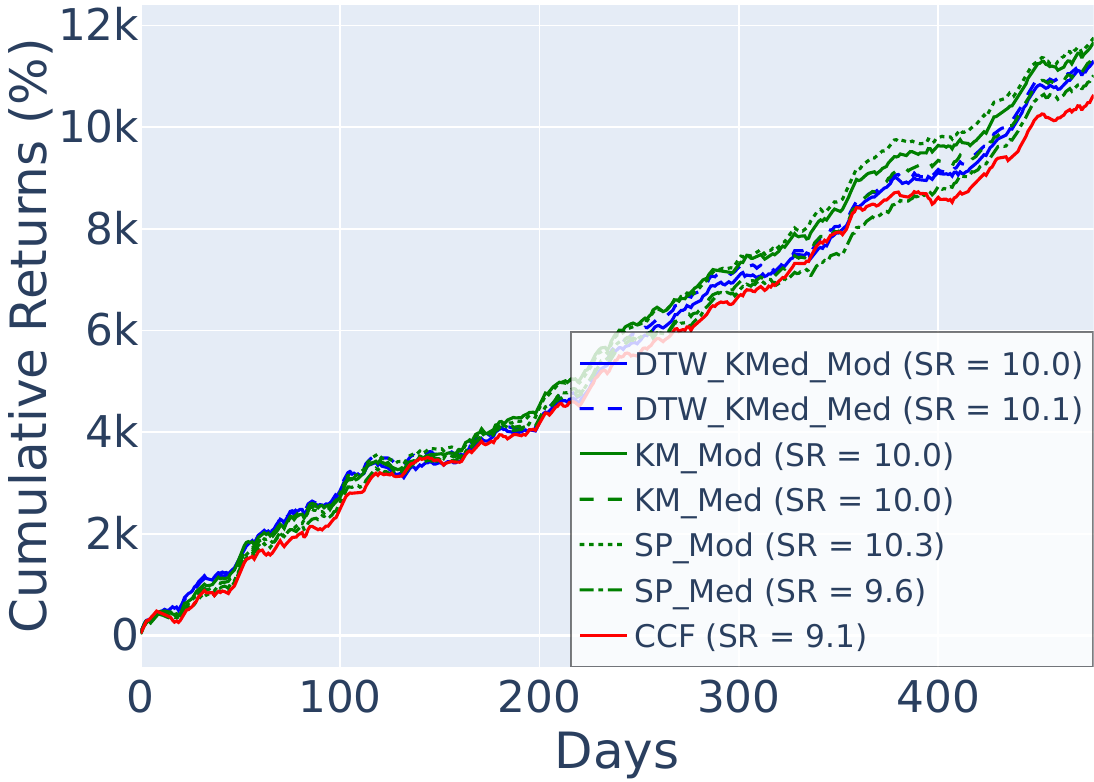}

    &
      \includegraphics[width=\linewidth,trim=0cm 0cm 0cm 0cm,clip]{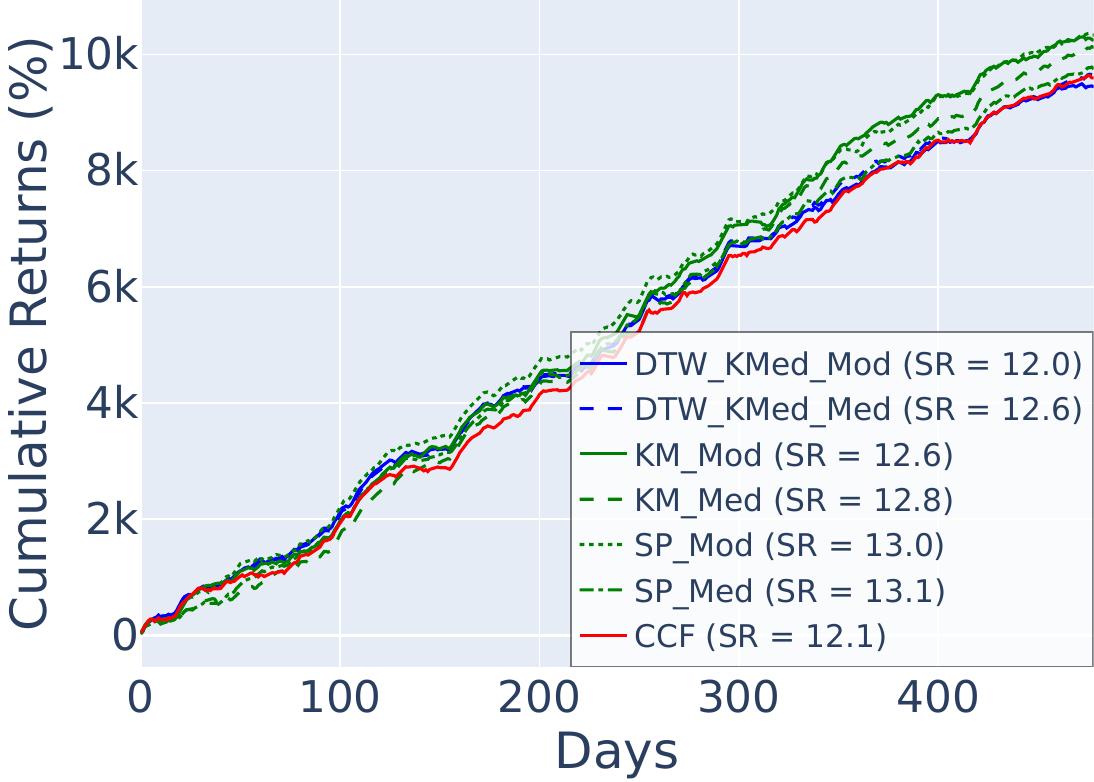}

      \includegraphics[width=\linewidth,trim=0cm 0cm 0cm 0cm,clip]{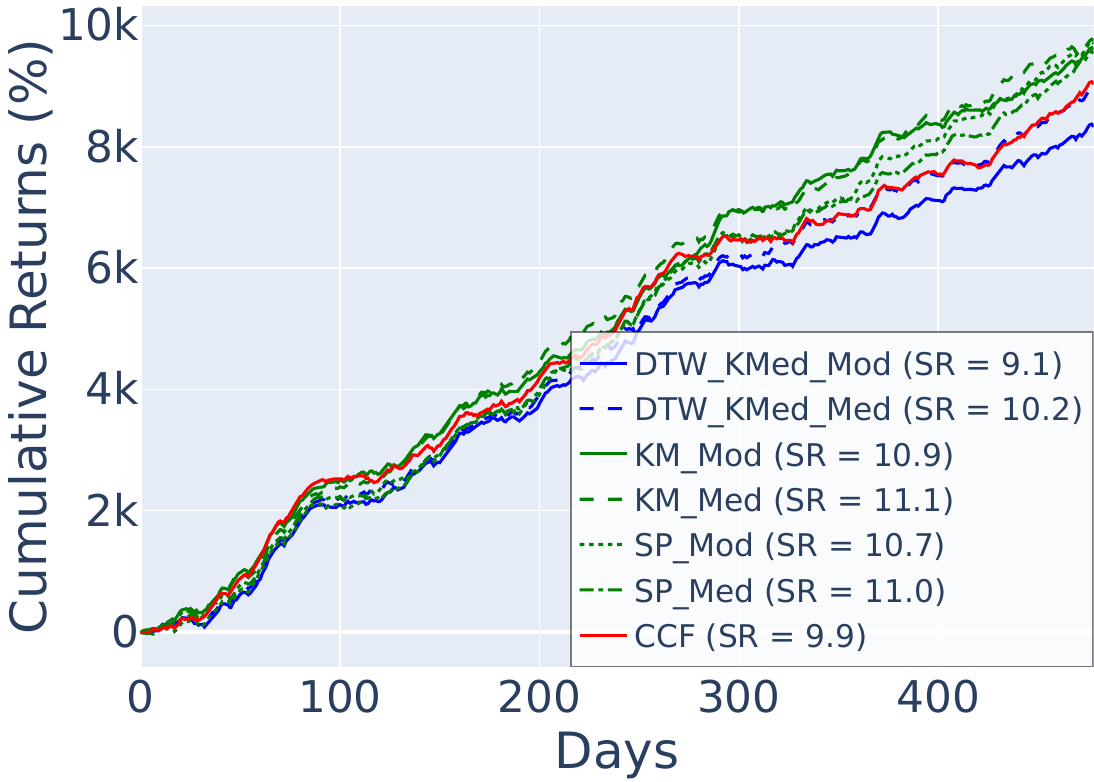}

        \includegraphics[width=\linewidth,trim=0cm 0cm 0cm 0cm,clip]{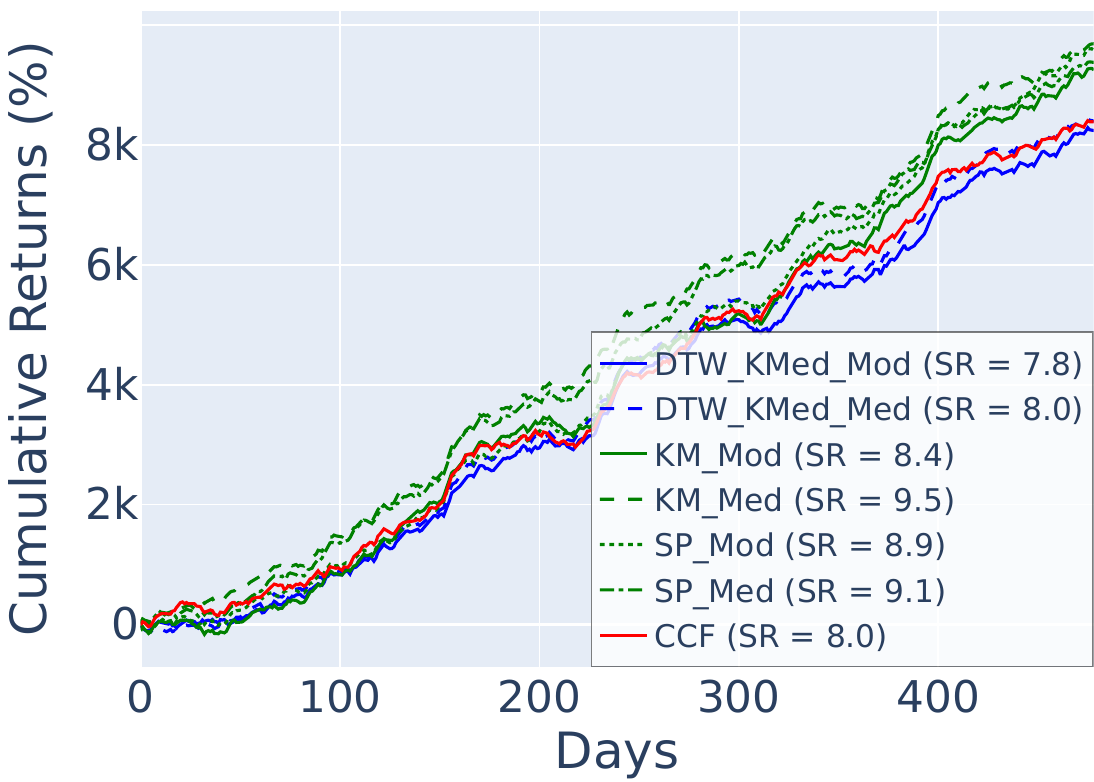}
    
    \\
    \multicolumn{1}{c}{$k=1$} & \multicolumn{1}{c}{$k=2$} & \multicolumn{1}{c}{$k=1$}  & \multicolumn{1}{c}{$k=2$} \\
    \end{tabular}}
\captionof{figure}{Left: $G_\beta$ strategy. Right: $D_\alpha$ strategy. The synthetic data set cumulative PnL experiment has been set with the values $n = 120$, $p = 1$, $\delta = 1$, and $\alpha = 0.25$. From top panel
to bottom panel, low noise $\sigma = 1$,  Medium noise $\sigma = 1.5$, and high noise $\sigma = 2$.}
\label{tab:lag_noise_pnl}
\end{table}

%%%%%%%%%%%%%%%%%%%%%%%%%%%%%%%%%%%%%

\begin{table}[htbp]
  \centering
  \resizebox{\linewidth}{!}{
    \begin{tabular}{p{4cm}p{4cm}|p{4cm}p{4cm}}
      \multicolumn{2}{c}{\textbf{$G_\beta$ strategy}}&\multicolumn{2}{c}{\textbf{$D_\alpha$ strategy}}\\
      \cmidrule(lr){1-2} \cmidrule(lr){3-4}\\
      \multicolumn{1}{c}{\textbf{Homogeneous Setting}}    &  \multicolumn{1}{c}{\textbf{Heterogeneous Setting}}  &      \multicolumn{1}{c}{\textbf{Homogeneous Setting}}    &  \multicolumn{1}{c}{\textbf{Heterogeneous Setting}}  \\

      \includegraphics[width=\linewidth,trim=0cm 0cm 0cm 2cm,clip]{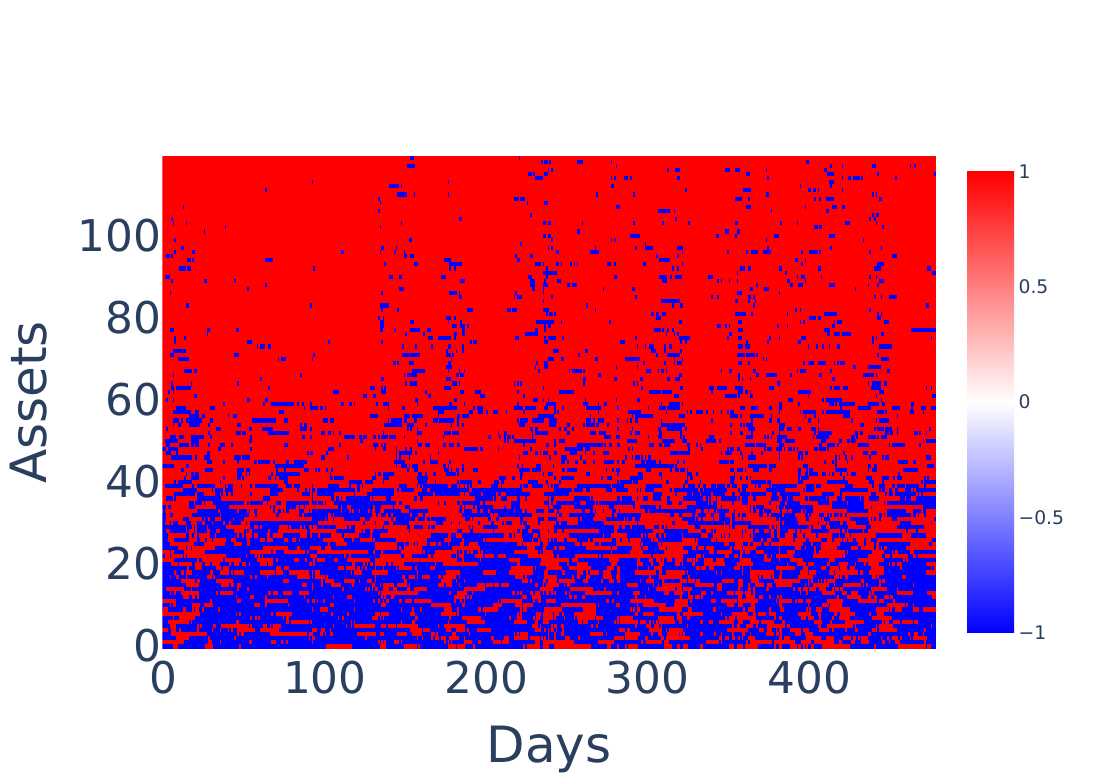}

      \includegraphics[width=\linewidth,trim=0cm 0cm 0cm 2cm,clip]{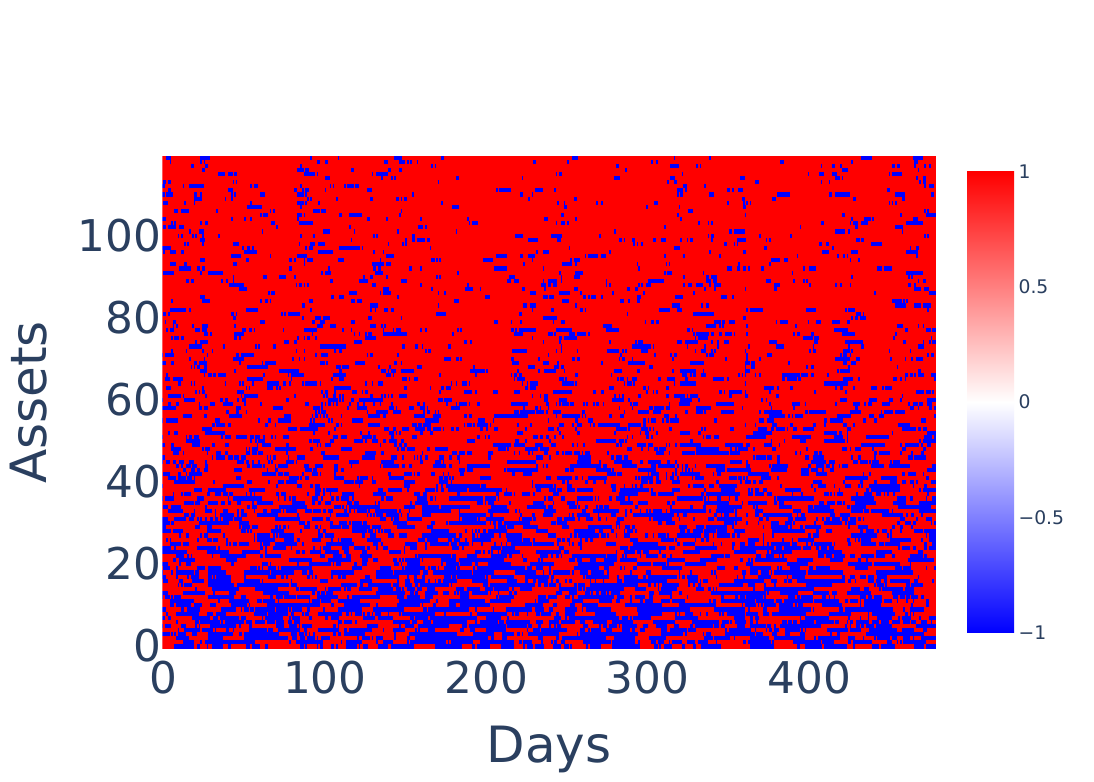}

      \includegraphics[width=\linewidth,trim=0cm 0cm 0cm 2cm,clip]{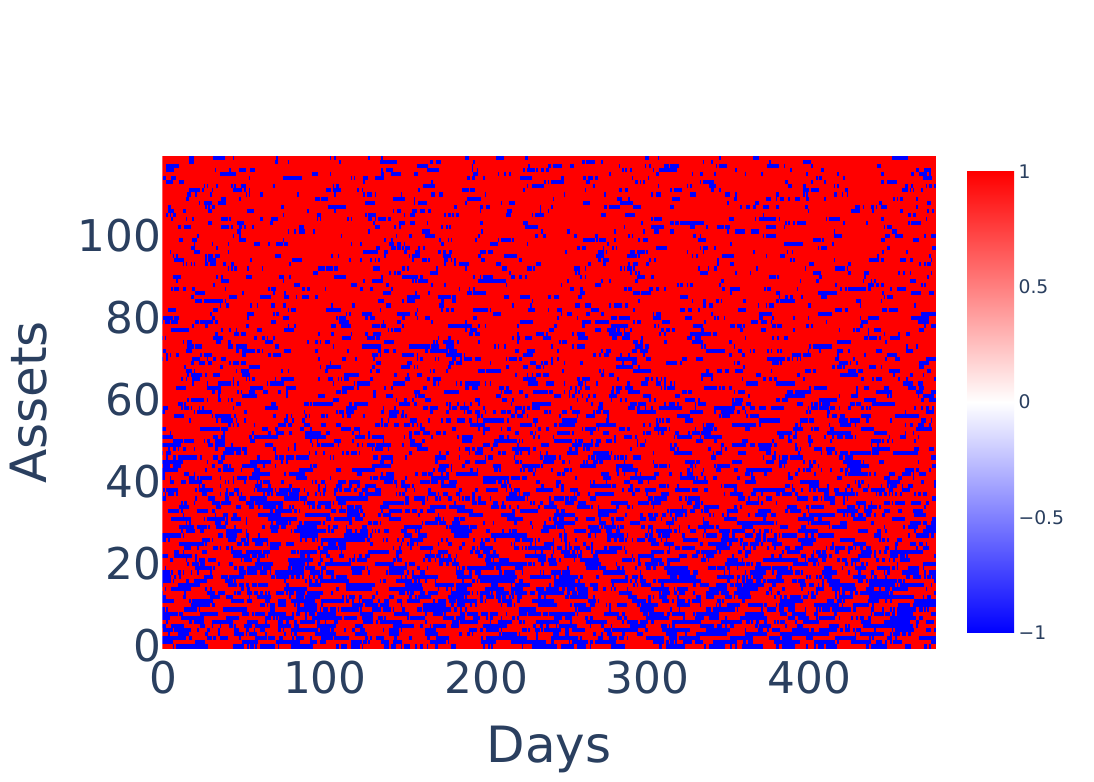}

    &
      \includegraphics[width=\linewidth,trim=0cm 0cm 0cm 2cm,clip]{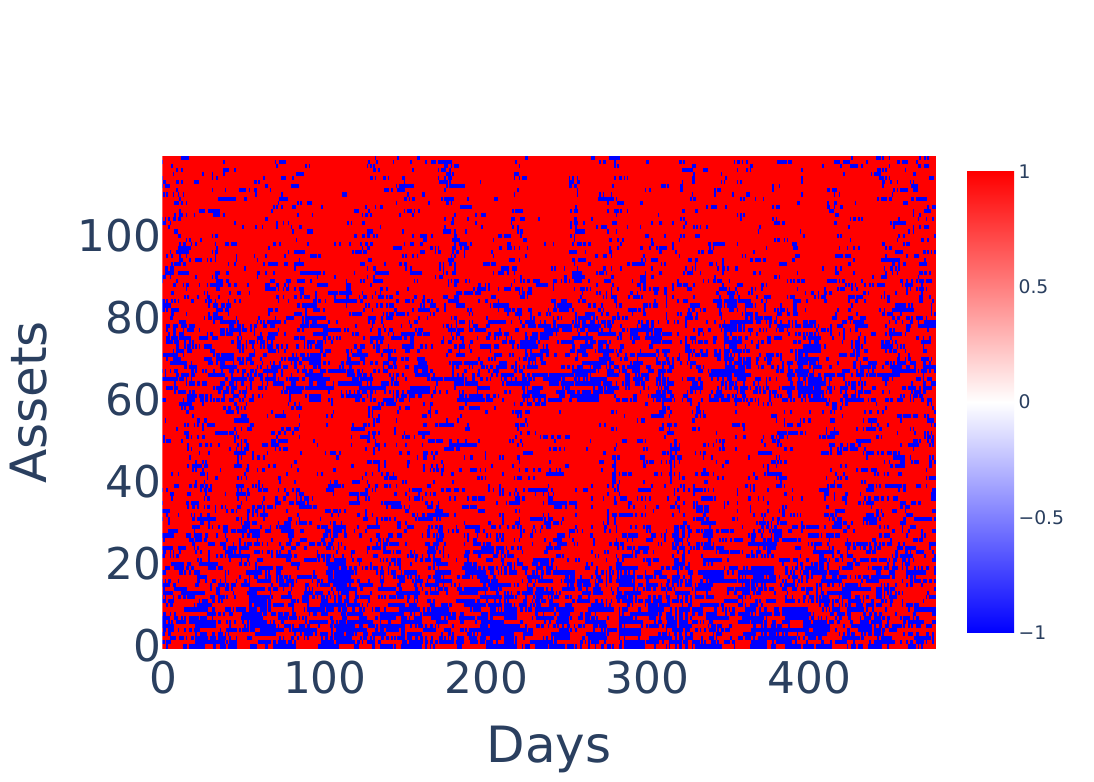}

      \includegraphics[width=\linewidth,trim=0cm 0cm 0cm 2cm,clip]{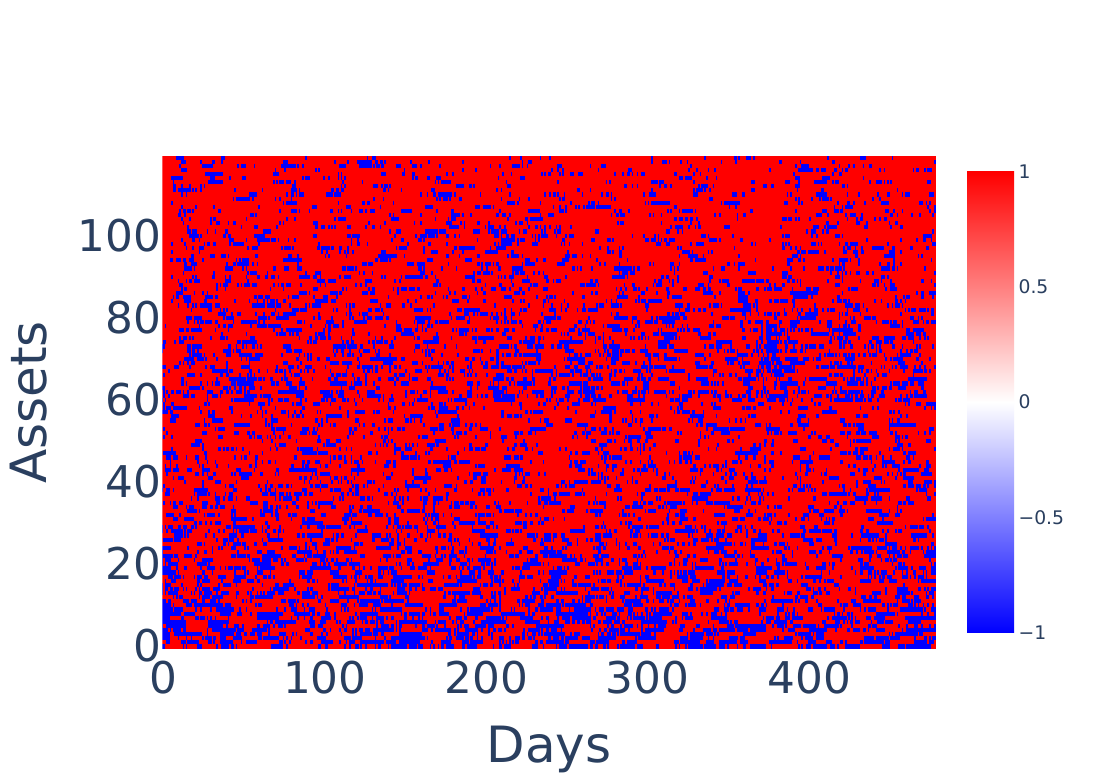}

      \includegraphics[width=\linewidth,trim=0cm 0cm 0cm 2cm,clip]{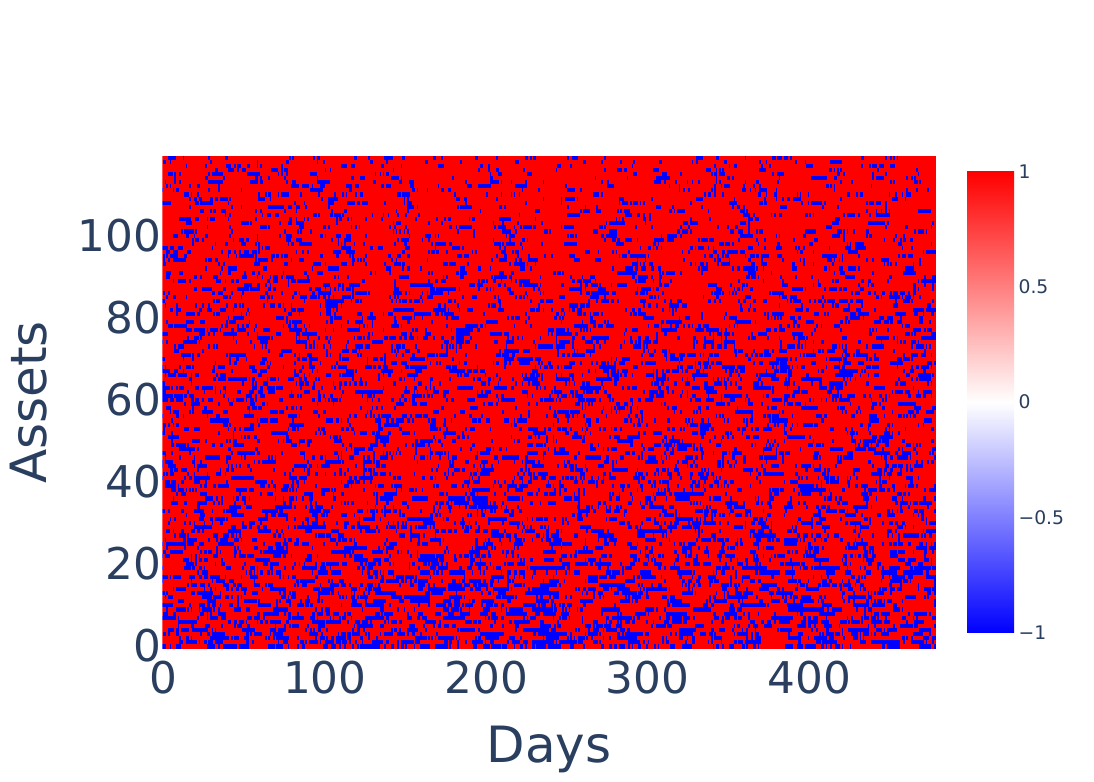}
&
     \includegraphics[width=\linewidth,trim=0cm 0cm 0cm 2cm,clip]{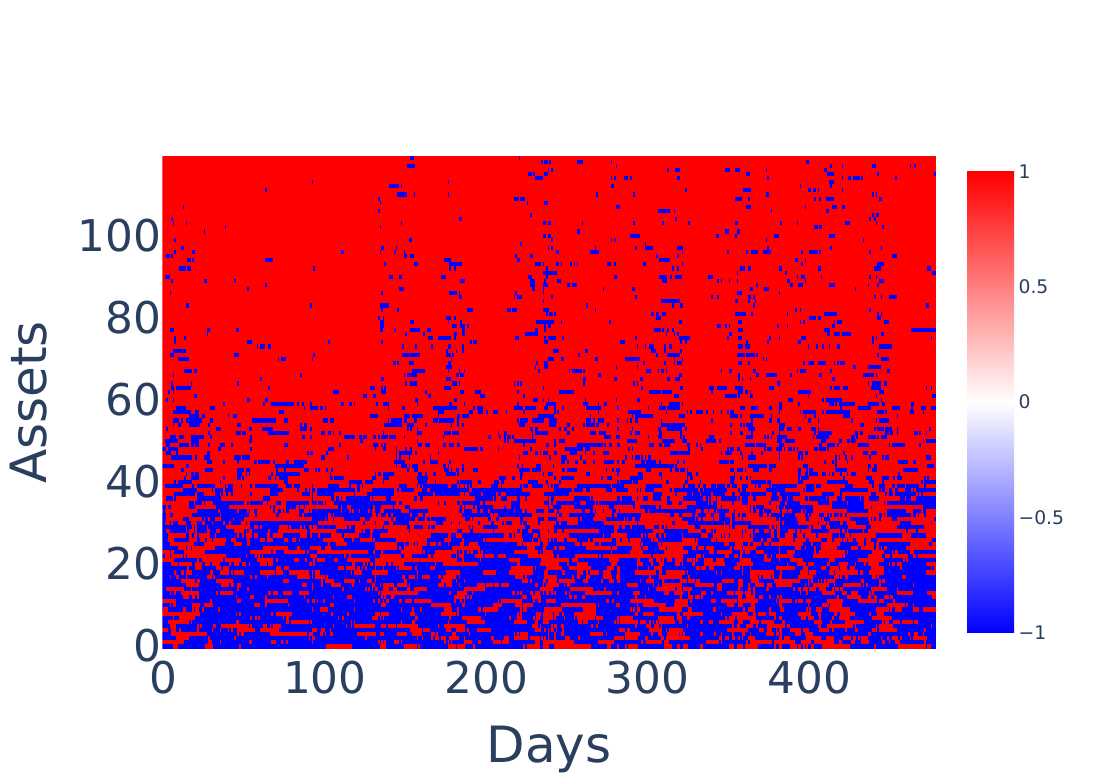}

      \includegraphics[width=\linewidth,trim=0cm 0cm 0cm 2cm,clip]{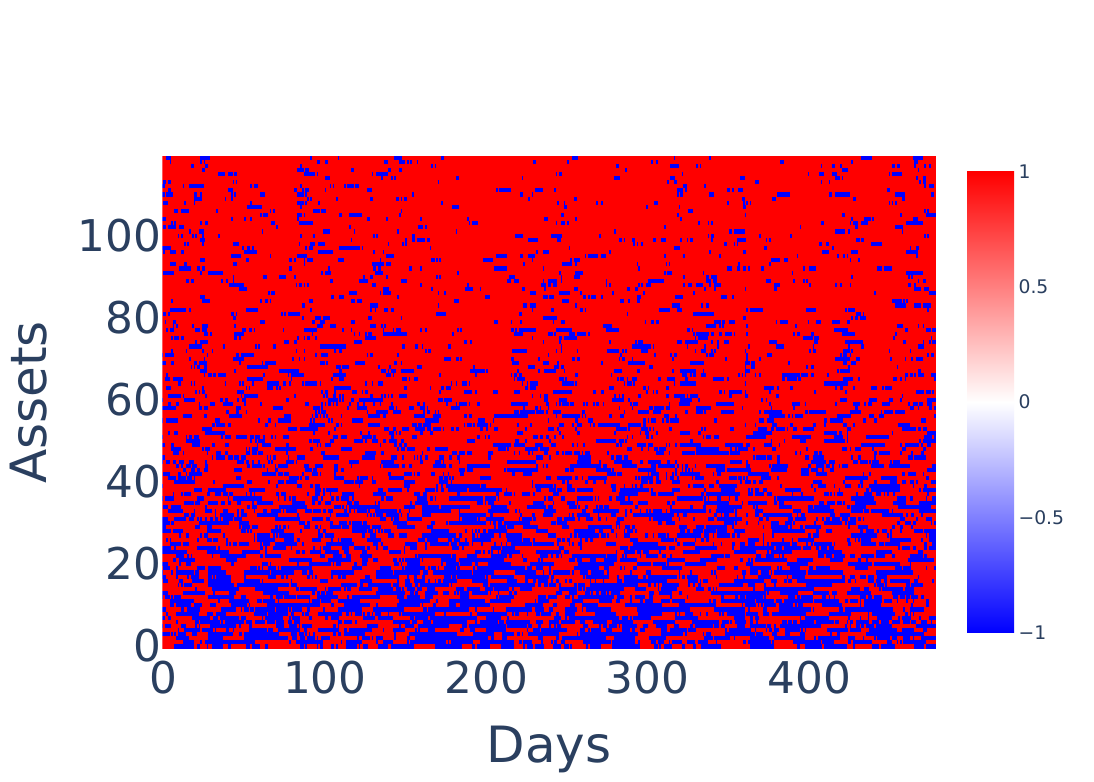}

      \includegraphics[width=\linewidth,trim=0cm 0cm 0cm 2cm,clip]{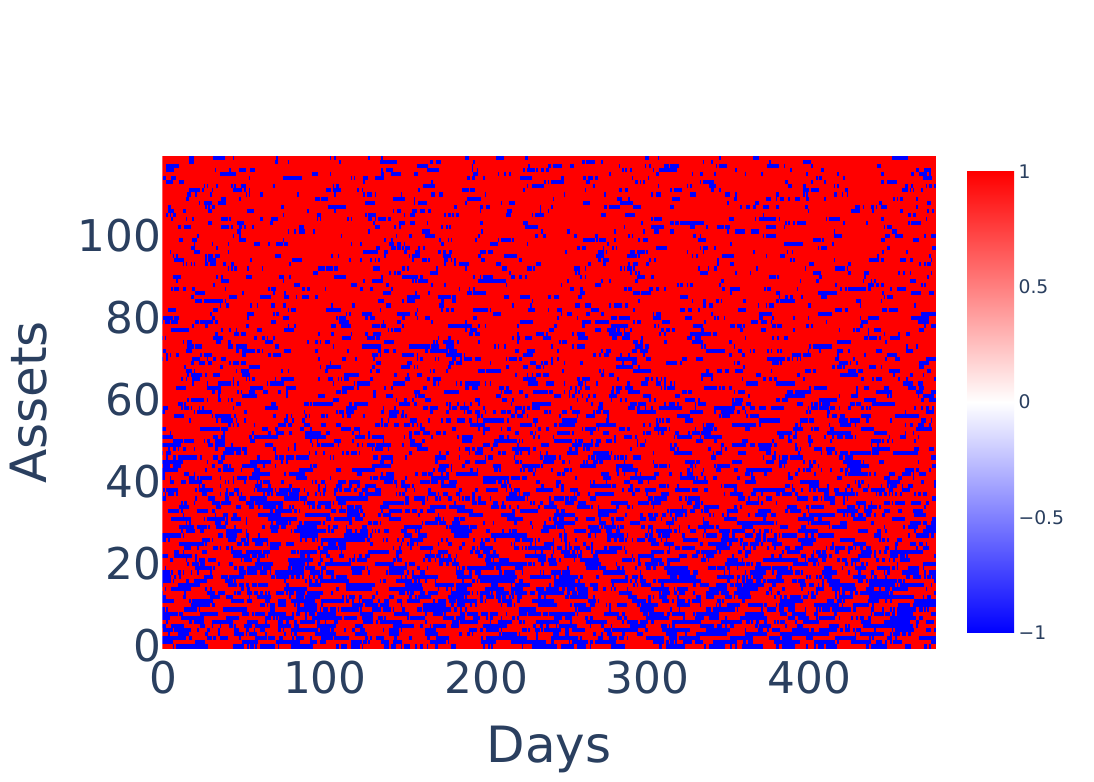}

    &
      \includegraphics[width=\linewidth,trim=0cm 0cm 0cm 2cm,clip]{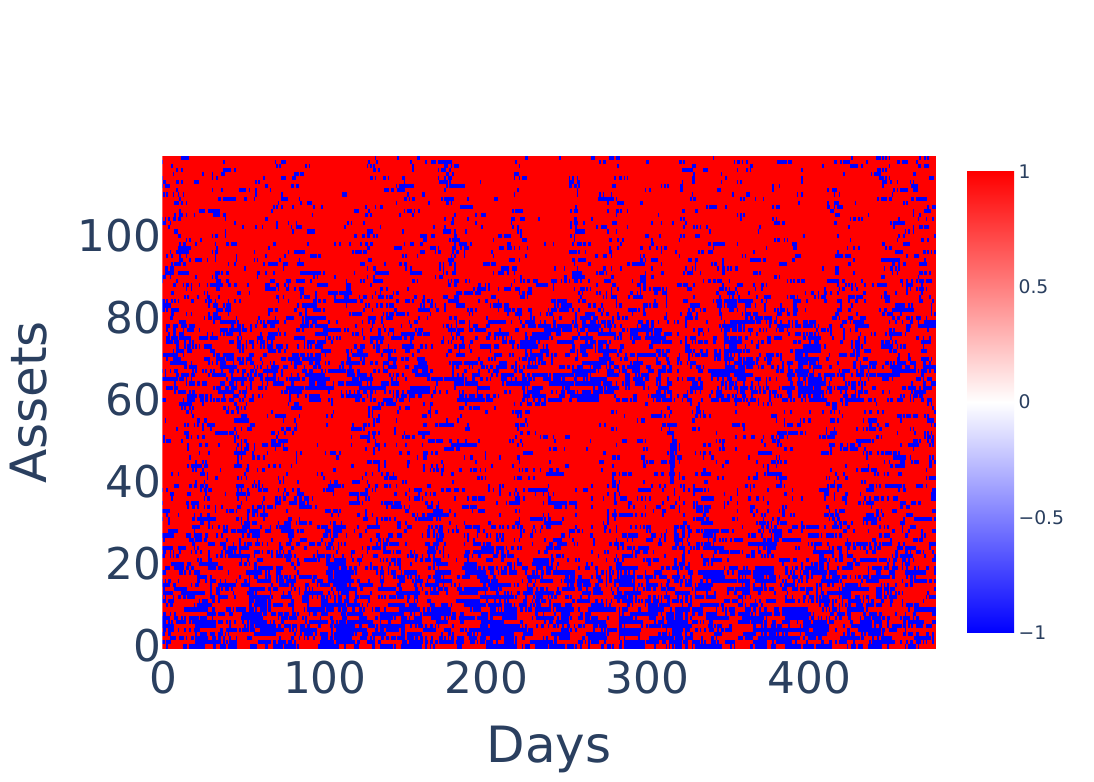}

      \includegraphics[width=\linewidth,trim=0cm 0cm 0cm 2cm,clip]{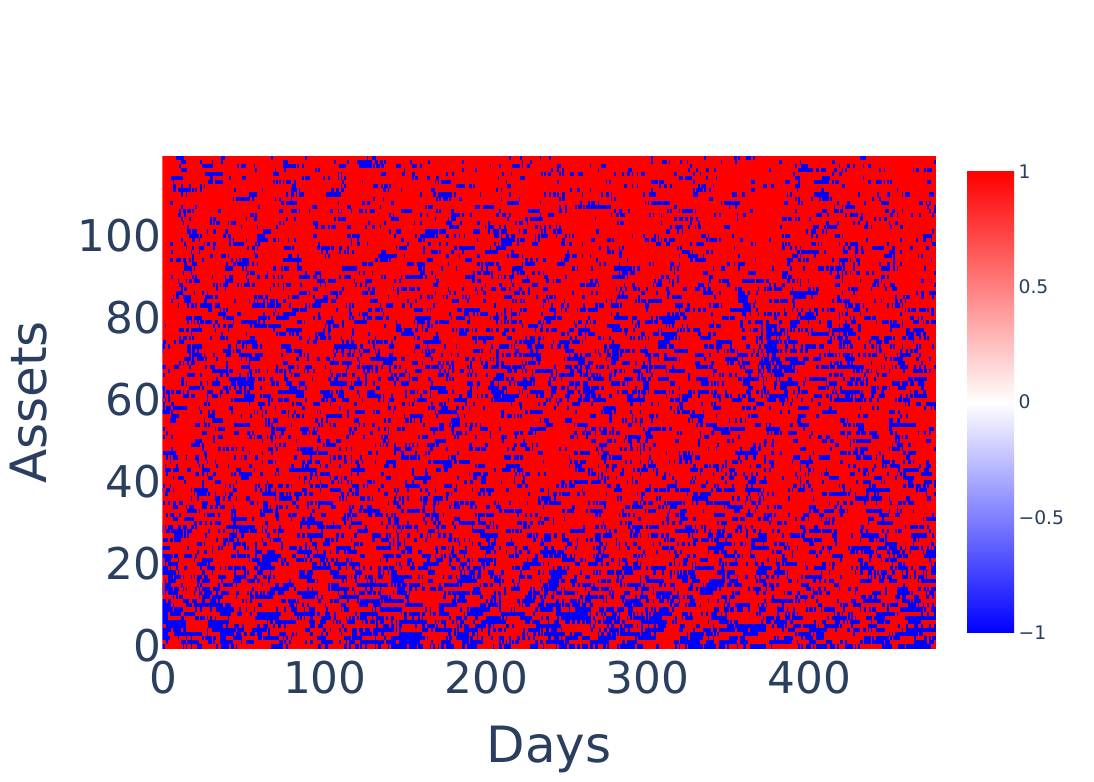}

      \includegraphics[width=\linewidth,trim=0cm 0cm 0cm 2cm,clip]{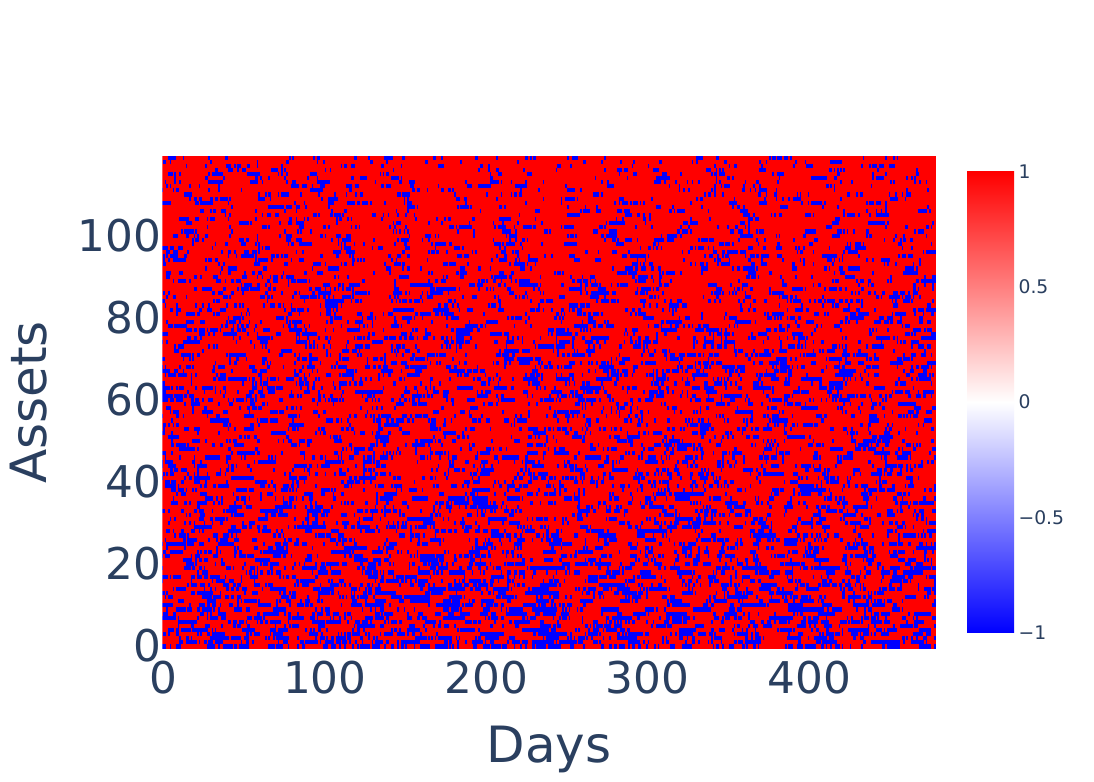}

    \\
        \multicolumn{1}{c}{$k=1$} & \multicolumn{1}{c}{$k=2$} & \multicolumn{1}{c}{$k=1$}  & \multicolumn{1}{c}{$k=2$} \\
    \end{tabular}}

\captionof{figure}{DTW\_KMod\_Mod: Left: $G_\beta$ strategy. Right: $D_\alpha$ strategy. $G_\beta$ are colored in blue, and $D_\alpha$ are colored in red. The synthetic data set has been set with the values $n = 120$, $p = 1$, $\delta = 1$, and $\alpha = 0.25$. From top panel
to bottom panel, low noise $\sigma = 1$,  Medium noise $\sigma = 1.5$, and high noise $\sigma = 2$.}
\label{tab:mod_noise_pnl}
\end{table}

\begin{table}[htbp]
  \centering
  \resizebox{\linewidth}{!}{
    \begin{tabular}{p{4cm}p{4cm}|p{4cm}p{4cm}}
      \multicolumn{2}{c}{\textbf{$G_\beta$ strategy}}&\multicolumn{2}{c}{\textbf{$D_\alpha$ strategy}}\\
      \cmidrule(lr){1-2} \cmidrule(lr){3-4}\\
      \multicolumn{1}{c}{\textbf{Homogeneous Setting}}    &  \multicolumn{1}{c}{\textbf{Heterogeneous Setting}}  &      \multicolumn{1}{c}{\textbf{Homogeneous Setting}}    &  \multicolumn{1}{c}{\textbf{Heterogeneous Setting}}  \\

      \includegraphics[width=\linewidth,trim=0cm 0cm 0cm 2cm,clip]{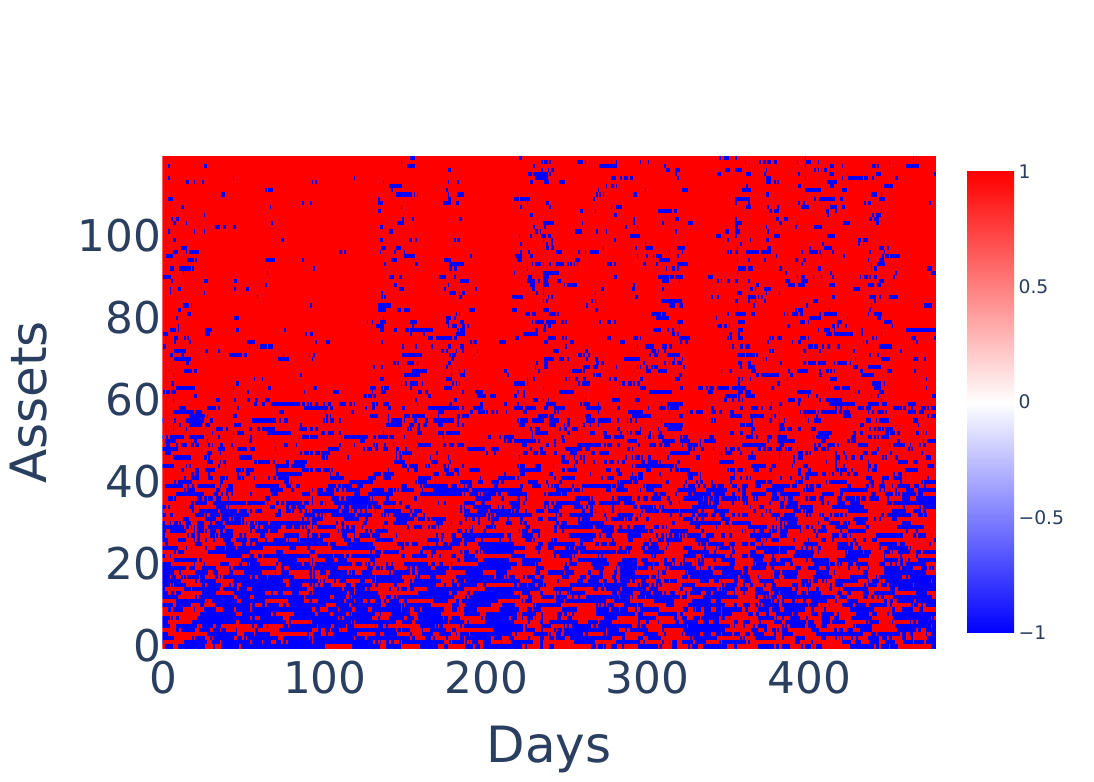}

      \includegraphics[width=\linewidth,trim=0cm 0cm 0cm 2cm,clip]{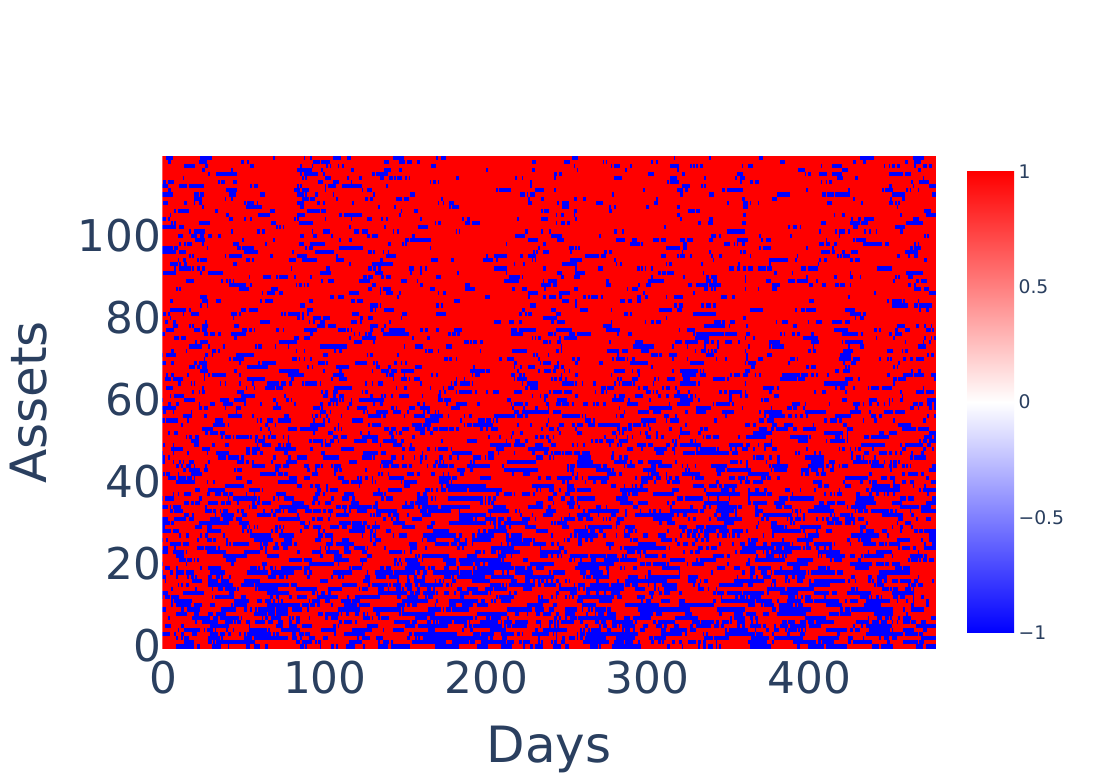}

      \includegraphics[width=\linewidth,trim=0cm 0cm 0cm 2cm,clip]{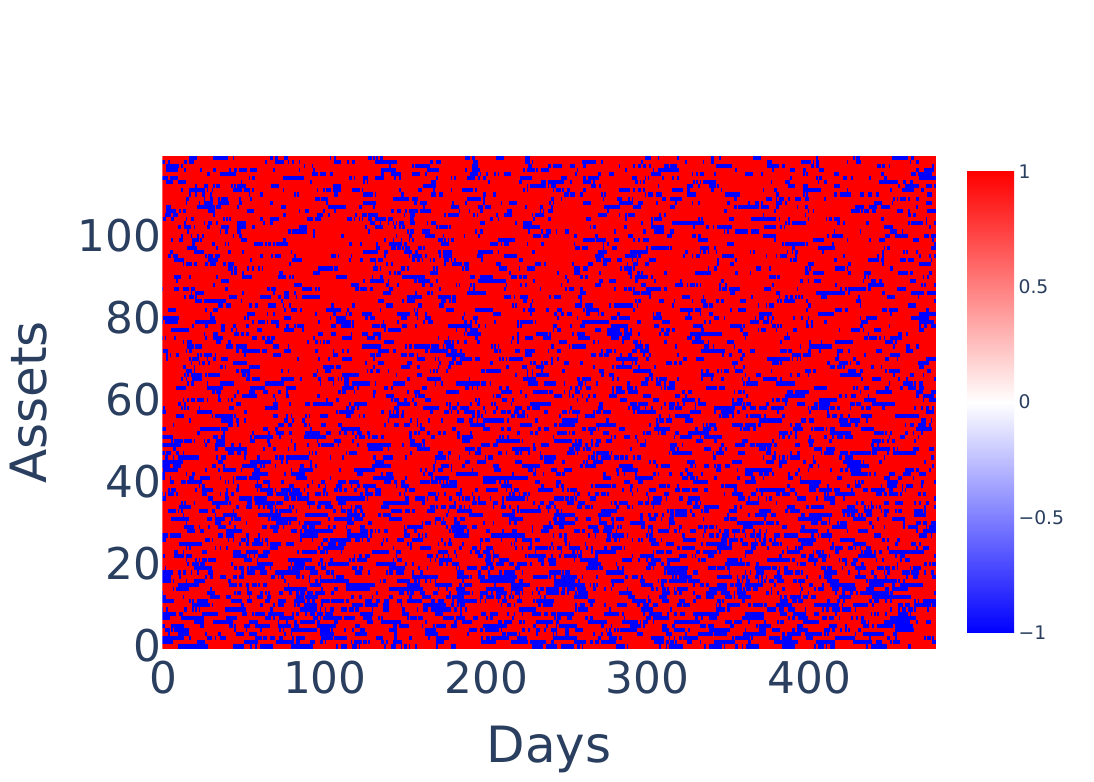}

    &
      \includegraphics[width=\linewidth,trim=0cm 0cm 0cm 2cm,clip]{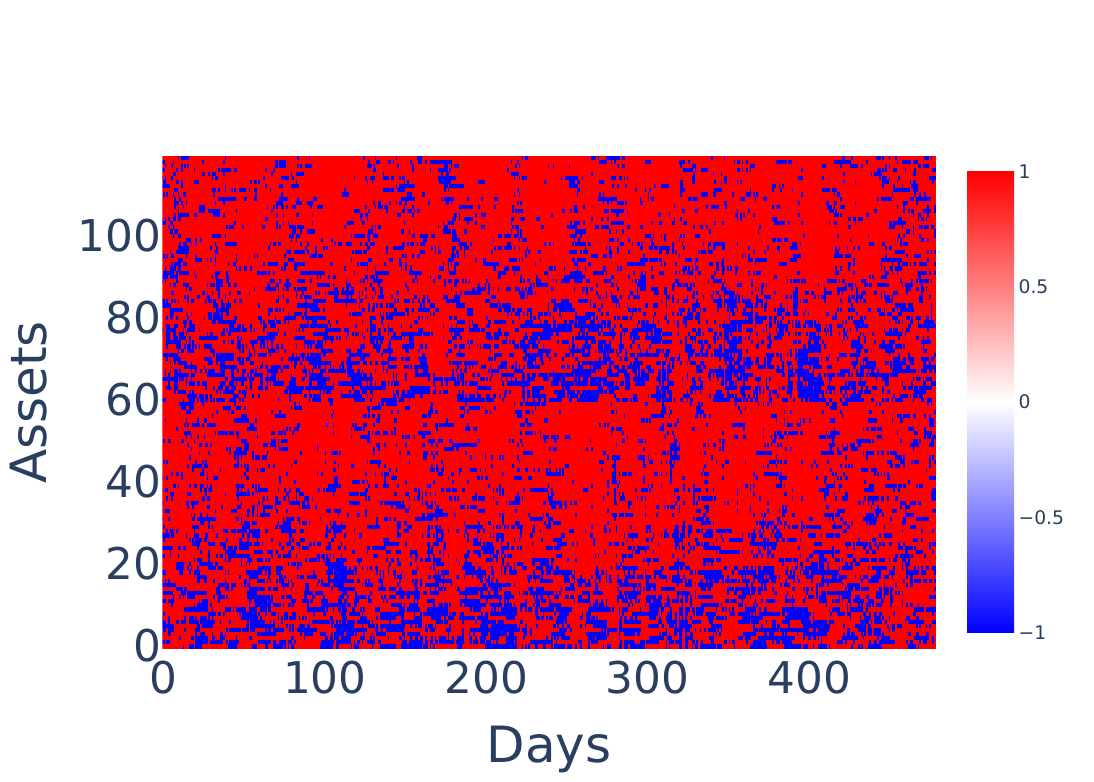}

      \includegraphics[width=\linewidth,trim=0cm 0cm 0cm 2cm,clip]{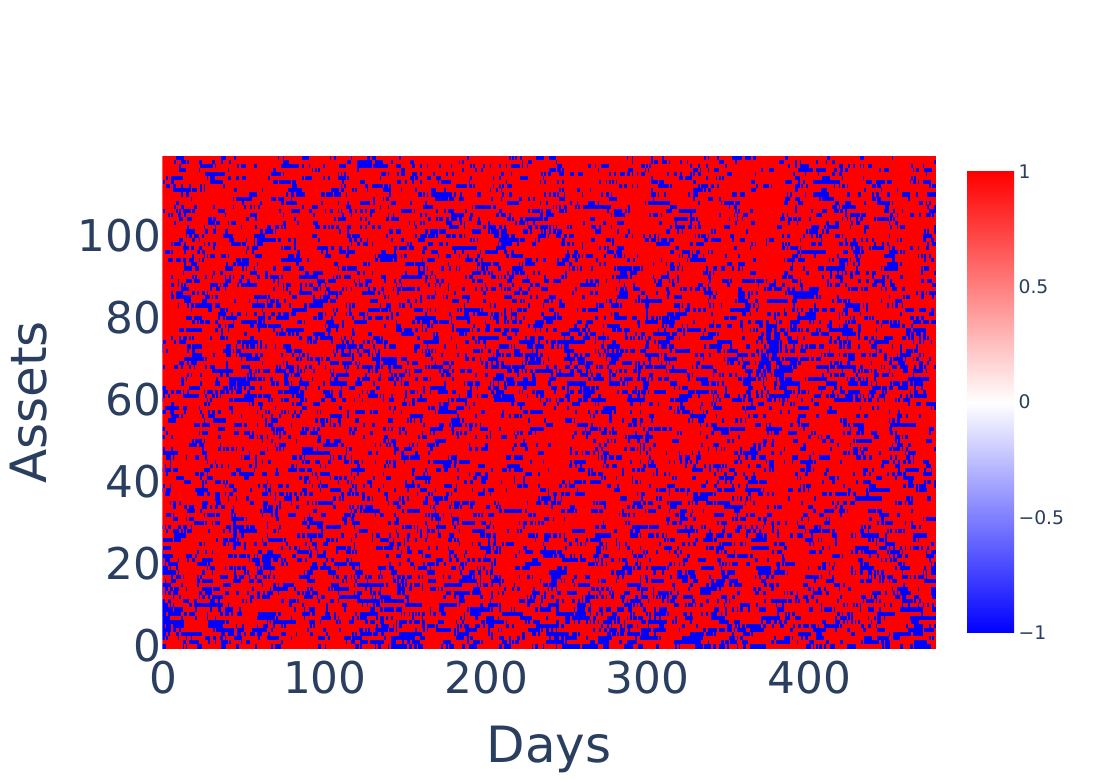}

      \includegraphics[width=\linewidth,trim=0cm 0cm 0cm 2cm,clip]{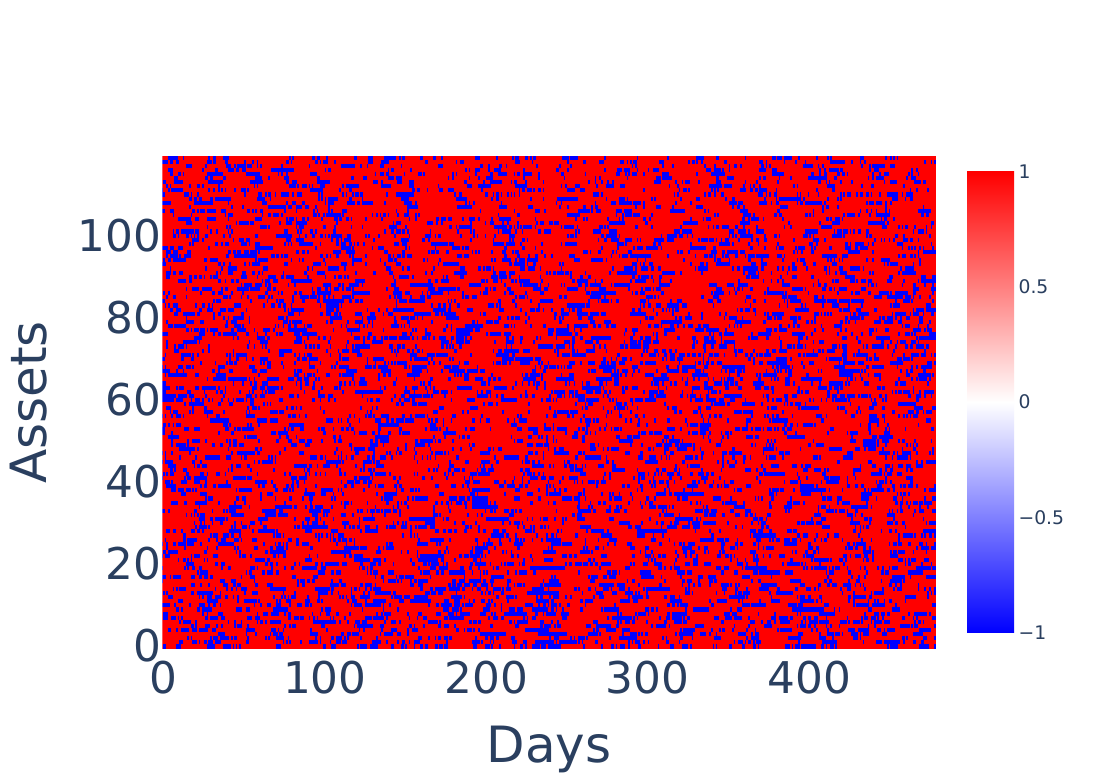}
&
     \includegraphics[width=\linewidth,trim=0cm 0cm 0cm 2cm,clip]{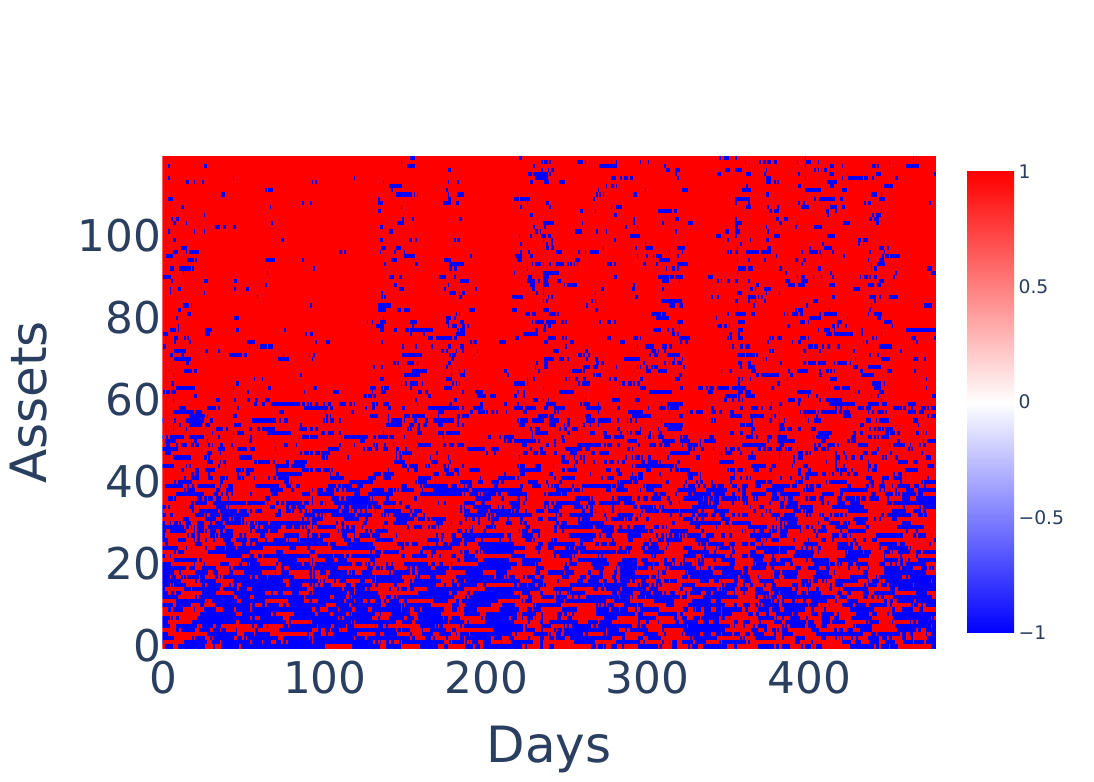}

      \includegraphics[width=\linewidth,trim=0cm 0cm 0cm 2cm,clip]{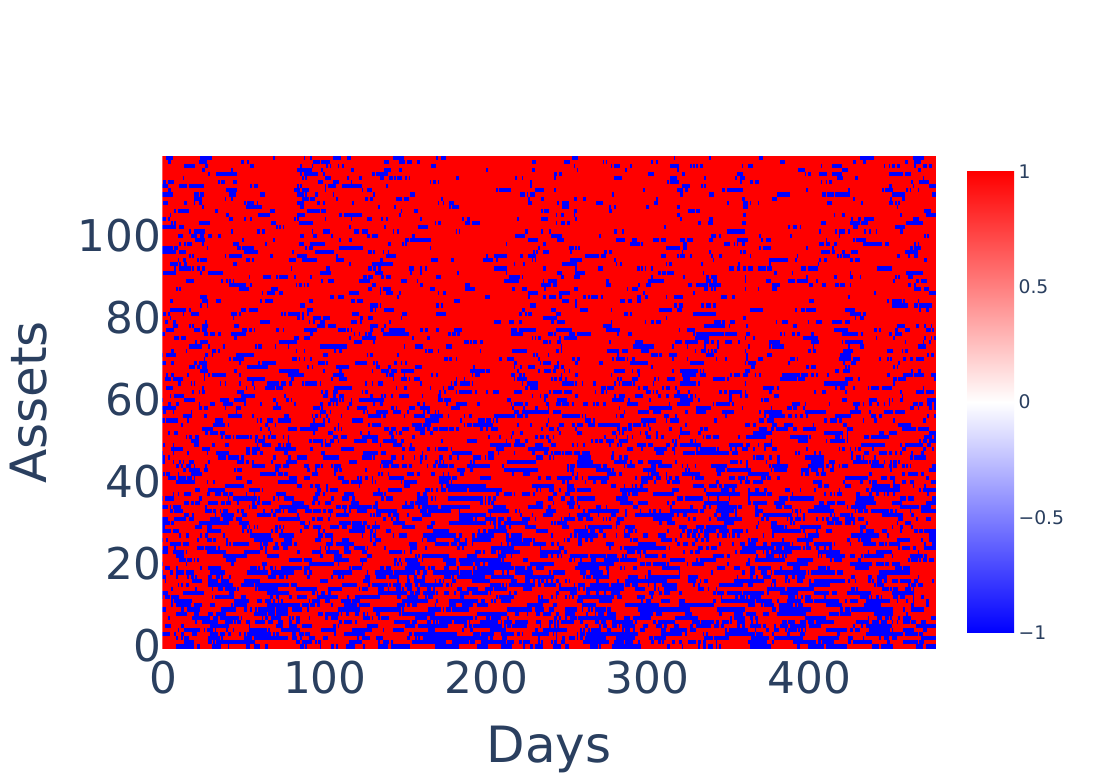}

      \includegraphics[width=\linewidth,trim=0cm 0cm 0cm 2cm,clip]{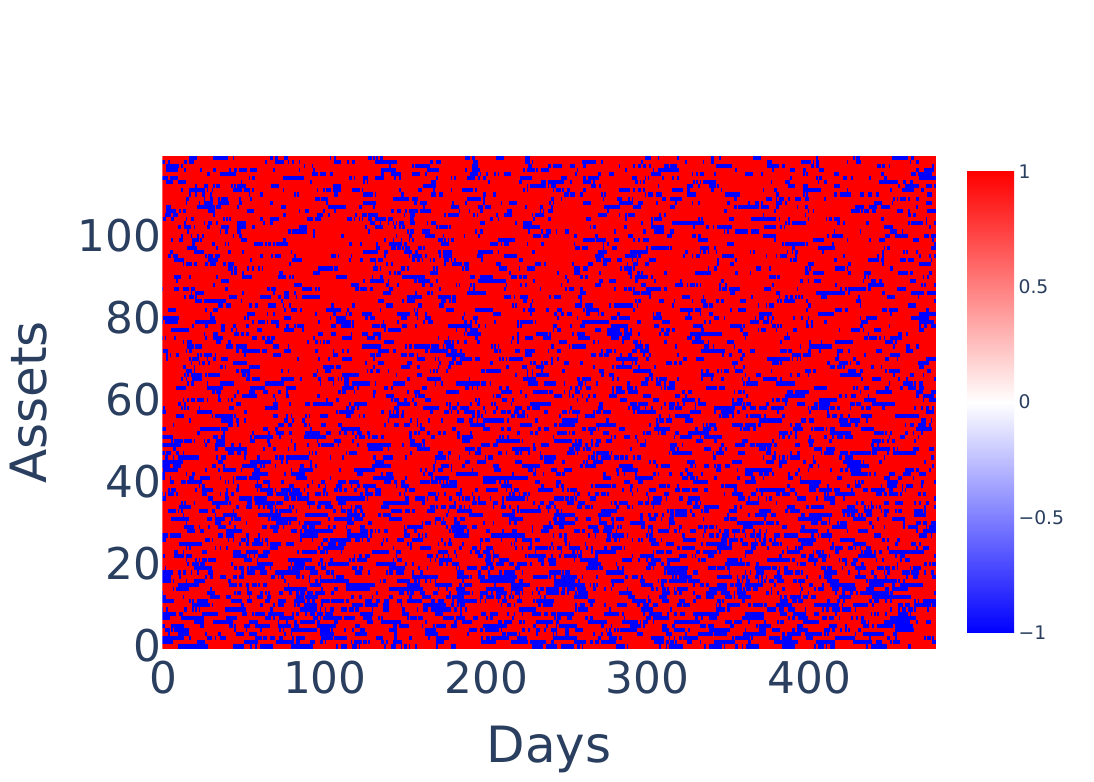}

    &
      \includegraphics[width=\linewidth,trim=0cm 0cm 0cm 2cm,clip]{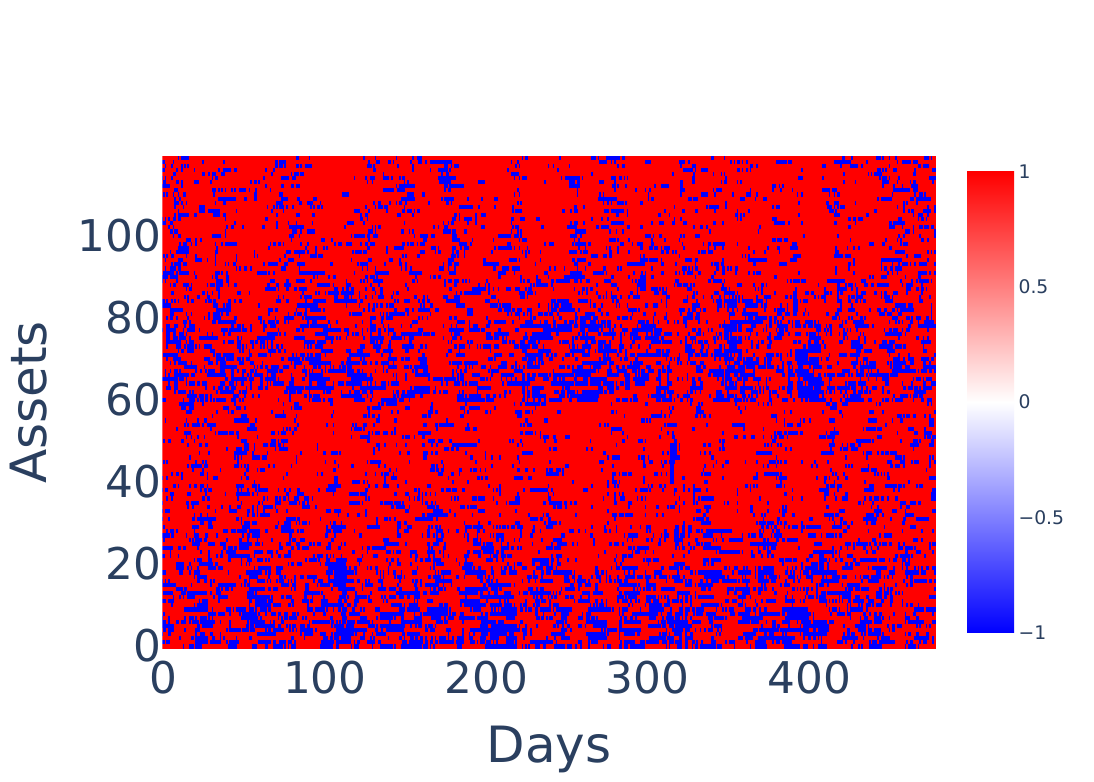}

      \includegraphics[width=\linewidth,trim=0cm 0cm 0cm 2cm,clip]{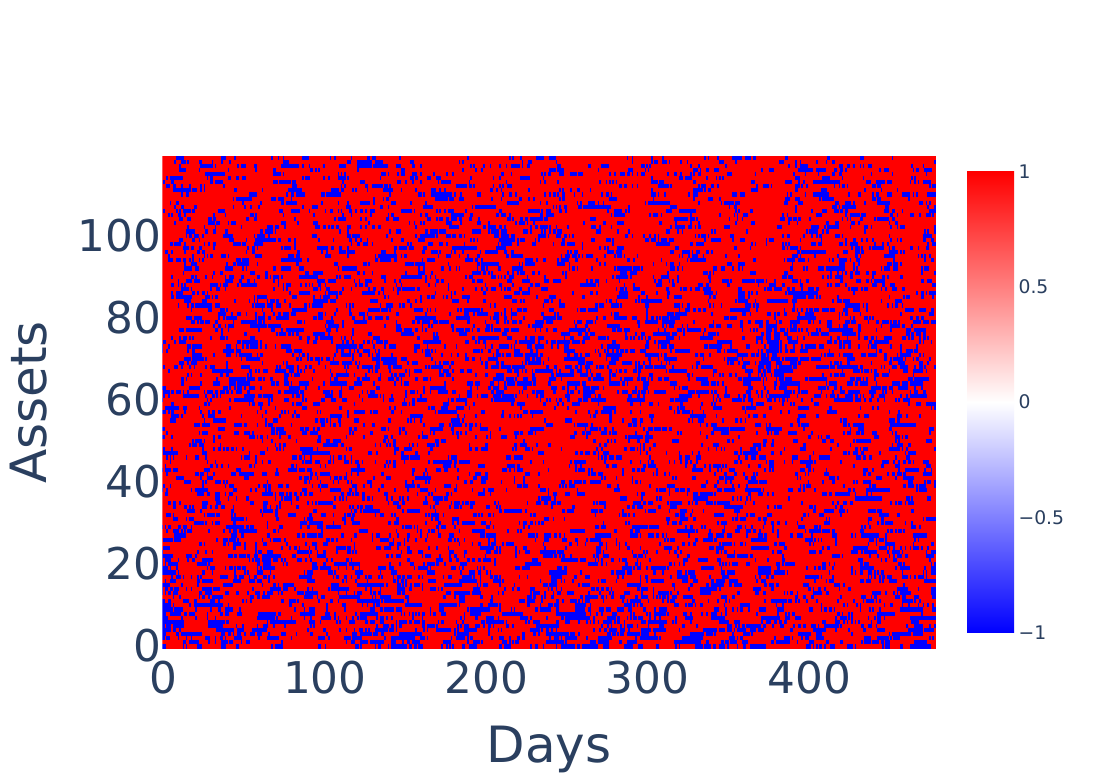}

      \includegraphics[width=\linewidth,trim=0cm 0cm 0cm 2cm,clip]{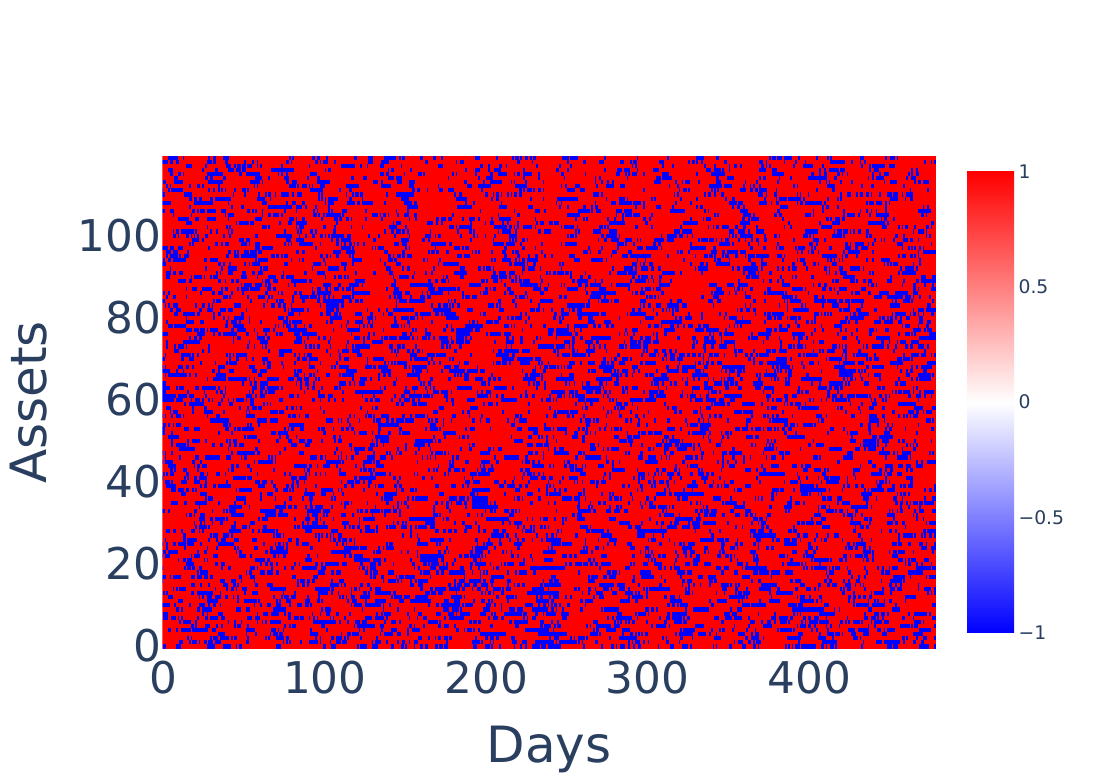}

    \\
           \multicolumn{1}{c}{$k=1$} & \multicolumn{1}{c}{$k=2$} & \multicolumn{1}{c}{$k=1$}  & \multicolumn{1}{c}{$k=2$} \\
    \end{tabular}}

\captionof{figure}{DTW\_KMod\_Med: Left: $G_\beta$ strategy. Right: $D_\alpha$ strategy. $G_\beta$ are colored in blue, and $D_\alpha$ are colored in red. The synthetic data set has been set with the values $n = 120$, $p = 1$, $\delta = 1$, and $\alpha = 0.25$. From top panel
to bottom panel, low noise $\sigma = 1$,  Medium noise $\sigma = 1.5$, and high noise $\sigma = 2$.}
\label{tab:med_noise_pnl}
\end{table}

\newpage
\subsection{Futures data set details}
\label{sec:pinnacle}

Tables [\ref{tab: Grains}, \ref{tab: Meats}, \ref{tab: Foodfibr}, \ref{tab: Metals}, \ref{tab: Indexes}, \ref{tab: Bonds}, \ref{tab: Currency}, \ref{tab: Oils}] show the futures contracts we used and its description from the Pinnacle Data Corp CLC Database. The data is based on ratio-adjusted methods, which removes the contract-to-contract gap, yet it will not go negative as it reduces the size of the price bars if they go lower and increases them if they go higher.

\begin{table}[!htbp]
  \centering
  \caption{Grains}
    \begin{tabular}{lccccc}
        \toprule
          \textbf{Identifier}    & \textbf{Description}  \\
        \midrule
            KW  & WHEAT, KC \\
            MW  & WHEAT, MINN \\
            NR  & ROUGH RICE \\
            W\_  & WHEAT, CBOT \\ 
            ZC  & CORN, Electronic \\
            ZL  & SOYBEAN OIL, Electronic \\
            ZM  & SOYBEAN MEAL, Electronic \\
            ZO  & OATS, Electronic \\ 
            ZR  & ROUGH RICE, Electronic \\
            ZS  & SOYBEANS, Electronic \\
            ZW  & WHEAT, Electronic \\
        \bottomrule
    \end{tabular}
\label{tab: Grains}
\end{table}

\begin{table}[!htbp]
  \centering
  \caption{Meats}
    \begin{tabular}{lccccc}
        \toprule
          \textbf{Identifier}    & \textbf{Description}  \\
        \midrule
            DA  & MILK III, Comp. \\
            ZF  & FEEDER CATTLE, Electronic \\
            ZT  & LIVE CATTLE, Electronic \\
            ZZ  & LEAN HOGS, Electronic \\
        \bottomrule
    \end{tabular}
\label{tab: Meats}
\end{table}

\begin{table}[!htbp]
  \centering
  \caption{Wood fibre}
    \begin{tabular}{lccccc}
        \toprule
          \textbf{Identifier}    & \textbf{Description}  \\
        \midrule
            LB  & LUMBER \\
        \bottomrule
    \end{tabular}

\label{tab: Foodfibr}
\end{table}

\begin{table}[!htbp]
  \centering
  \caption{Metals}
    \begin{tabular}{lccccc}
        \toprule
          \textbf{Identifier}    & \textbf{Description}  \\
        \midrule
            ZG  & GOLD, Electronic \\
            ZI  & SILVER, Electronic \\
            ZP  & PLATINUM, electronic \\
            ZA  & PALLADIUM, electronic \\
            ZK  & COPPER, electronic \\
        \bottomrule
    \end{tabular}
\label{tab: Metals}
\end{table}

\begin{table}[!htbp]
  \centering
  \caption{Indexes}
    \begin{tabular}{lccccc}
        \toprule
          \textbf{Identifier}    & \textbf{Description}  \\
        \midrule
            AX  & GERMAN DAX INDEX \\
            CA  & CAC40 INDEX \\
            DX  & US DOLLAR INDEX \\
            EN  & NASDAQ, MINI \\
            ES  & S \& P 500, MINI \\
            GI  & GOLDMAN SAKS C. I. \\
            LX  & FTSE 100 INDEX \\
            MD  & S \& P 400 (Mini electronic) \\
            NK  & NIKKEI INDEX \\
            SC  & S \& P 500, composite \\
        \bottomrule
    \end{tabular}
\label{tab: Indexes}
\end{table}

\begin{table}[!htbp]
  \centering
  \caption{Bonds}
    \begin{tabular}{lccccc}
        \toprule
         \textbf{Identifier}    & \textbf{Description}  \\
        \midrule
            DT &  EURO BOND (BUND) \\
            FB &  T-NOTE, 5yr composite \\
            GS &  GILT, LONG  BOND \\
            SS & STERLING, SHORT \\
            TY & T-NOTE, 10yr composite \\
            TU & T-NOTES, 2yr composite \\
            US & T-BONDS, composite \\
            UB & EURO BOBL \\
            UZ & EURO SCHATZ \\
        \bottomrule
    \end{tabular}
\label{tab: Bonds}
\end{table}

\begin{table}[!htbp]
  \centering
  \caption{Currency}
    \begin{tabular}{lccccc}
        \toprule
          \textbf{Identifier}    & \textbf{Description}  \\
        \midrule
            AN & AUSTRALIAN \$\$\, composite\\ 
            BN & BRITISH POUND, composite \\
            CN & CANADIAN \$\$\, composite \\
            EC & EURODOLLAR, composite \\
            FN & EURO, composite \\
            JN & JAPANESE YEN, composite \\
            MP & MEXICAN PESO \\
            SN & SWISS FRANC, composite \\
        \bottomrule
    \end{tabular}
\label{tab: Currency}
\end{table}

\begin{table}[!htbp]
  \centering
  \caption{Oils}
    \begin{tabular}{lccccc}
        \toprule
          \textbf{Identifier}    & \textbf{Description}  \\
        \midrule
            ZB  & RBOB, Electronic \\
            ZH  & HEATING OIL, electronic \\
            ZN  & NATURAL GAS, electronic \\
            ZU  & CRUDE OIL, Electronic \\
        \bottomrule
    \end{tabular}
\label{tab: Oils}
\end{table}

\clearpage

%\newpage

%\input{supplementary}

\end{document}